\documentclass{aa}

\usepackage{graphicx}
\usepackage{epsfig}
\usepackage{natbib}
\usepackage{txfonts}

\newcommand{\pun}[1]{\mbox{\rm\,#1}} 

\newcommand{\logg}{\ensuremath{\log g}}

\newcommand{\mlp}{\ensuremath{\alpha_{\mathrm{MLT}}}}

\newcommand{\moh}{\ensuremath{[\mathrm{M/H}]}}

\newcommand{\aofe}{\ensuremath{[\mathrm{\alpha/Fe}]}}

\newcommand{\Teff}{\ensuremath{T_{\mathrm{eff}}}}
\newcommand{\tauross}{\ensuremath{\tau_{\mathrm{Ross}}}}

\newcommand{\beq}{\begin{equation}}
\newcommand{\eeq}{\end{equation}}

\newcommand{\var}[1]{{\ensuremath{\sigma^2_{#1}}}}

\newcommand{\xtmean}[1]{\ensuremath{\left\langle #1\right\rangle}}

\newcommand{\eref}[1]{\mbox{(\ref{#1})}}

\newcommand{\MoH}{\ensuremath{\left[\mathrm{M}/\mathrm{H}\right]}}

\newcommand{\COBOLD}{{\tt CO$^5$BOLD}}
\newcommand{\LHD}{{\tt LHD}}
\newcommand{\PHOENIX}{{\tt PHOENIX}}
\newcommand{\ATLAS}{{\tt ATLAS}}

\newcommand{\MARCS}{{\tt MARCS}}

\newcommand{\LINFOR}{{\tt Linfor3D}}
\newcommand{\STAGGER}{{\tt STAGGER}}

\newcommand{\tex}{\ensuremath{T_\mathrm{e}}}
\newcommand{\tm}{\ensuremath{\overline{T}}}
\newcommand{\delt}{\ensuremath{\Delta T}}
\newcommand{\deltex}{\ensuremath{\Delta\tex}}

\begin{document}

   \title{Three-dimensional hydrodynamical \COBOLD\ model atmospheres of red giant stars\\}

   \subtitle{III. Line formation in the atmospheres of giants located close to the base of RGB}

   \author{V.~Dobrovolskas\inst{1}
           \and
           A.~Ku\v{c}inskas\inst{2}
           \and
           M.~Steffen\inst{3,5}
           \and
           H.-G.~Ludwig\inst{4}
           \and
           D.~Prakapavi\v{c}ius\inst{2}
           \and
           J.~Klevas\inst{2}
           \and\\
           E.~Caffau\inst{4,5}
           \and
           P.~Bonifacio\inst{5}
           }

   \institute{
           Vilnius University Astronomical Observatory, M. K. \v{C}iurlionio 29, Vilnius LT-03100, Lithuania\\
           \email{Vidas.Dobrovolskas@ff.vu.lt}
           \and
           Vilnius University Institute of Theoretical Physics and Astronomy, A. Go\v{s}tauto 12, Vilnius LT-01108, Lithuania\\
           \email{arunas.kucinskas,jonas.klevas,dainius.prakapavicius@tfai.vu.lt}
           \and
           Leibniz-Institut f\"ur Astrophysik Potsdam, An der Sternwarte 16, D-14482 Potsdam, Germany\\
           \email{msteffen@aip.de}
           \and
           Landessternwarte -- Zentrum f\"ur Astronomie der Universit\"at Heidelberg, K\"{o}nigstuhl 12, D-69117 Heidelberg, Germany\\
           \email{hludwig,elcaffau@lsw.uni-heidelberg.de}
           \and
           GEPI, Observatoire de Paris, CNRS, Universit\'{e} Paris Diderot, Place Jules Janssen, 92190 Meudon, France\\
           \email{piercarlo.bonifacio@obspm.fr}
           }

   \date{Received XXX; accepted XXX}

  \abstract
   {}
   {We utilize state-of-the-art three-dimensional (3D) hydrodynamical and classical 1D stellar model atmospheres to study the influence of convection on the formation properties of various atomic and molecular spectral lines in the atmospheres of four red giant stars, located close to the base of the red giant branch, RGB ($\Teff\approx5000$\,K, $\log g=2.5$), and characterized by four different metallicities, $\moh=0.0,-1.0,-2.0,-3.0$.}
   {The role of convection in the spectral line formation is assessed with the aid of abundance corrections, i.e., the differences in abundances predicted for a given equivalent width of a particular spectral line with the 3D and 1D model atmospheres. The 3D hydrodynamical and classical 1D model atmospheres used in this study were calculated with the \COBOLD\ and 1D \LHD\ codes, respectively. Identical atmospheric parameters, chemical composition, equation of state, and opacities were used with both codes, therefore allowing a strictly differential analysis of the line formation properties in the 3D and 1D models.}
   {We find that for lines of certain neutral atoms, such as \ion{Mg}{i}, \ion{Ti}{i}, \ion{Fe}{i}, and \ion{Ni}{i}, the abundance corrections strongly depend both on metallicity of a given model atmosphere and the line excitation potential, $\chi$. While abundance corrections for all lines of both neutral and ionized elements tend to be small at solar metallicity ($\leq\pm0.1$\,dex), for lines of neutral elements with low ionization potential and low-to-intermediate $\chi$ they quickly increase with decreasing metallicity, reaching in their extremes to $-0.6\dots-0.8$\,dex. In all such cases the large abundance corrections are due to horizontal temperature fluctuations in the 3D hydrodynamical models. Lines of neutral elements with higher ionization potentials ($E_{\rm ion}\gtrsim10$\,eV) generally behave very similarly to lines of ionized elements characterized with low ionization potentials ($E_{\rm ion}\lesssim6$\,eV). In the latter case, the abundance corrections are small (generally, $\leq\pm0.1$\,dex) and are caused by approximately equal contributions from the horizontal temperature fluctuations and differences between the temperature profiles in the 3D and 1D model atmospheres. Abundance corrections of molecular lines are very sensitive to  metallicity of the underlying model atmosphere and may be larger (in absolute value) than $\sim-0.5$\,dex at $\moh=-3.0$ ($\sim-1.5$\,dex in the case of CO). At fixed metallicity and excitation potential, the abundance corrections show little variation within the wavelength range studied here, $400-1600$\,nm. We also find that an approximate treatment of scattering in the 3D model calculations (i.e., ignoring the scattering opacity in the outer, optically thin, atmosphere) leads to the abundance corrections that are altered by less than $\sim0.1$\,dex, both for atomic and molecular (CO) lines, with respect to the model where scattering is treated as true absorption throughout the entire atmosphere, with the largest differences for the resonance and low-excitation lines.}
   {}


\authorrunning{Dobrovolskas et al.}
\titlerunning{Three-dimensional hydrodynamical \COBOLD\ model atmospheres of red giant stars III.}

  \maketitle

\section{Introduction}

All low and intermediate mass stars evolve through the red giant stage. Owing to their large numbers, red giants are important tracers of intermediate-age and old stellar populations, and their high luminosities make them accessible for observations both in and beyond the Milky Way. This makes them particularly suitable for use in spectroscopic studies that aim to understand the chemical evolution of their host populations. On the other hand, stars on the red giant branch (RGB) experience mixing episodes during which material with the chemical composition altered via nuclear reactions is brought to the stellar surface. Elemental abundances in the atmospheres of red giant stars may therefore provide a wealth of important information about the nucleosynthesis and mixing processes in stars and about the chemical evolution histories of their host populations.

To obtain information from stellar spectra about the abundances of the chemical species of interest one has to rely on stellar model atmospheres. Inevitably, the confidence with which one can derive chemical composition of stars is determined by the accuracy and physical realism of the underlying model atmospheres. Classical 1D model atmospheres are most widely used for such tasks today. Stellar atmospheres in these models are treated as static and 1D, so that resulting model structures are functions of the radial coordinate alone. Convection in such models is described by using the mixing-length theory \citep[MLT; see, e.g.,][]{BV58,CM91}, along with a number of free parameters, such as the mixing-length parameter, \mlp, and microturbulence velocity. Thanks to their static 1D nature and various simplifications involved, such models are fast to compute, which allowed large model grids to be produced using a variety of 1D model atmosphere codes, such as \ATLAS\ \citep[][]{CK03}, \MARCS\ \citep[][]{GEK08}, and \PHOENIX\ \citep[][]{BH05}, for various combinations of stellar atmosphere parameters. Despite numerous simplifications, 1D model atmospheres have aided in gaining enormous amounts of information about the processes taking place in stellar atmospheres and deeper interiors.

Since stellar atmospheres are neither static nor 1D, a large increase in computational power is required  to include more realistic 3D time-dependent physics into stellar atmosphere modeling. Mostly because of such computational complexities the advance of 3D hydrodynamical model atmospheres and their applications in stellar abundance work has been relatively slow. Despite this, there has been a notable increase in the applications of the 3D hydrodynamical model atmospheres to study stars and stellar populations during the recent years \citep[see, e.g.,][]{ANT00,CLS08,GHB09,BBL10}. This is partly related to the fact that a number of studies have shown that differences in the elemental abundances inferred from the same spectral line strength with the classical 1D and 3D hydrodynamical model atmospheres may be significant, especially at lower metallicities where they may reach $-1.0$\,dex \citep[see, e.g.,][]{CAT07,DKL10,IKL10}. Obviously, such differences in the obtained abundances may lead to important implications, such as when reconstructing most likely scenarios of chemical evolution of stellar populations in the Galaxy and beyond.

To take a closer look at the role of convection in the spectral line formation in the atmospheres of red giant stars, we started a systematic study of the internal structures and observable properties of red giant stars by using 3D hydrodynamical \COBOLD\ model atmospheres for this purpose. In the first two papers in this series \citep[][Papers~I and II hereafter]{LK12,KSL13}, we investigated the role of convection on the atmospheric structures and spectral line formation in a red giant star located close to the RGB tip ($\Teff\approx3660$\,K, $\log g=1.0$, $\moh=0.0$). In this study we extend our previous work and focus on the properties of spectral line formation in somewhat warmer red giants at several different metallicities ($\moh=0.0,-1.0,-2.0,-3.0$) located close to the RGB base ($\Teff\approx5000$\,K, $\log g=2.5$).

The paper is structured as follows. In Sect.~2 we describe the 3D hydrodynamical and classical 1D model atmospheres used in this study, and outline the details of spectral line synthesis calculations. In this section we also introduce the concept of abundance corrections, which is used throughout the paper to assess the differences in the abundances predicted for the same line strength with the 3D and 1D model atmospheres. The abundance corrections obtained for various neutral and ionized atoms and for molecules are discussed in Sect.~3, together with the role of scattering and the importance of the mixing-length parameter, \mlp, used with the 1D model atmospheres. Finally, summary and conclusions are provided in Sect.~4.

\section{Stellar atmosphere models and spectral line synthesis calculations}

In this study we utilized the \COBOLD\ and \LHD\ model atmosphere codes to calculate, respectively, the 3D hydrodynamical and classical 1D atmosphere models of red giants located close to the RGB base. Both \COBOLD\ and \LHD\ simulations shared identical atmospheric parameters, chemical composition, opacities, and equation of state, in order to limit the discrepancies in the model predictions to  differences in the underlying model physics. Such differential comparison of the model predictions allows us to assess the importance of 3D hydrodynamical effects on the spectral line formation, and enables us to determine the differences between the abundances of different chemical elements derived with the 3D and 1D model atmospheres. In what follows below we briefly describe the details of model atmosphere and spectral line synthesis calculations.

\subsection{3D hydrodynamical \COBOLD\ stellar model atmospheres\label{sect:3D_models}}

The 3D hydrodynamical models used in this study are part of the CIFIST model atmosphere grid which will eventually cover stars on the main sequence, subgiant, and red giant branches \citep{LCS09}. \COBOLD\ solves the coupled nonlinear equations of compressible hydrodynamics in an external gravity field, together with radiative transfer equation \citep[][]{FSL12}. The model atmospheres were computed using the ``box-in-a-star'' setup, i.e., the part of the stellar atmosphere modeled was small compared to the size of a star itself. The simulations were performed on a Cartesian grid of $160\times160\times200$ grid points in $x,y,z$ direction, respectively. The model box had open upper and lower boundaries (matter was allowed to enter and leave the simulation box freely), and periodic boundaries in the horizontal direction (matter leaving the box on one side was allowed to enter it again from the opposite side). We used monochromatic opacities from the \MARCS\ stellar atmosphere package \citep{GEK08} which we grouped into a smaller number of opacity bins using the opacity binning technique \citep{nordlund82,ludwig92,LJS94,vogler04,VBS04}, with five opacity bins for the $\MoH=0.0$ model and six bins for the $\MoH=-1.0$, $-2.0$, and $-3.0$ models. Solar-scaled elemental abundances used were those from \citet{AGS05}. It is important to stress that we also applied a constant enhancement in the alpha-element abundances of $\aofe=+0.4$ for the models at metallicities $\moh\leq-1.0$. All model simulations were performed under the assumption of local thermodynamic equilibrium, LTE. Scattering was treated as true absorption and magnetic fields were neglected.

\begin{table}[tb]
\caption{Parameters of the \COBOLD\ red giant models used in this work.}
\label{table:atm-par}
\centering
\begin{tabular}{ccccc}
\hline
$\langle\Teff\rangle$,  & $\log g$, & \moh & Grid dimension, & Resolution,   \\
    K                   &   CGS     &      &     Mm          & grid points   \\
\hline
 4970 & 2.5 &   0    & 573\(\times\)573\(\times\)243 & 160\(\times\)160\(\times\)200\\
 4990 & 2.5 & \(-1\) & 573\(\times\)573\(\times\)245 & 160\(\times\)160\(\times\)200\\
 5020 & 2.5 & \(-2\) & 584\(\times\)584\(\times\)245 & 160\(\times\)160\(\times\)200\\
 5020 & 2.5 & \(-3\) & 573\(\times\)573\(\times\)245 & 160\(\times\)160\(\times\)200\\
\hline
\end{tabular}
\end{table}

Parameters of the individual 3D model atmospheres used in this work are provided in Table~\ref{table:atm-par}. One may notice that each model is characterized by a slightly different average effective temperature, $\xtmean{\mbox{\Teff}}$. This is because effective temperature is not the input parameter for calculating 3D model atmosphere with the \COBOLD\ code. Instead, one sets the value of entropy of inflowing gas at the bottom of the model atmosphere. This determines radiative flux at the outer boundary and thus - the effective temperature of a given model. Another important aspect is that radiative flux leaving the model atmosphere is subject to random spatial and temporal fluctuations due to stochastic nature of convection, which causes the effective temperature to fluctuate too. Therefore, the effective temperature of the 3D hydrodynamical model atmosphere can not be set in advance and should be fine-tuned by adjusting the entropy inflow at the lower boundary of the model atmosphere. Note, however, that differences between the average \Teff\ of individual 3D model atmospheres are small ($<30$\,K) while the effective temperatures of all models are very close to the target value of $\Teff=5000$\,K (Table~\ref{table:atm-par}).

Since 3D spectral line synthesis calculations are very computationally demanding, we selected a smaller subset of 3D atmospheric structures at different instants of time (snapshots) to be used in all 3D line synthesis work. At all metallicities, these subsets consisted of 20 snapshots spaced with a step of $\sim7$\,hours over the interval of $\sim4.7-6.5$\,days in stellar time. Since the convective time scales in the red giant models studied here are equal to $\sim5-15$\,hours (as measured by the Brunt-Vais\"{a}l\"{a} time scale), such spacing of the 3D snapshots allows to consider them statistically independent. Snapshots of the 3D model were selected so that statistical properties of the 20-snapshot ensemble would best reproduce the corresponding properties of the entire model run, such as the average effective temperature and its standard deviation, mean velocity at optical depth unity, mean velocity profile and mean mass velocity profile.

In this study we also used average $\xtmean{\mbox{3D}}$ models. At each metallicity, these model atmospheres were computed by horizontally averaging each 3D atmospheric structure in the subset of twenty 3D model snapshots selected for the line synthesis calculations, as described above (the fourth power of temperature was averaged on surfaces of equal optical depth). The fourth power of temperature was chosen over the first because in the former case one may expect to retain better representation of the radiative flux throughout the model atmosphere. On the other hand, we found that differences in the $\xtmean{\mbox{3D}}$ abundances obtained with the models computed by averaging the fourth and first powers of temperature were always small (for example, in case of \ion{Fe}{i} and \ion{Fe}{ii} lines with $\chi=0-6$\,eV these differences were typically significantly below 0.05\,dex). We refer the reader to Appendix~\ref{sect:AA} where we provide some further thoughts regarding the choice of temperature averaging scheme. Every $\xtmean{\mbox{3D}}$ model is a 1D structure, hence the average $\xtmean{\mbox{3D}}$ models do not contain information about the horizontal fluctuations of thermodynamic quantities and velocity fields. Therefore, they can be used to estimate the role of horizontal  fluctuations on, e.g., the spectral line formation, which can be done by comparing the predictions of the 3D and $\xtmean{\mbox{3D}}$ models.

\subsection{1D \LHD\ model atmospheres\label{sect:1D_models}}

\begin{figure}[tb]
\centering
\includegraphics[width=9cm]{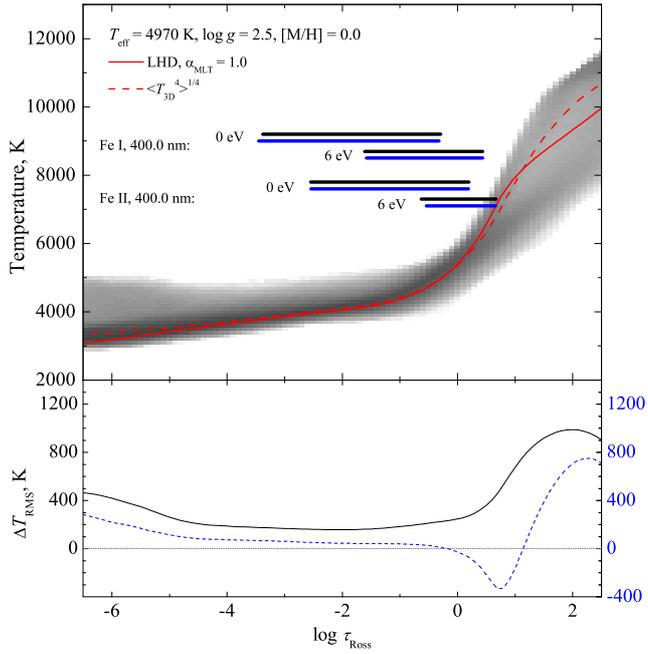}
   \caption{\textbf{Upper panel}: temperature profiles in the red giant model with $\Teff/\log g/\moh=4970/2.5/0.0$, plotted versus the Rosseland optical depth, \tauross, and shown for the following model atmospheres: 3D (density plot), $\xtmean{\mbox{3D}}$ (dashed line), and 1D (solid line). Horizontal bars mark the approximate location of the Fe I and Fe II line formation regions in the 3D (black) and 1D (blue) atmosphere models, at $\lambda=400$\,nm and $\chi=0$ and 6\,eV (bars mark the regions where the equivalent width, $W$, of a given spectral line grows from 5\% to 95\% of its final value). \textbf{Lower panel}: RMS horizontal temperature fluctuations in the 3D model (solid line), and difference between the temperature profiles of the $\xtmean{\mbox{3D}}$ and 1D models (dashed line), shown as functions of the Rosseland optical depth. In both panels, all quantities related to the 3D and $\xtmean{\mbox{3D}}$ models were obtained using the subset of twenty 3D model snapshots utilized in the 3D spectral line synthesis calculations (see Sect.~\ref{sect:3D_models}).
   }
   \label{fig:Ttaumm00}
\end{figure}

\begin{figure}[tb]
\centering
\includegraphics[width=9cm]{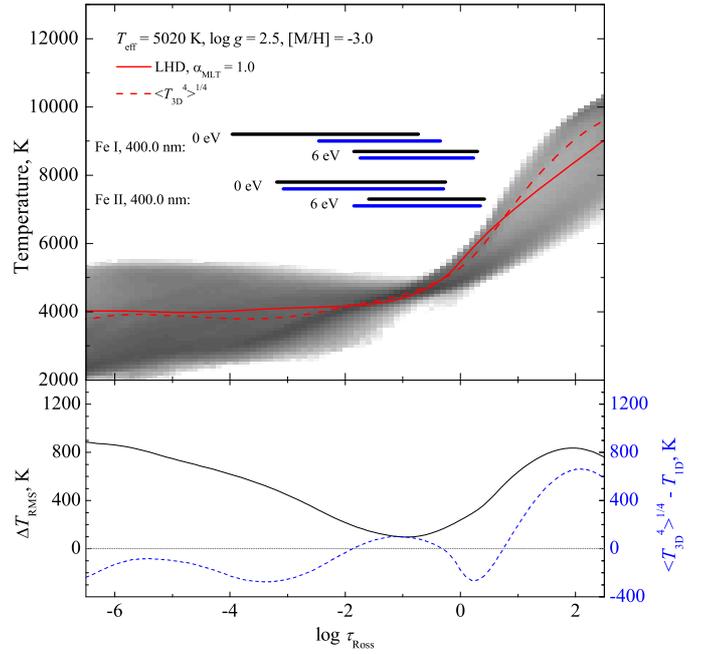}
   \caption{Same as in Fig.~\ref{fig:Ttaumm00} but for the red giant model with $\Teff/\log g/\moh=5020/2.5/-3.0$.
   }
   \label{fig:Ttaumm30}
\end{figure}

1D models were calculated using the \LHD\ code \citep[see, e.g.,][]{CLS08}. We recall that both 3D and 1D models used in this work shared identical atmospheric parameters (Table~\ref{table:atm-par}), chemical composition, equation of state, and opacities (see Sect.~\ref{sect:3D_models} for details). Convection in the \LHD\ models was treated according to the formalism of the mixing-length theory by \citet{M78}. The mixing-length parameter used throughout the \LHD\ simulations was $\mlp=1.0$. In principle, the use of 1D models calculated with different mixing-length parameters results in different abundance corrections. However, in the case of the giants studied here it is only relevant for high-excitation atomic lines (see Sect.~\ref{sect:alpha-mlp}). Temperature stratifications of the 3D and 1D models at [M/H] = 0.0 and --3.0 are shown in Figs.~\ref{fig:Ttaumm00} and \ref{fig:Ttaumm30}, respectively.

\subsection{Spectral line synthesis\label{sect:line_synth}}

In order to investigate the influence of convection on the spectral line formation we utilized fictitious spectral lines. For these lines, their central wavelengths, excitation potentials, $\chi$, and oscillator strengths, $\log gf$, were freely chosen to cover the range of values of existing spectral lines. Such approach allows us to study the connections between the atomic line parameters and line formation properties in the presence of realistically modeled convection, and has been already applied successfully in a number of studies \citep[e.g.,][Paper~II]{SH03,CAT07,DKL10,IKL10}.

In this study, equivalent widths of synthetic spectral lines were fixed to $W \leq 0.5$\,pm. Because these lines are very weak, they should be confined to the linear part of the curve of growth and thus their strengths should be independent of the choice of microturbulence velocity, $\xi_{\rm mic}$, used in the spectral line synthesis with the $\xtmean{\mbox{3D}}$ and 1D model atmospheres. In such situation, the differences in the line strengths predicted by the 3D and 1D model atmospheres can be attributed to purely 3D hydrodynamical effects, allowing investigation of their importance in the spectral line formation.

Spectral line synthesis calculations for all chemical elements studied in this work were done using the \LINFOR\ \footnote{http://www.aip.de/\textasciitilde mst/linfor3D\_main.html} code, which solves the radiative transfer equation under the assumption of LTE. In order to speed up spectral line synthesis calculations, we used only every third point of the original 3D model structure along the $x$ and $y$ spatial directions. We verified that such simplification has left the resulting elemental abundances essentially unchanged (i.e., with respect to the calculations performed using the full model box). The differences were always below 0.003\,dex in the case of \MoH=$-3$ metallicity model and \ion{Fe}{i} lines. In our simulations the radiative transfer was solved for 3 vertical and 4 azimuthal directions. Line synthesis calculations covered the Rosseland optical depth range of $\log \tau_{\rm Ross} = -6\dots2$, in steps of $\Delta\log\tau_{\rm Ross}=0.08$. Line profiles were computed with a typical wavelength resolution of $130-150$ points per profile. In case of the 3D model atmospheres, spectral line profiles of every chemical element studied here were calculated for each of the twenty 3D model snapshots, at each of the four metallicities. The obtained individual spectral line profiles were then co-added to obtain the final 3D line profile at each metallicity. Similarly, in case of the average 3D models line profiles were calculated for every horizontally averaged 3D model structure from the 20-snapshot ensemble; the final $\xtmean{\mbox{3D}}$ line profile was then obtained by co-adding all twenty individual line profiles. All 1D line profiles were also calculated using the \LINFOR\ code. A microturbulence velocity of $\xi_{\rm mic}=1.0$\,km/s was used in all $\xtmean{\mbox{3D}}$ and 1D spectral synthesis calculations, although we note again that the choice of $\xi_{\rm mic}$ does not change the line strength of very weak fictitious lines studied here.

One may envision that, alternatively, all 3D model snapshots could be averaged to produce a single $\xtmean{\mbox{3D}}$ model for the synthesis of $\xtmean{\mbox{3D}}$ spectral lines, i.e., instead of using individual averaged 3D models for each 3D snapshot as we did in this study. Our test calculations showed, however, that in case of \ion{Fe}{i} and \ion{Fe}{ii} lines with $\chi=0-6$\,eV differences in the line profiles computed using these two different approaches would lead to differences in the elemental abundances of less than $0.03$\,dex (in case of \ion{Fe}{i} lines they were always smaller than 0.01\,dex). Therefore, for the purposes of present study the choice in the procedure used to compute the $\xtmean{\mbox{3D}}$ line profiles is of minor importance.

\subsection{3D--1D abundance corrections\label{sect:abund_corr}}

The role of 3D hydrodynamical effects in spectral line formation was studied by utilizing 3D--1D abundance corrections. The 3D--1D abundance correction, $\Delta_{\rm 3D-1D}$, is the difference in the abundance, $A({\rm X_{i}})$, of a chemical element $\rm X_{i}$ obtained at a given equivalent width from the 3D and 1D curves of growth, $\Delta_{\rm 3D-1D}\equiv A({\rm X_{i}})_{\rm 3D} - A({\rm X_{i}})_{\rm 1D}$ \citep[see, e.g.,][]{CLS11}. The 3D--1D abundance correction is sensitive to the effects related to both the horizontal temperature fluctuations and differences in temperature stratifications of the 3D and 1D model atmospheres. The size of these effects can be measured separately, by utilizing information content of the $\xtmean{\mbox{3D}}$ models. Since the $\xtmean{\mbox{3D}}$ model does not contain information about the horizontal fluctuations of thermodynamic quantities and velocity fields, the $\Delta_{\rm 3D-\langle3D\rangle}\equiv A({\rm X_{i}})_{\rm 3D}-A({\rm X_{i}})_{\rm \langle3D\rangle}$ correction can be used to measure the effect of horizontal temperature fluctuations. Similarly, the $\Delta_{\rm \langle3D\rangle-1D}\equiv A({\rm X_{i}})_{\rm \langle3D\rangle}-A({\rm  X_{i}})_{\rm 1D}$ correction provides information about the role of differences between the temperature profiles of the average $\xtmean{\mbox{3D}}$ and 1D models. Obviously, the final abundance correction is a sum of the two constituents, $\Delta_{\rm 3D-1D}\equiv\Delta_{\rm 3D-\langle3D\rangle}+\Delta_{\rm \langle3D\rangle-1D}$.

3D--1D abundance corrections were calculated for a number of neutral and singly ionized elements: \ion{Li}{i}, \ion{C}{i}, \ion{O}{i}, \ion{Na}{i}, \ion{Mg}{i}, \ion{Al}{i}, \ion{Si}{i}, \ion{Si}{ii}, \ion{K}{i}, \ion{Ca}{i}, \ion{Ca}{ii}, \ion{Ti}{i}, \ion{Ti}{ii}, \ion{Fe}{i}, \ion{Fe}{ii}, \ion{Ni}{i}, \ion{Zn}{i}, \ion{Zr}{i}, \ion{Zr}{ii}, \ion{Ba}{ii}, and \ion{Eu}{ii}. These elements allow to sample different nucleosynthetic channels: proton capture (Na, Al, K), triple-$\alpha$ (carbon), $\alpha$-chain captures (O, Mg, Si, Ca, Ti), nuclear statistical equilibrium (Ti, Fe, Ni, Zn), slow neutron capture (Zn, Zr, Ba), rapid neutron capture (Ba, Eu). Lithium is a special element in this context, in a sense that it was produced during the Big-Bang nucleosynthesis and by several other processes afterwards, including cosmic ray spallation in the interstellar medium and several stellar processes. We also obtained abundance corrections for several molecules: CH, CO, C$_2$, NH, CN, and OH which are important tracers of CNO elements.

The 3D--1D abundance corrections for neutral and ionized atoms were calculated at three wavelengths, $\lambda=400$, 850 and 1600\,nm. The two shorter wavelengths were chosen to bracket the typical blue and red limits of modern optical spectrographs, such as UVES/GIRAFFE@VLT, HIRES@Keck, HDS@SUBARU, while the longer approximately coincides with the $H$-band of near-infrared spectrographs, such as CRIRES@VLT, NIRSPEC@Keck. The two longest wavelengths also coincide with the maximum and minimum absorption of the $\rm H^{-}$ ion at $\sim850$ and $\sim1640$\,nm, respectively, while the bluest marks the spectral region where $\rm H^{-}$ absorption becomes progressively smaller and opacities from various metals become increasingly more important. In case of the molecules, we used the wavelengths of real molecular lines/bands in the blue part of the spectrum (see Paper~II for detailed description of the wavelength selection). In case when real molecular bands were not available at these short wavelengths, abundance corrections were obtained at $\lambda=400$\,nm instead. The abundance corrections were computed at four excitation potentials, $\chi=0$, 2, 4, and 6\,eV, for both neutral and ionized atoms, while for the molecules they were obtained at $\chi=0$, 1, 2, and 3\,eV.

\onlfig{
\begin{figure*}[tb]
\centering
\includegraphics[width=12.9cm]{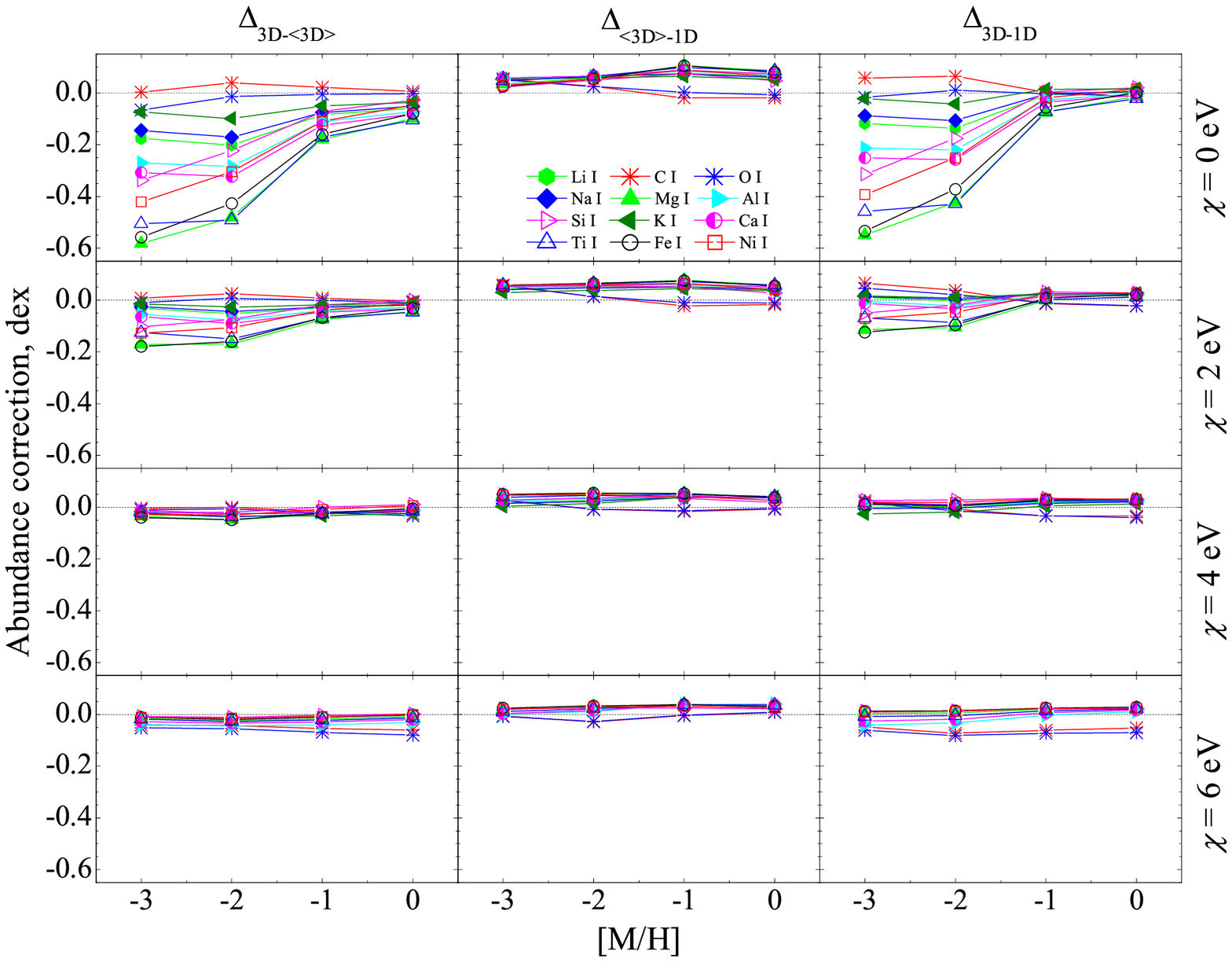}
   \caption{The same as in Fig.~\ref{fig:dabuAtomw850} but at $\lambda=400$\,nm.
   }
   \label{fig:dabuAtomw400}
\end{figure*}
}

Finally, we note that some species studied here have either only a few observable spectral lines or their lines may be simply too weak to be detected (as it is the case with, e.g., \ion{Li}{i}, \ion{C}{i}, \ion{O}{i}). We should stress that such elements were included in the present analysis because we aimed to understand the general properties of spectral line formation in the presence of convection, especially the trends of abundance corrections in a wider range of atomic line parameters. Obviously, in all such cases the exact values of abundance corrections in a wide atomic parameter range may only be of academic interest.

\begin{figure*}[tb]
\centering
\includegraphics[width=12.9cm]{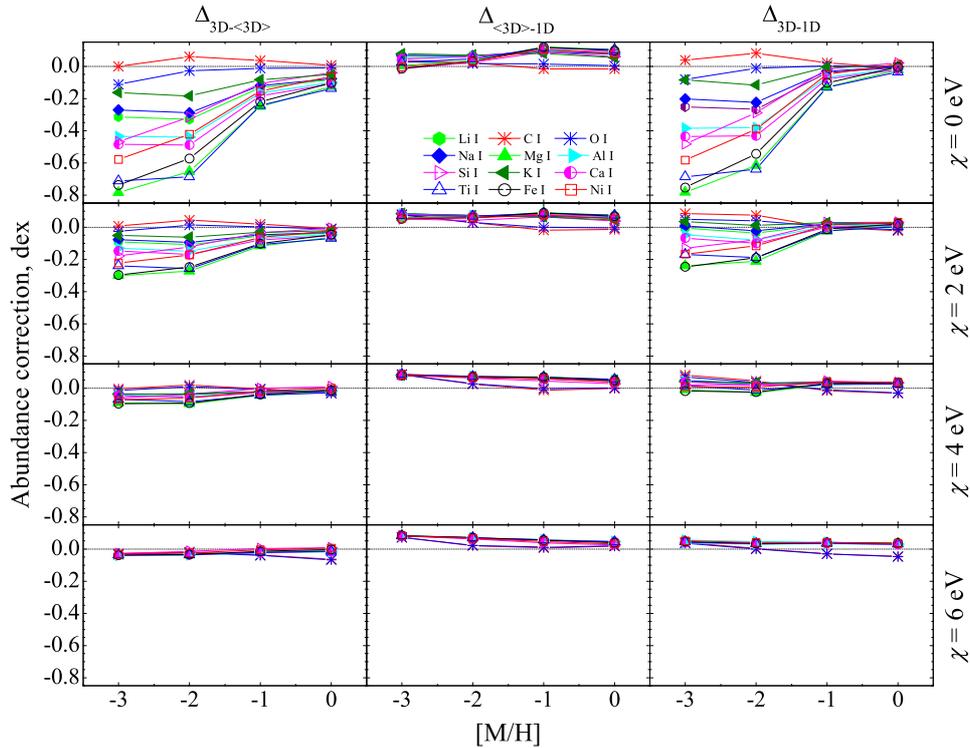}
   \caption{Abundance corrections for spectral lines of neutral atoms plotted versus metallicity at $\lambda=850$\,nm: $\Delta_{\rm 3D-\langle3D\rangle}$ (left column), $\Delta_{\rm \langle3D\rangle-1D}$ (middle column), and $\Delta_{\rm 3D-1D}$ (right column). Corrections are given at several different excitation potentials, as indicated on the right side of each row. Abundance corrections at $\lambda=400$\,nm and $\lambda=1600$\,nm (Fig.~\ref{fig:dabuAtomw400} and Fig.~\ref{fig:dabuAtomw1600}, respectively) are available in the online version only.
           }
      \label{fig:dabuAtomw850}
\end{figure*}

\onlfig{
\begin{figure*}[tb]
\centering
\includegraphics[width=13cm]{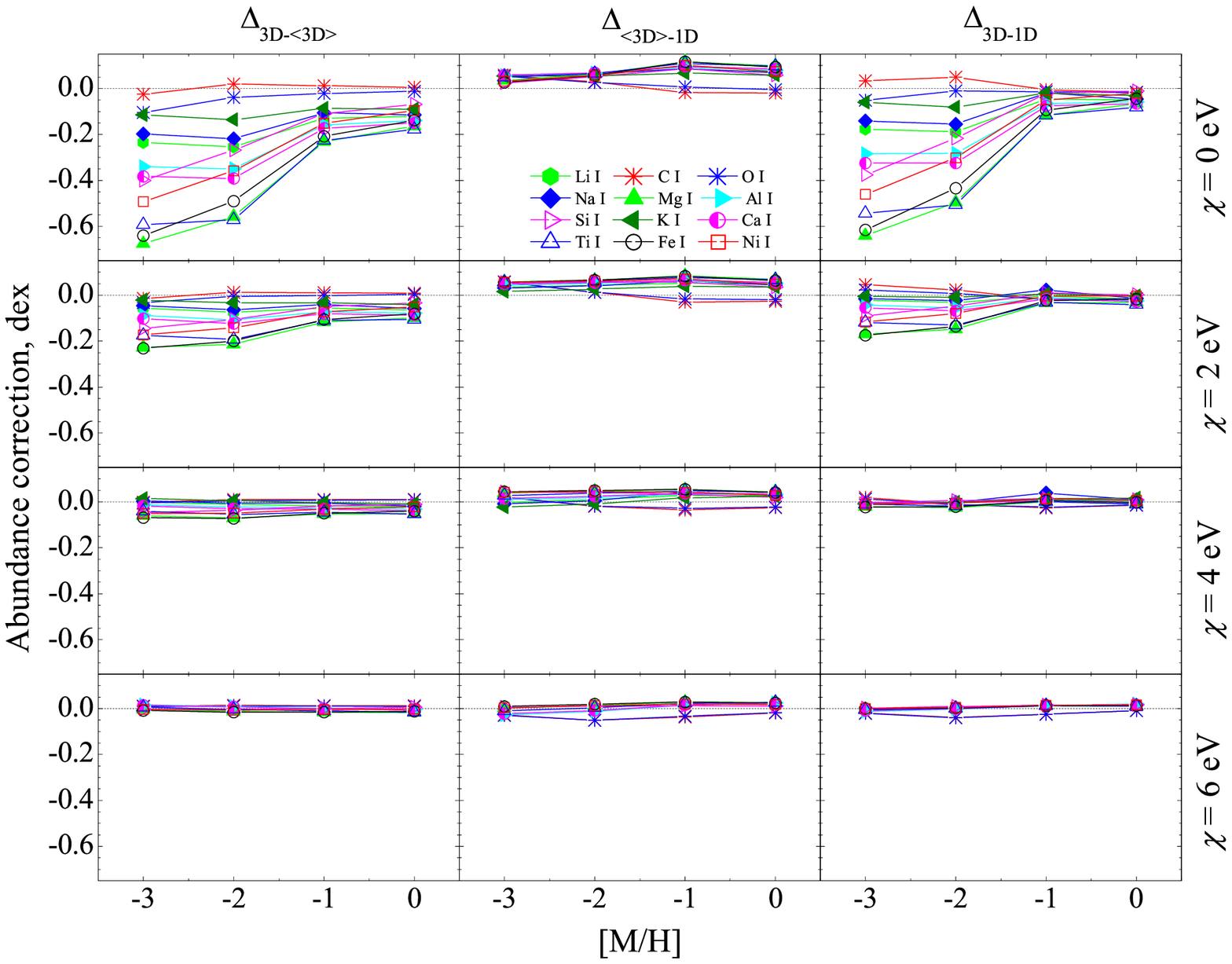}
   \caption{The same as in Fig.~\ref{fig:dabuAtomw850} but at $\lambda=1600$\,nm.
           }
      \label{fig:dabuAtomw1600}
\end{figure*}
}

\onlfig{
\begin{figure*}[tb]
\centering
\includegraphics[width=13cm]{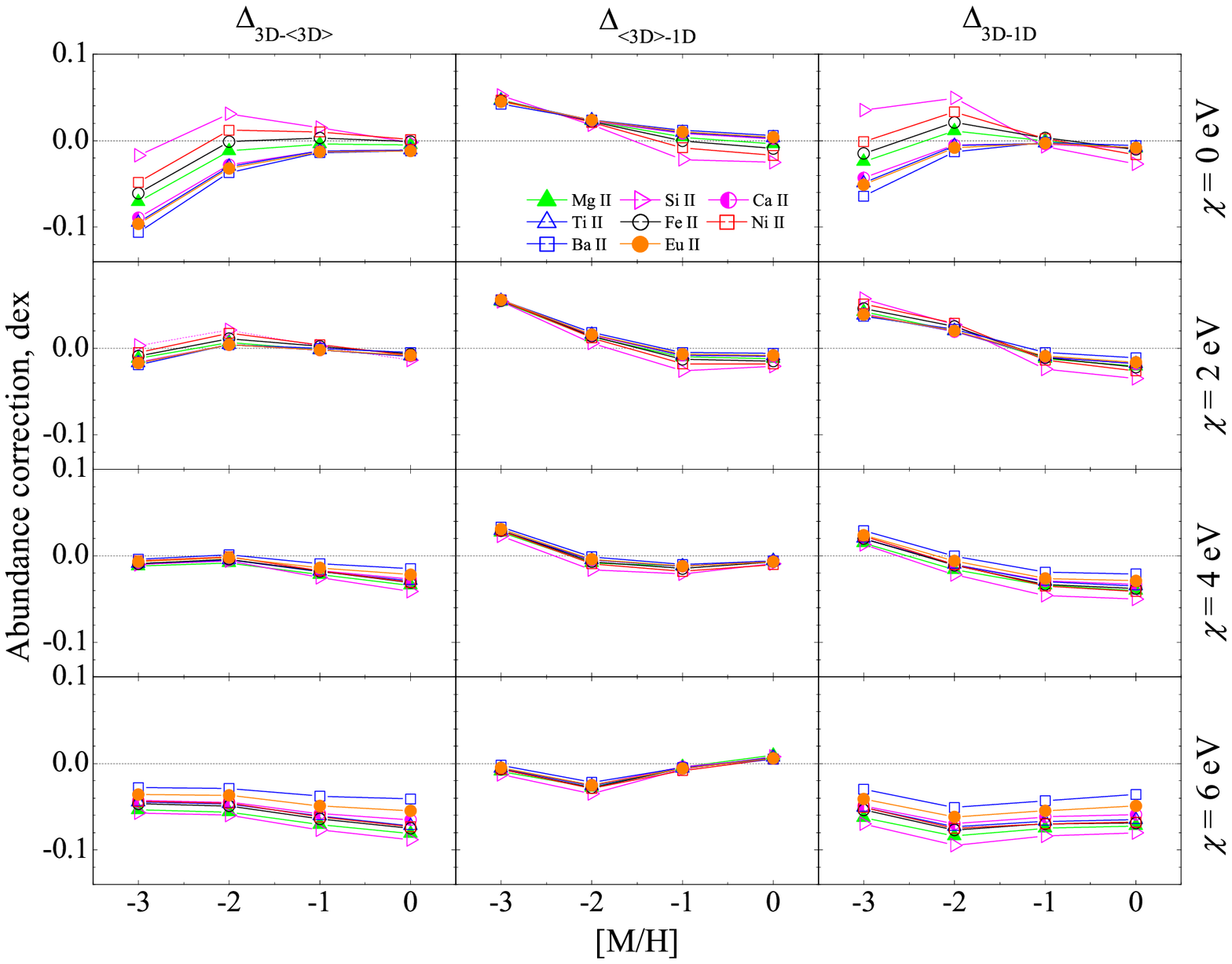}
   \caption{The same as in Fig.~\ref{fig:dabuIonw850} but at $\lambda=400$\,nm.
           }
      \label{fig:dabuIonw400}
\end{figure*}
}

\begin{figure*}[tb]
\centering
\includegraphics[width=13cm]{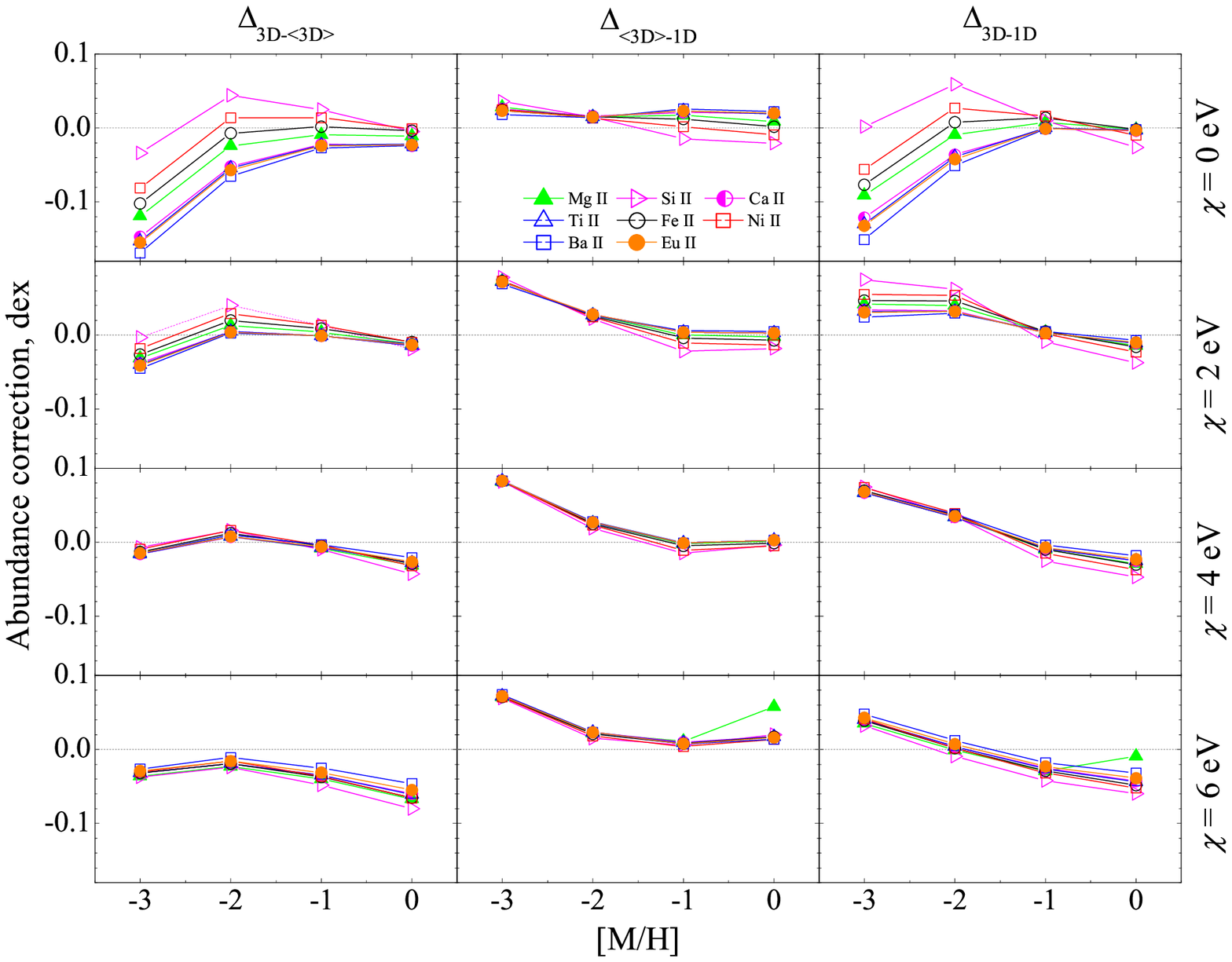}
   \caption{The same as in Fig.~\ref{fig:dabuAtomw400} but for lines of ionized atoms, plotted versus metallicity and excitation potential at $\lambda=850$\,nm. Abundance corrections at $\lambda=400$\,nm and $\lambda=1600$\,nm (Fig.~\ref{fig:dabuIonw400} and Fig.~\ref{fig:dabuIonw1600}, respectively) are available in the online version only.
           }
      \label{fig:dabuIonw850}
\end{figure*}

\onlfig{
\begin{figure*}[tb]
\centering
\includegraphics[width=13cm]{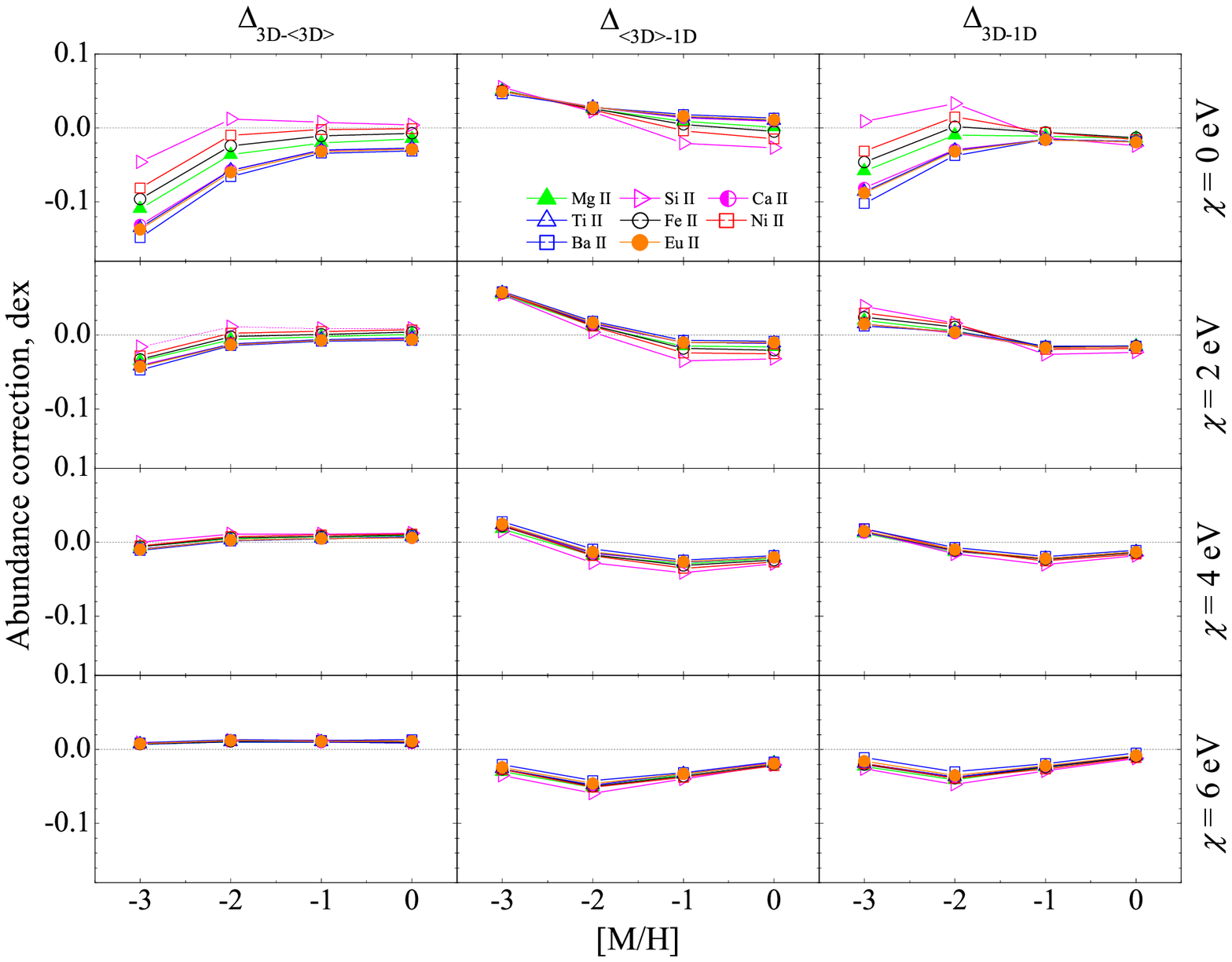}
   \caption{The same as in Fig.~\ref{fig:dabuIonw850} but at $\lambda=1600$\,nm.
           }
      \label{fig:dabuIonw1600}
\end{figure*}
}

\onlfig{
\begin{figure*}[tb]
\centering
\includegraphics[width=15cm]{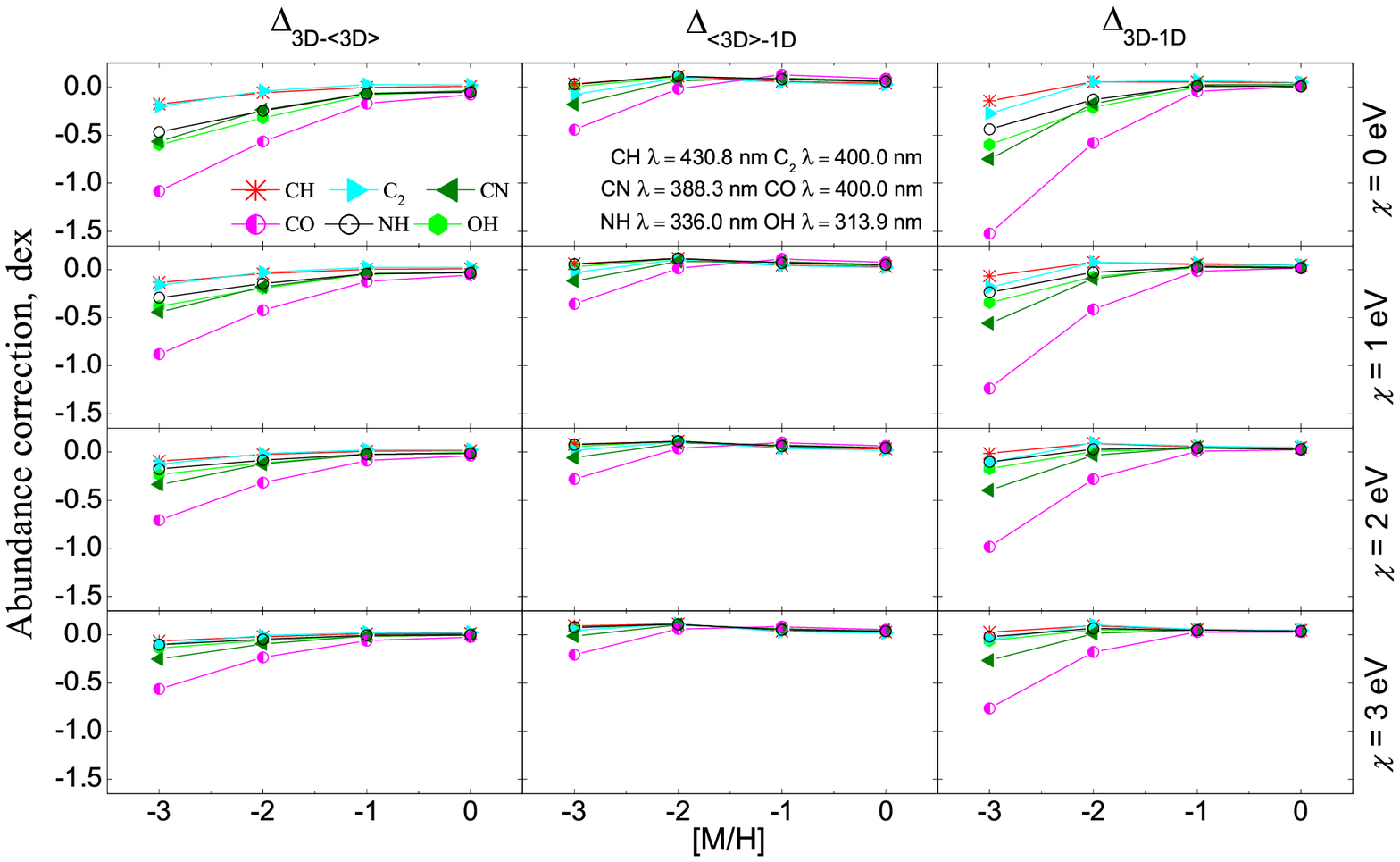}
   \caption{The same as in Fig.~\ref{fig:dabumolw850} but at $\lambda=400$\,nm.
           }
      \label{fig:dabumolw400}
\end{figure*}
}

\begin{figure*}[tb]
\centering
\includegraphics[width=15cm]{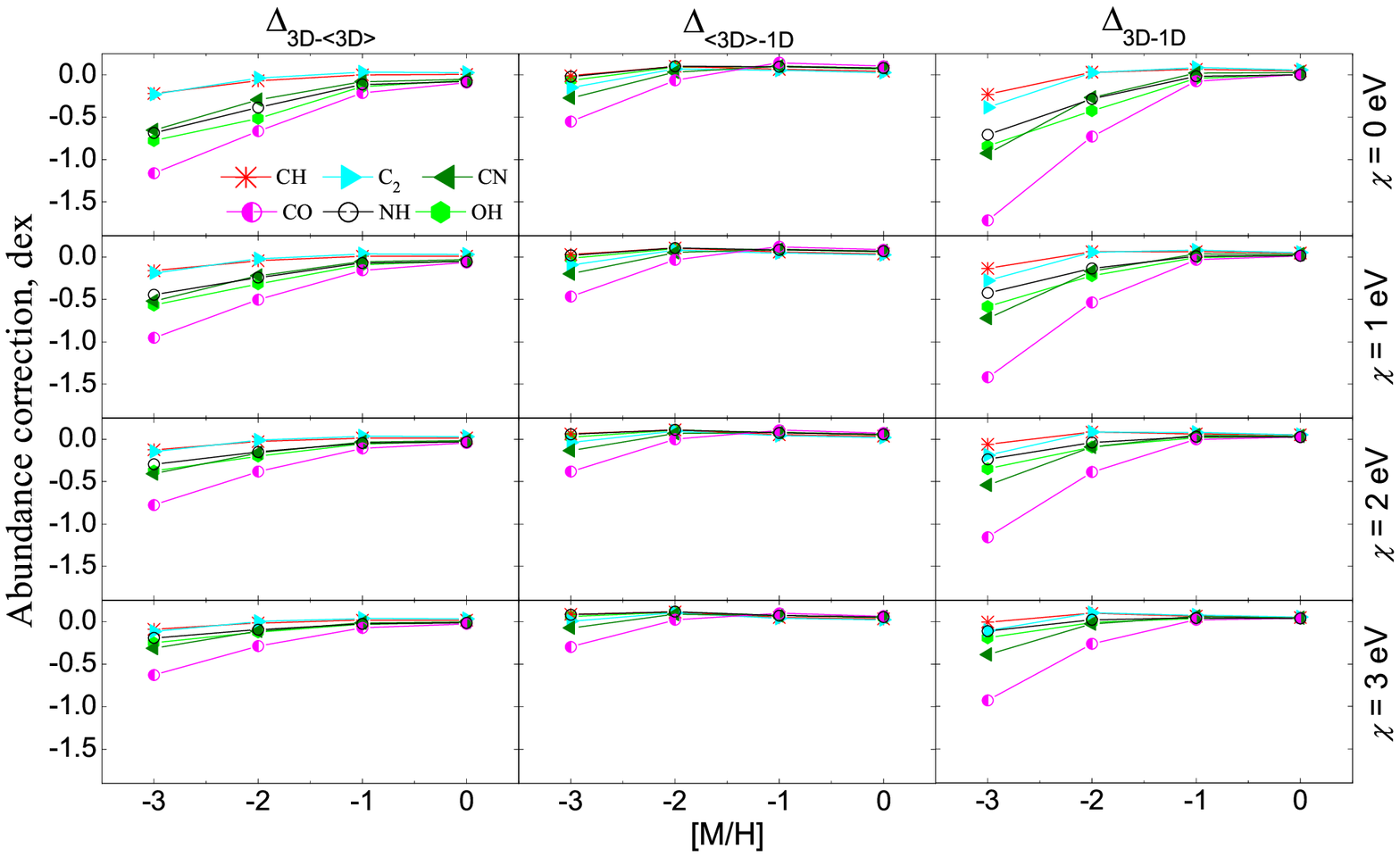}
   \caption{The same as in Fig.~\ref{fig:dabuAtomw400} but for molecular lines plotted versus metallicity and excitation potential at $\lambda=850$\,nm. Abundance corrections at $\lambda=400$\,nm and $\lambda=1600$\,nm (Fig.~\ref{fig:dabumolw400} and Fig.~\ref{fig:dabumolw1600}, respectively) are available in the online version only.
           }
      \label{fig:dabumolw850}
\end{figure*}

\onlfig{
\begin{figure*}[tb]
\centering
\includegraphics[width=15cm]{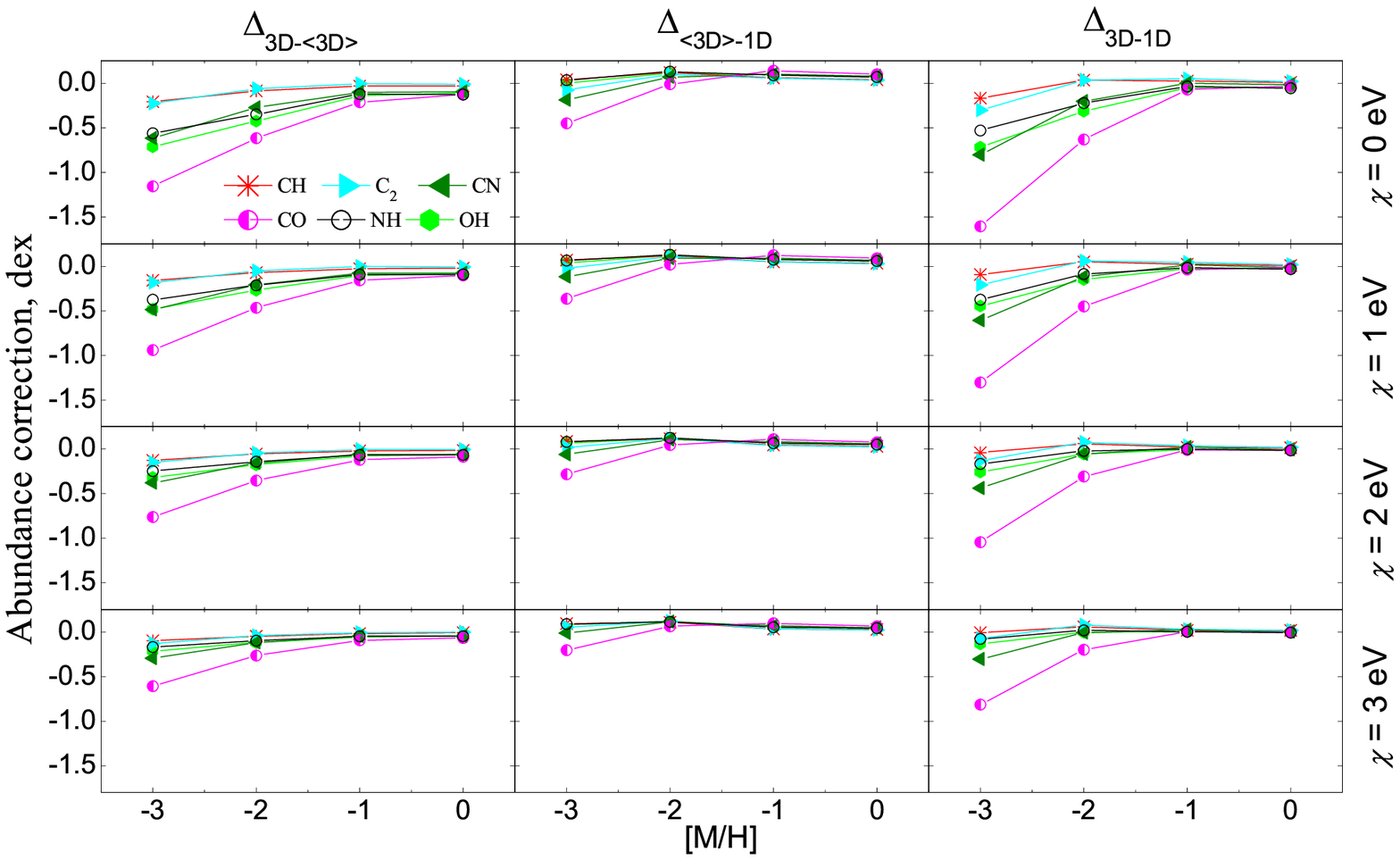}
   \caption{The same as in Fig.~\ref{fig:dabumolw850} but at $\lambda=1600$\,nm.
           }
      \label{fig:dabumolw1600}
\end{figure*}
}

\section{Results and discussion}

\subsection{Abundance corrections for lines of neutral atoms\label{sect:acn}}

The 3D--1D abundance corrections for neutral atoms are plotted in Fig.~\ref{fig:dabuAtomw400}, \ref{fig:dabuAtomw850}, and \ref{fig:dabuAtomw1600} for $\lambda=400$, 850, and 1600\,nm, respectively. Each figure shows three abundance corrections, $\Delta_{\rm 3D-\langle3D\rangle}$, $\Delta_{\rm \langle3D\rangle-1D}$, and $\Delta_{\rm 3D-1D}$, plotted versus metallicity at four different line excitation potentials, $\chi=0$, 2, 4, and 6\,eV.

For some elements, one may notice a strong dependence of the abundance corrections on metallicity: corrections are small at $\moh=0.0$ and $-1.0$ but they grow quickly with decreasing metallicity and for certain elements may reach to $-0.8$\,dex at $\moh=-3.0$ (Fig.~\ref{fig:dabuAtomw400}--\ref{fig:dabuAtomw1600}). Abundance corrections are largest at the lowest excitation potentials, $\chi=0,2$\,eV, but they quickly decrease with increasing $\chi$: corrections then become both small (less than $\pm0.1$\,dex) and essentially independent of metallicity. Such behavior is defined by the atomic properties of chemical elements and the location of line formation regions associated with particular spectral lines. At all metallicities, lines with lower excitation potentials form in the outer atmospheric layers, but their formation regions shift deeper into the atmosphere with increasing $\chi$ (Fig.~\ref{fig:Ttaumm00}-\ref{fig:Ttaumm30}). At solar metallicity, differences between the temperature profiles of the $\xtmean{\mbox{3D}}$ and 1D model atmospheres are small and change very little throughout the entire model atmosphere. Similarly, horizontal temperature fluctuations, as measured by their RMS value ($\Delta T_{\rm RMS} = \sqrt{\langle(T - T_0)^2\rangle_{x,y,t}}$, here the angle brackets denote temporal and horizontal averaging on surfaces of equal optical depth, and $T_0=\langle T \rangle_{x,y,t}$, is the depth-dependent average temperature) are not large either (i.e., compared with their extent at lower metallicities) and change little with optical depth (Fig.~\ref{fig:Ttaumm00}). Therefore, at solar metallicity all three abundance corrections, $\Delta_{\rm 3D-\langle3D\rangle}$, $\Delta_{\rm \langle3D\rangle-1D}$, and $\Delta_{\rm 3D-1D}$, are small and nearly independent of the line excitation potential. On the other hand, differences in the temperature profiles of the $\xtmean{\mbox{3D}}$ and 1D models are larger in the outer atmosphere of low metallicity models (Fig.~\ref{fig:Ttaumm30}). Horizontal temperature fluctuations are also largest in the outer atmosphere, besides, they increase rapidly with decreasing metallicity. Consequently, at $\moh<-1.0$ the abundance corrections for most elements are largest for low-excitation lines, i.e., those that form farthest in the atmosphere. For such lines, $\Delta_{\rm 3D-\langle3D\rangle}$ correction is significantly larger than $\Delta_{\rm \langle3D\rangle-1D}$, especially at lowest metallicities. Also, the two abundance corrections are nearly always of opposite sign, thus the absolute value of the total correction, $|\Delta_{\rm 3D-1D}|$, is smaller than the sum $|\Delta_{\rm 3D-\langle3D\rangle}| + |\Delta_{\rm \langle3D\rangle-1D}|$.

One may notice however, that the size of $\Delta_{\rm 3D-1D}$ corrections at a given low metallicity may be very different for different elements, ranging roughly from $-0.8$\,dex to +0.1\,dex at $\moh=-3.0$ (Fig.~\ref{fig:dabuAtomw400}--\ref{fig:dabuAtomw1600}). Such differences are caused by the interplay of ionization and excitation. Elements with small ionization potential (such as \ion{Li}{i}, \ion{Na}{i}, \ion{K}{i}) are nearly completely ionized throughout the entire atmosphere, at all metallicities. Therefore, neutral atoms of such elements are in their minority ionization stage. In all such cases the line opacity, $\kappa_\ell$, can be roughly approximated as $\kappa_\ell \sim 10^{\theta\,(E_{\rm ion}-\chi)}$, where $\theta=5040/T$ and $E_{\rm ion}$ is the ionization energy of a given element (see Paper~II, Appendix~A, eq.~A5). In such cases, the line opacity becomes a very sensitive function of the temperature at low $\chi$. Consequently, for lines with low $\chi$ large temperature fluctuations in the outer atmospheric layers at low metallicities cause large variations in the line strength, which translate into large abundance corrections $\Delta_{\rm 3D-\langle3D\rangle}$, and thus $\Delta_{\rm 3D-1D}$. On the other hand, neutral elements with high $E_{\rm ion}$, such as \ion{C}{i}, \ion{O}{i}, are in their majority ionization stage. In such case, the excitation potential dominates over the ionization energy and thus the high-excitation lines become most sensitive to the temperature variations. In fact, the dependence of all abundance corrections for \ion{C}{i} and \ion{O}{i} on both \moh\ and $\chi$ is very similar to those of ionized elements.

Abundance corrections in the infrared wavelength range are very similar to those obtained either at $\lambda=400$ or 850\,nm and for certain elements may reach to $\Delta_{\rm 3D-1D}=-0.7$\,dex. This is in contrast to the results obtained by us for significantly cooler red giant at $\Teff=3600$\,K, $\log g=1.0$, and $\moh=0.0$ for which the abundance corrections at 1600\,nm were significantly smaller than those in the optical wavelength range (Paper~II). Obviously, the latter does not hold for the red giants studied here since their abundance corrections are large at all wavelengths (see Appendix~\ref{sect:AB} for a detailed discussion). These results therefore suggest that the use of 3D hydrodynamical models may in fact be necessary for the abundance diagnostics with lines of neutral atoms in red giants at low metallicities.

\subsection{Abundance corrections for lines of ionized atoms\label{sect:aci}}

The 3D--1D abundance corrections for ionized atoms are shown in Fig.~\ref{fig:dabuIonw400}, \ref{fig:dabuIonw850}, and \ref{fig:dabuIonw1600}, at $\lambda=400$, 850, and 1600\,nm, respectively. As in the case with neutral atoms, we provide three abundance corrections, $\Delta_{\rm 3D-\langle3D\rangle}$, $\Delta_{\rm \langle3D\rangle-1D}$, and $\Delta_{\rm 3D-1D}$, plotted versus metallicity at four different line excitation potentials, $\chi=0$, 2, 4, and 6\,eV.

At all \moh\ and $\chi$ studied here, the abundance corrections for lines of ionized atoms are confined to the range of $\sim\pm0.1$\,dex and show little sensitivity to changes in both metallicity and excitation potential (i.e., if compared to trends seen with lines of neutral atoms). Lines of ionized atoms form significantly deeper in the atmosphere where both the horizontal temperature fluctuations (which determine the size of $\Delta_{\rm 3D-\langle3D\rangle}$ correction) \textit{and} the differences between temperature profiles of the $\xtmean{\mbox{3D}}$ and 1D models (which influence the size of $\Delta_{\rm \langle3D\rangle-1D}$ correction) are smallest at all metallicities and change little with \moh. This leads to small abundance corrections that are insensitive to changes in \moh\ and $\chi$. On the other hand, elements with lower ionization energies ($E_{\rm ion}<6$\,eV) are nearly completely ionized throughout the entire atmosphere of red giants studied here. For lines of such ionized elements it is the excitation potential that determines the line opacity, $\kappa_\ell$, and thus the strengths of high-excitation lines are most sensitive to temperature fluctuations (see Paper~II, Appendix B). Since temperature fluctuations are in fact smallest at the depths where such lines form, and because this holds at all metallicities, this too leads to $\Delta_{\rm 3D-1D}$ corrections that are small and show little variation with either metallicity or excitation potential (though, as expected, they increase slightly with $\chi$).

It is worthwhile noting that $\Delta_{\rm 3D-\langle3D\rangle}$ and $\Delta_{\rm \langle3D\rangle-1D}$ corrections are often of opposite sign, especially at the lowest metallicities, thus their sum leads to somewhat smaller total abundance corrections, $\Delta_{\rm 3D-1D}$. Since $\kappa_\ell$ is a highly nonlinear function of temperature, horizontal temperature fluctuations produce larger line opacities leading to stronger lines and thus, negative $\Delta_{\rm 3D-\langle3D\rangle}$ corrections. On the other hand, since the temperature of the $\xtmean{\mbox{3D}}$ model is generally lower than that of the 1D model in the line forming regions, lines of ionized species are weaker in $\xtmean{\mbox{3D}}$ than in 1D. This leads to positive $\Delta_{\rm \langle3D\rangle-1D}$ corrections. As in the case of neutral atoms, abundance corrections at $\lambda=1600$\,nm are comparable to those obtained at 400 and 850\,nm and may reach to $-0.1$\,dex at $\moh=-3.0$.

Qualitatively, the dependence of abundance corrections on metallicity and excitation potential seen in Figs.~\ref{fig:dabuAtomw400}--\ref{fig:dabuIonw850} for neutral and ionized atoms is very similar to the trends found by \citet{CAT07}, who computed abundance corrections for red giants with atmospheric parameters nearly identical to those used in our study. One obvious difference between the results obtained in the two studies is that our abundance corrections are somewhat smaller. It is possible, however, that this discrepancy may be traced back to differences between the underlying 3D model atmospheres, i.e., the {\tt STAGGER} code used by \citet{CAT07} and \COBOLD\ code utilized in our study. Indeed, the two codes use different opacities and opacity binning techniques, different equations of state, and so forth. One should also note a major difference in the 1D model atmospheres used as reference. While \LHD\ models use the same opacities, microphysics, and numerical schemes used by \COBOLD, the \MARCS\ 1D model atmosphere used as reference by \citet{CAT07} uses opacities and EOS different from those used in their 3D model.

\subsection{Abundance corrections for molecular lines\label{sect:dabumol}}

The 3D--1D abundance corrections for molecular lines are shown in Fig.~\ref{fig:dabumolw400}, \ref{fig:dabumolw850}, and \ref{fig:dabumolw1600}, at the bluest wavelength, 850\,nm, and 1600\,nm, respectively. We provide three abundance corrections, $\Delta_{\rm 3D-\langle3D\rangle}$, $\Delta_{\rm \langle3D\rangle-1D}$, and $\Delta_{\rm 3D-1D}$, plotted versus metallicity at four different excitation potentials, $\chi=0$, 1, 2, and 3\,eV.

As in the case of abundance corrections for neutral atoms, corrections for the molecules are strongly dependent on metallicity and are largest at the lowest metallicities. While for the majority of molecules the total abundance correction, $\Delta_{\rm 3D-1D}$, generally does not exceed $-1.0$\,dex at $\moh=-3.0$, in case of CO it reaches to $-1.5$\,dex. The large total abundance corrections are mostly due to the large $\Delta_{\rm 3D-\langle3D\rangle}$ corrections, which, as in the case of neutral and ionized atoms, indicates the importance of horizontal temperature fluctuations in the formation of molecular lines. The contribution due to differences in the average 3D and 1D temperature profiles, however, is also non-negligible, somewhat in contrast to the much less important $\Delta_{\rm \langle3D\rangle-1D}$ abundance corrections in the case of neutral and ionized atoms.

The trends in the abundance corrections for molecules seen in Fig.~\ref{fig:dabumolw400}--\ref{fig:dabumolw1600} are determined by a subtle interplay of a number of different factors (see Appendix~\ref{sect:AC} for a more extended discussion). To briefly summarize, as in the case of the cool red giant studied in Paper~II CO is the most abundant molecule in the atmospheric layers just above the optical surface (Fig.\,\ref{fig:frac-molec}). Since most of the carbon is locked into CO in the cooler regions ($T \le 4000$~K), only minor amounts of C are available to form other carbon-bearing molecules. In fact, this makes their number densities a highly nonlinear function of the local temperature. Since the line opacity is directly proportional to the number density of a given molecule, large horizontal temperature fluctuations (especially in the lowest metallicity models) should potentially lead to substantial abundance corrections.

In the following, we focus on the lowest metallicity model, $\moh=-3.0$, where the amplitude of the horizontal temperature fluctuations in the line forming regions is high and produces large variations of the molecular number densities. Therefore, the most pronounced molecular abundance corrections $\Delta_{\rm 3D-\langle3D\rangle}$ are derived for $\moh=-3.0$, $\chi=0.0$~eV (upper left panel of Fig.\,\ref{fig:dabumolw400}). We note that the magnitude of the correction is not simply a monotonic function of the dissociation energy of the molecule, $D_0$. A monotonic increase of $|\Delta_{\rm 3D-\langle3D\rangle}|$ with $D_0$ would be expected if \emph{all} molecules were just minority species (see Eq.\,\ref{eqn:C1}, Appendix~\ref{sect:AC}, and Eqs.\, C.4, C.5 in Paper~II). In the atmosphere under investigation here, however, the temperature dependence of the number densities of the carbon-bearing molecules is influenced by the strong coupling with the formation of CO. In the hotter parts of the atmosphere, a temperature increase always leads to a destruction of the molecules, while in the cooler regions, where atomic carbon is a minority species, temperature increase leads to the destruction of CO, and in turn, to a larger concentration of atomic carbon that enables an enhanced formation of carbon-bearing molecules, which thus exhibit a reversed temperature sensitivity.

Such subtle interplay leads to the situation where it is difficult to predict the magnitude of the 3D corrections from the basic molecular properties. Here we rather interpret the numerical results provided by the spectrum synthesis calculations.

As is evident from Fig.\,\ref{fig:nd-molec} (upper left panel), the number density of CO at any given optical depth in the atmosphere experiences the broadest range of fluctuations in the 3D model ($\moh=-3.0$). This is due to the fact that CO has the highest dissociation energy, $D_0=11.1$~eV. Moreover, CO forms rather high in the atmosphere (Fig.\,\ref{fig:cf-molec}) where the largest horizontal temperature fluctuations are encountered. Consequently, the line opacity of CO experiences the largest variations in the line forming region, such that the average line opacity, $\langle \kappa_\ell(T) \rangle$ is significantly larger than the line opacity corresponding to the average temperature $\kappa_\ell(\langle T\rangle)$, leading to the largest (most negative) abundance corrections.

CH ($D_0=3.5$~eV) and C$_2$ ($D_0=6.3$~eV) show the smallest (least negative) 3D corrections. Unexpectedly, the corrections for these two molecules are almost identical, although their dissociation energies are rather different. The explanation for this unexpected behavior is related to the fact that the temperature dependence of the line opacity at constant optical depth $\tau_{\rm Ross}=-2$ changes sign, both for CH and C$_2$ (see middle panels of Fig.~\ref{fig:nd-molec} in Appendix~\ref{sect:AC}). In the high-temperature part ($\theta < 1.3$), the slope $\partial{\log \kappa_\ell}/\partial{\theta}$ is more positive for C$_2$ than for CH, in accordance with the different values of $D_0$. But in the low-temperature regions ($\theta > 1.3$), the slope is more negative for C$_2$ than for CH, because $X($C$_2)$ depends on $X^2($C$)$, while $X($CH$)$ depends only linearly on $X($C$)$. This compensatory effect apparently leads to very similar effective amplification factors (see Appendix~\ref{sect:AB}) $A_\ell = \langle \kappa_\ell(T) \rangle / \kappa_\ell(\langle T\rangle)$ for both molecules, and hence to similar abundance corrections.

The other molecules, CN ($D_0=7.8$~eV), NH ($D_0=3.5$~eV) and OH ($D_0=4.4$~eV) occupy an intermediate position in the range of abundance corrections. NH and OH have similar molecular properties and line formation regions, and hence their corrections are similar, too. However, it is surprising that the corrections for NH and OH fall in the same range as those for CN, which has a much higher dissociation energy. Again, the explanation is related to the change of sign of the slope $\partial{\log \kappa_\ell}/\partial{\theta}$ in the case of CN, due to
the coupling with CO at low temperatures, which is missing in the case of NH and OH, where the slope is therefore essentially constant over the whole temperature range (see Fig.~\ref{fig:nd-molec}, lower panels).

These results qualitatively agree with those obtained by \citet{CAT07}, in a sense that the magnitude of abundance corrections for the molecules increases significantly with decreasing metallicity. However, quantitatively the results are different: abundance corrections at the lowest metallicity probed in our study ($\MoH=-3.0$) are on average by $0.5-0.6$\,dex smaller compared with the corrections derived by \citet{CAT07} using a model with similar atmospheric parameters. For example, for NH molecular line at $\lambda=336$\,nm and $\chi=0$\,eV excitation potential we obtain $\Delta_{\rm 3D-1D} \approx -0.5$\,dex while \citet{CAT07} gets $\Delta_{\rm 3D-1D} \approx -1.0$\,dex. Apparently, these discrepancies are due to differences in the temperature structure of the 3D models used in the two studies. Indeed, in the case of the \citet{SN98} models used by \citet{CAT07} the difference between the average $\xtmean{\mbox{3D}}$ and 1D temperature stratifications may reach to $\sim1000$\,K at $\log\tau_{\rm Ross}\approx-4$ for the $\MoH=-3.0$ model, whereas in our case these differences do not exceed  $\sim300$\,K at $\log \tau_{\rm Ross}<-1$ (Fig~\ref{fig:Ttaumm30}). We would like to stress once more that \citet{CAT07} compare their 3D results with those obtained using 1D \MARCS\ models; clearly, such approach is less self-consistent since differences in, e.g., equation of state, opacities used with the two model atmospheres may affect the resulting 3D--1D abundance corrections. The reason for the different temperature structures is presently not well understood but some possible reasons are discussed below.


\subsection{Scattering and spectral line formation}

The \COBOLD\ models treat scattering as true absorption which may be seen as rather crude approximation. Indeed, as it was discussed recently by \citet[][]{CHA11}, the treatment of scattering may have a significant impact on the thermal structure of 3D model atmospheres. In their study, coherent isotropic scattering was implemented in the model atmosphere code and the obtained model structures were compared with those where continuum scattering was treated either as true absorption or scattering opacity neglected in the optically thin regions. These tests have shown that in the former case the resulting temperature profiles were significantly warmer with respect to those calculated with coherent isotropic scattering. It has been thus argued by \citet{CHA11} that the different thermal structures obtained with the \STAGGER\ and \COBOLD\ codes may be due to the different treatment of scattering. Interestingly, the authors also found that in case when scattering opacity was neglected in the optically thin regions the resulting temperature profile was very similar to that obtained when scattering was properly included into the radiative transfer calculations.

\begin{figure}[tb]
\centering
\includegraphics[width=9cm]{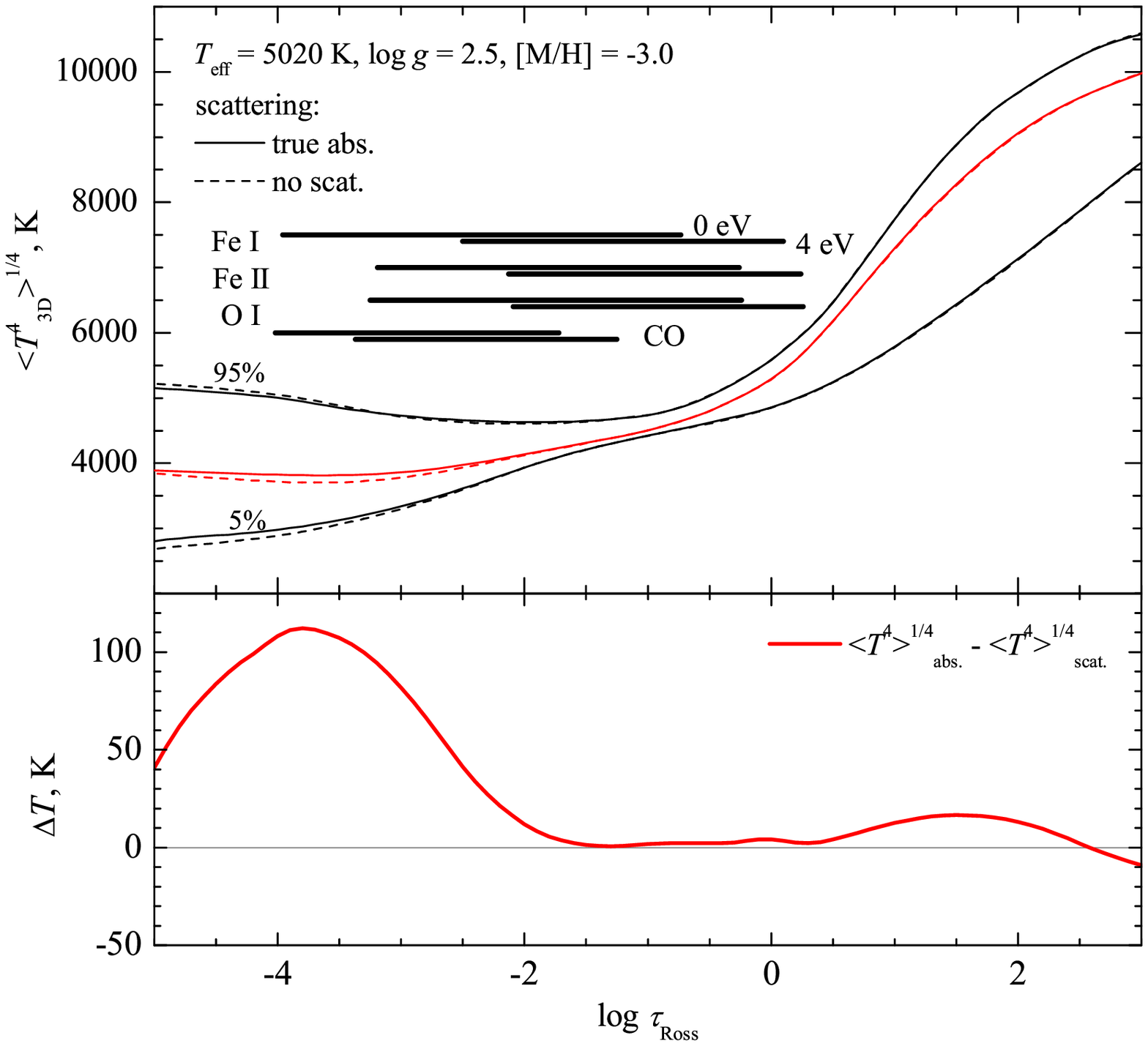}
   \caption{\textbf{Top panel.} Temperature profiles of the average $\xtmean{\mbox{3D}}$ model ($\Teff=5020$\,K, $\log g=2.5$, and $\moh=-3.0$) calculated with scattering treated as true absorption (red solid line) and with scattering opacity neglected at small $\tau_{\rm Ross}$ (red dashed line). Black lines mark the 5th and 95th percentile of the temperature distribution in the 3D model with scattering treated as true absorption (solid lines) and scattering opacity neglected in optically thin regions (dashed lines). Horizontal bars indicate the approximate location where lines of several trace elements form at $\lambda=400$\,nm and $\chi=0$ and 4\,eV and of the CO molecule at $\lambda=400$\,nm and $\chi=0$ and 3\,eV (bars mark the regions where the equivalent width, $W$, of a given spectral line grows from 5\% to 95\% of its final value). \textbf{Bottom panel.} Difference in the temperature profiles corresponding to the average $\xtmean{\mbox{3D}}$ models calculated with the two different treatments of scattering.
   }
   \label{fig:scat_temp}
\end{figure}

\begin{figure*}[tb]
\centering
\includegraphics[width=14cm]{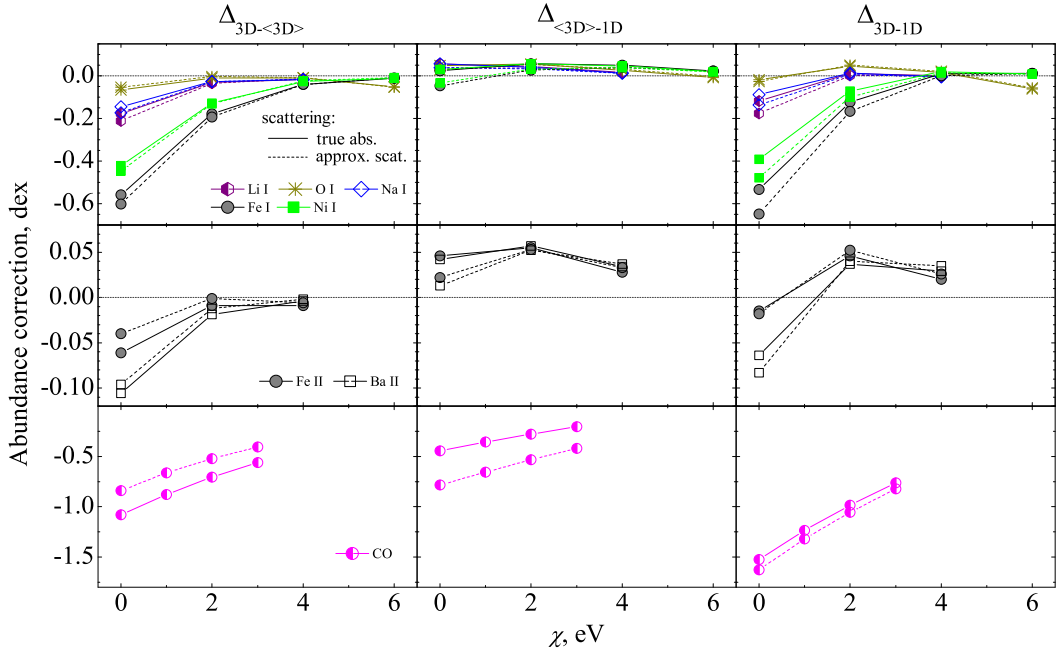}
   \caption{Abundance corrections obtained for selected elements and for the CO molecule with the model atmospheres computed using different treatments of scattering (plotted versus the excitation potential $\chi$, for lines at $\lambda=400$\,nm). For each element, two cases are shown: (a) with scattering treated as true absorption (solid lines), and (b) with scattering opacity neglected in the optically thin regions (dashed lines). Atmospheric parameters of the model atmospheres are $\Teff=5020$, $\log g=2.5$, and $\moh=-3.0$. Three types of abundance corrections are shown: $\Delta_{\rm 3D-\langle3D\rangle}$ (left column), $\Delta_{\rm \langle3D\rangle-1D}$ (middle column), and $\Delta_{\rm 3D-1D}$ (right column).
   }
   \label{fig:scat_ac}
\end{figure*}

The role of scattering in the \COBOLD\ model atmospheres has been recently studied by \citet{LS12} who compared thermal structures of the standard \COBOLD\ models (i.e., those calculated with continuum scattering treated as true absorption) and the \COBOLD\ models computed with continuum scattering opacity left out in the optically thin regions. Surprisingly, temperature profiles in the two models were different by only $\sim120$\,K at $\tauross=-4.0$, in contrast to $\sim600$\,K obtained by \citet[][]{CHA11}. While the exact cause of this difference is still unclear, \citet[][]{LS12} have suggested that they may be due to different procedures used to compute binned opacities used with the two types of models.

Obviously, it is important to understand the consequences that the differences in the treatment of scattering may have on the temperature stratification in the model atmosphere which, in turn, may influence the spectral line formation. We therefore calculated a number of fictitious lines of several chemical elements (\ion{Li}{i}, \ion{O}{i}, \ion{Na}{i}, \ion{Fe}{i}, \ion{Fe}{ii}, \ion{Ni}{i}, and \ion{Ba}{ii}), by utilizing the \COBOLD\ model in which scattering opacity was neglected in the optically thin regions \citep[taken from][]{LS12}. According to the reasoning provided in \citet[][]{CHA11}, the thermal structure of such models should be very similar to that in the models calculated with an exact treatment of scattering. Therefore, the comparison of line formation properties in these and standard models (i.e., those in which scattering is treated as true absorption) may allow to assess the importance of indirect effects of scattering on the spectral line formation via its influence on the temperature profiles. Since scattering becomes increasingly more important at low \MoH\ where both line and continuum opacities are significantly reduced, the lowest metallicity \COBOLD\ model was used for these tests ($\Teff=5020$\,K, $\log g =2.5$, and $\moh=-3.0$). Spectral line synthesis was performed using 20 fully relaxed 3D snapshots of this test model, using the procedure identical to that utilized with the standard \COBOLD\ models (see Sect.~\ref{sect:line_synth}). The abundance corrections obtained with this and the standard model are shown in Fig.~\ref{fig:scat_ac}.

The obtained results suggest that differences in the treatment of scattering within the \COBOLD\ setup have only minor influence on the spectral line strengths. The abundance corrections obtained for the standard \COBOLD\ model and the one in which scattering opacity was neglected in the optically thin regions differ by less than 0.1\,dex, both for neutral atoms and ions (see Fig.~\ref{fig:scat_ac}). For models with scattering opacity neglected, slightly lower temperature in the outer atmospheric layers (Fig.~\ref{fig:scat_temp}) leads to somewhat larger abundance corrections for the low-excitation spectral lines of neutral atoms ($\sim$0.1 dex). Since higher excitation lines form deeper in the atmosphere where differences in the thermal profiles are smaller, the influence of differences in the treatment of scattering becomes negligible for such lines, with changes in the abundance corrections of less than $0.01$\,dex at $\chi>4$\,eV. Lines of ionized elements form deep in the atmosphere too, thus, irrespective of their excitation potential, differences in the treatment of scattering do not affect their line strengths.

The situation may be slightly different in case of molecular lines. The formation of such lines extends into the outer atmosphere where the differences in the treatment of scattering may have an impact on the resulting line strengths. Indeed, in case of CO the differences in the abundance corrections components $\Delta_{\rm 3D-\langle3D\rangle}$ and $\Delta_{\rm \langle3D\rangle-1D}$ calculated using the two treatments of scattering may reach to $\sim-0.2$\,dex and $\sim0.3$\,dex, respectively (Fig.~\ref{fig:scat_ac}). However, due to their opposite sign the full 3D--1D abundance correction $\Delta_{\rm 3D-1D}$ becomes smaller, $\sim0.1$\,dex. Nevertheless, such differences may still be important in stellar abundance work.

Our results therefore suggest that the treatment of scattering may be important in case of the low-excitation lines with $\chi\leq2$\,eV and molecular lines. For elements such as sodium, where frequently only resonance or low-excitation lines are available for the abundance diagnostics, different recipes in the treatment of scattering may lead to systematic abundance differences of up to $0.1$\,dex at $\moh=-3.0$. On the other hand, strengths of high-excitation spectral lines, as well as lines of ionized elements, seem to be little affected by the choice of scattering prescription.

\begin{figure}[tb]
\centering
\includegraphics[width=8cm]{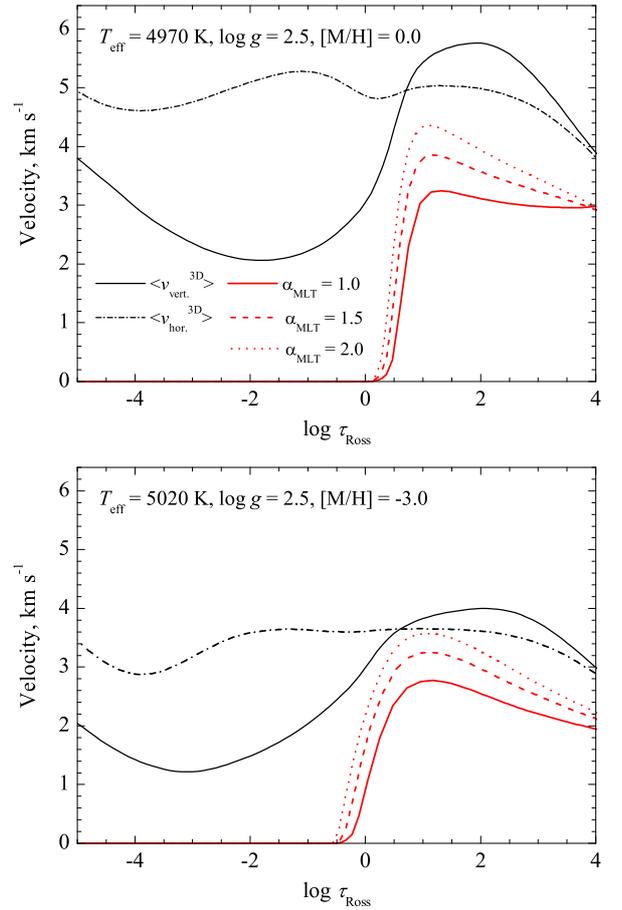}
   \caption{Velocity profiles of the LHD models (red/gray lines) at $\MoH=0.0$ (top panel) and $\MoH=-3.0$ (bottom panel) computed using three different mixing-length parameters, $\mlp=1, 1.5, 2.0$. Vertical and horizontal velocity profiles of the average $\xtmean{\mbox{3D}}$ model (computed on the $\log \tau_{\rm Ross}$ iso-surfaces are shown as solid and dot-dashed black lines, respectively.
             }
      \label{fig:lhd_vel}
\end{figure}

\subsection{Influence of the mixing-length parameter \mlp\ on the abundance corrections\label{sect:alpha-mlp}}

According to the Schwarzschild criterion for the onset of convection, \LHD\ models predict that at Solar metallicity convective flux should be zero at around and above the optical depth unity (Fig.~\ref{fig:lhd_vel}, top panel). The situation is slightly different at $\moh=-3.0$ where, because of the lower opacity, convection in the \LHD\ models reaches into layers above the optical surface, with slightly different extension for different choices of the mixing-length parameter, \mlp\ (Fig.~\ref{fig:lhd_vel}, bottom panel). The majority of spectral lines used in the abundance analysis have $\chi\leq4$\,eV and typically form in the atmospheric layers above $\log \tau_{\rm Ross} = 0.0$. Such lines should therefore be insensitive to the choice of \mlp, especially at solar metallicity. However, certain exceptions may occur in case of lines characterized by very high excitation potential that form at $\log \tau_{\rm Ross}\sim0.0$ or slightly below.

\begin{figure*}[tb]
\centering
\includegraphics[width=14cm]{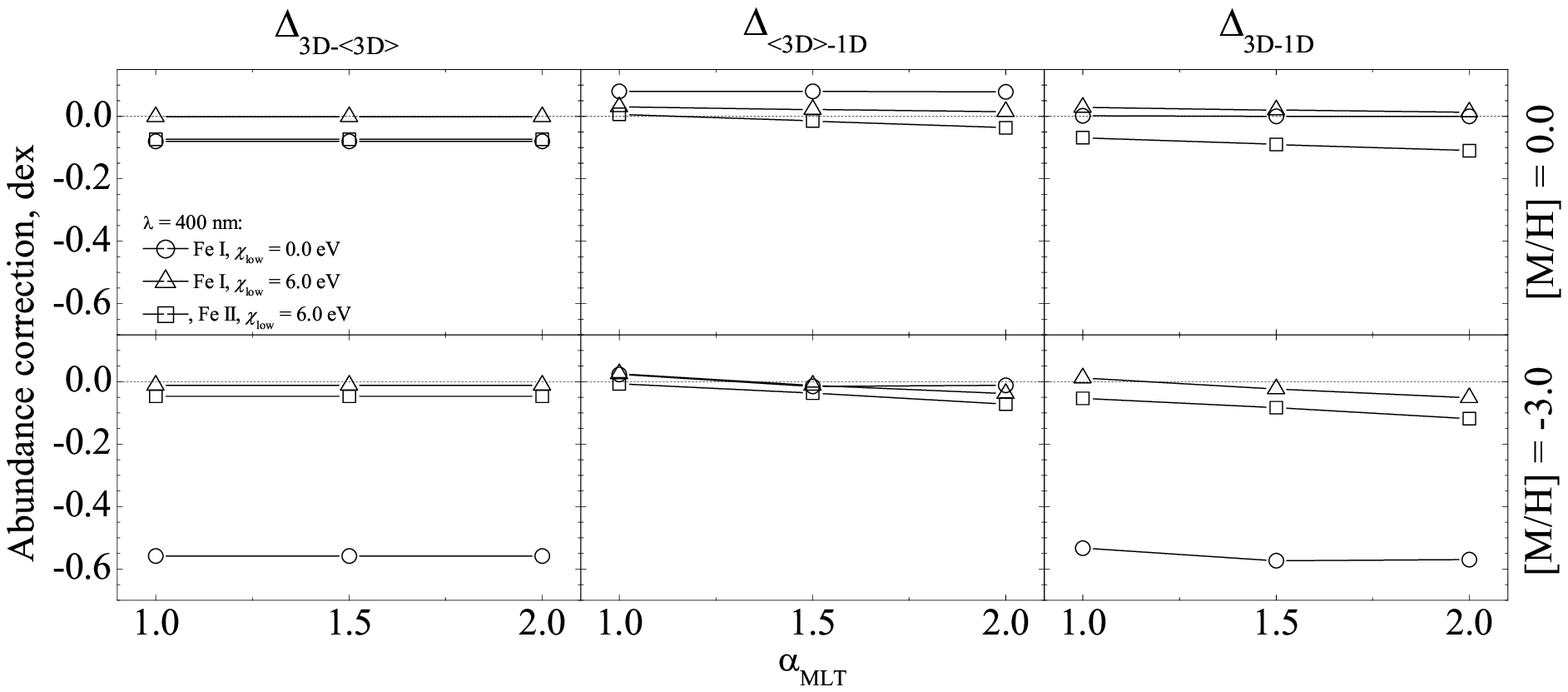}
   \caption{Abundance corrections for two fictitious lines of neutral iron ($\chi=0$ and 6\,eV) and one of ionized iron ($\chi=6$\,eV) at 400\,nm, plotted versus the mixing-length parameter \mlp\ used with the 1D LHD model atmospheres at $\MoH=0.0$ (top row) and $\MoH=-3.0$ (bottom row). Three types of abundance corrections are shown: $\Delta_{\rm 3D-\langle3D\rangle}$ (left column), $\Delta_{\rm \langle3D\rangle-1D}$ (middle column), and $\Delta_{\rm 3D-1D}$ (right column). Note that $\Delta_{\rm 3D-\langle3D\rangle}$ abundance correction is not influenced by the choice of \mlp\ (there is no comparison 1D model atmosphere involved); we nevertheless show all three abundance corrections to provide an indication of the absolute size of abundance corrections involved and the range of their variations with \mlp.
           }
      \label{fig:mlp_dabu}
\end{figure*}

To check the influence of the choice of \mlp\ used with the classical 1D models on the abundance corrections, we therefore made several test calculations using \LHD\ model atmospheres computed with $\mlp=1.0, 1.5, 2.0$. Abundance corrections were computed for several weak ($W<0.5$\,pm) fictitious lines of \ion{Fe}{i} ($\chi=0$ and 6\,eV) and \ion{Fe}{ii} ($\chi=6$\,eV), at $\MoH=0.0$ and $-3.0$. The results obtained show that in the case of \ion{Fe}{i} lines the dependence on \mlp\ at solar metallicity is indeed negligible, with the difference of abundance corrections for \mlp=1.0\ and 2.0 of less than 0.01\,dex ($\sim0.04$\,dex for \ion{Fe}{ii} $\chi=6$\,eV line, Fig.~\ref{fig:mlp_dabu}). These differences are somewhat larger at $\MoH=-3.0$ but in any case they are below $\sim0.04$ and $\sim0.07$\,dex for \ion{Fe}{i} lines with $\chi=0$ and 6\,eV, respectively. The differences are very similar for \ion{Fe}{ii} lines, too.

The variations in the abundance corrections with \mlp\ occur because the temperature profiles of the 1D models at different \mlp\ are slightly different at the optical depths where these spectral lines form. For example, temperature in the \LHD\ model with $\mlp=1.0$ is slightly higher than in the model with $\mlp=2.0$ at the optical depths where the \ion{Fe}{ii} lines form. This leads to stronger \ion{Fe}{ii} lines in the model with $\mlp=1.0$. Since the lines computed with the 3D models are generally stronger than those obtained in 1D, stronger 1D lines at $\mlp=1.0$ lead to slightly less negative abundance corrections with respect to those at $\mlp=2.0$.

These test results indicate that the choice of the mixing-length parameter used with the comparison 1D model atmospheres may be important in case of weak, higher-excitation spectral lines, which has also been discussed for the Sun by \citet{Caffau09}. One may expect that in case of stronger lines this dependence may become less pronounced, because such lines tend to form over a wider range of optical depth and typically extend into outer atmospheric layers which are insensitive to the choice of \mlp\ used in the 1D models. The results obtained here nevertheless indicate that this issue should be properly taken into account when computing abundance corrections for the higher-excitation spectral lines, especially at lower metallicities.

\section{Conclusions}

We have studied the influence of convection on the spectral line formation in the model atmospheres of red giant stars located in the lower part of the RGB ($\Teff\approx5000$\,K, $\log g=2.5$), at four different metallicities, $\moh=0.0, -1.0, -2.0, -3.0$. As in our previous studies \citep[e.g.][Paper~II]{DKL10,IKL10}, the influence of convection was studied by focusing on the 3D--1D abundance corrections, i.e., the differences in abundances predicted for the same line equivalent width by the 3D hydrodynamical \COBOLD\ and classical 1D \LHD\ model atmospheres. Abundance corrections were computed for a set of fictitious spectral lines of various astrophysically important neutral and ionized elements (\ion{Li}{i}, \ion{C}{i}, \ion{O}{i}, \ion{Na}{i}, \ion{Mg}{i}, \ion{Al}{i}, \ion{Si}{i}, \ion{Si}{ii}, \ion{K}{i}, \ion{Ca}{i}, \ion{Ca}{ii}, \ion{Ti}{i}, \ion{Ti}{ii}, \ion{Fe}{i}, \ion{Fe}{ii}, \ion{Ni}{i}, \ion{Zn}{i}, \ion{Zr}{i}, \ion{Zr}{ii}, \ion{Ba}{ii}, and \ion{Eu}{ii}), at three different wavelengths ($\lambda=400$, 850, and 1600\,nm) and four line excitation potentials ($\chi=0,2,4,6$\,eV). Only weak lines ($W<0.5$\,pm) were used in the analysis in order to avoid the influence of the microturbulence velocity used with the average $\xtmean{\mbox{3D}}$ and 1D model atmospheres.

Abundance corrections for the low-excitation lines of neutral atoms show a significant dependence on both metallicity and line excitation potential, especially at low metallicities where differences in the abundances obtained with the 3D and 1D model atmospheres, $\Delta_{\rm 3D-1D}$, may reach up to $-0.8$\,dex. The corrections are largest for elements with the lowest ionization potentials (generally, $E_{\rm ion} < 6$\,eV), i.e., those that are mostly ionized throughout the entire model atmosphere. The line opacity of the neutral atoms, $\kappa_\ell$, in this case is proportional to $\kappa_\ell\sim10^{(\theta \times (E_{\rm ion}-\chi))}$ which makes the low-excitation lines most temperature sensitive. Since the formation regions of low-excitation lines extend well into the outer atmosphere where temperature fluctuations are largest, this, in combination with the strong temperature sensitivity of the low excitation lines, leads to the largest abundance corrections. Note that differences between the average temperature profile of the 3D model and that of the 1D model are essentially depth-independent and typically do not exceed a few hundred K, irrespective of the metallicity. The abundance corrections arising due to these differences, $\Delta_{\rm \langle3D\rangle-1D}$, are modest, typically $\leq\pm0.1$\,dex, whereas corrections due to horizontal temperature fluctuations, $\Delta_{\rm 3D-\langle3D\rangle}$, may be significantly larger and reach to $-0.8$\,dex, which may lead to large total abundance corrections, $\Delta_{\rm 3D-1D}$. The corrections decrease quickly with increasing excitation potential and for lines with $\chi>2$\,eV they are confined to $\pm0.1$\,dex within the entire metallicity range. This is because lines with higher $\chi$ are less sensitive to temperature and, besides, they form deeper in the atmosphere where temperature fluctuations are significantly smaller. On the other hand, there is very little variation in the abundance corrections with both metallicity and excitation potential for the lines of neutral atoms of elements with high ionization potentials, i.e., those that are predominantly neutral throughout the entire atmosphere. In this case, it is the high-excitation lines that are most temperature sensitive, but since they form  deep in the atmosphere where temperature fluctuations are smaller the resulting abundance corrections are also small.

In case of lines of ionized atoms, the abundance corrections are small at all metallicities and excitation potentials ($\leq\pm0.1$\,dex) and show little variation with either \moh\ or $\chi$. For elements that are predominantly ionized it is the high-excitation lines that are most temperature sensitive. Such lines form deep in the atmosphere where the horizontal temperature fluctuations are small, this leads to small $\Delta_{\rm 3D-\langle3D\rangle}$ abundance corrections. Since their $\Delta_{\rm \langle3D\rangle-1D}$ corrections are never large, the total corrections are therefore significantly smaller than those for lines of, e.g., neutral atoms with low ionization potentials.

Abundance corrections of molecular lines generally depend strongly on the metallicity of the underlying model atmosphere and may may reach to $\sim -1.0$\,dex at $\moh=-3.0$ ($\sim-1.5$\,dex in case of CO). Since molecular lines are frequently used to study abundances of astrophysically important elements (such as CNO, for example), the results obtained here indicate that the usage of 3D hydrodynamical model atmospheres in such studies may be essential.

The obtained abundance corrections show little variation with wavelength (both for atoms and molecules), at least in the range of 400--1600\,nm. This is in contrast to what was obtained in case of cooler red giant located close to the RGB tip, for which the corrections at 1600\,nm were significantly smaller than in the optical wavelength range (Paper~II). This may indicate that for certain combinations of atmospheric parameters the effects of convection on the spectral line formation can be equally important over a wide wavelength range.

Rather surprisingly, we find that differences in the treatment of scattering in the radiative transfer calculations seem to have a rather small impact on the resulting thermal structures of the red giant atmospheres studied here, at least within the \COBOLD\ model setup. The differences in the temperature profiles obtained when scattering is neglected in the optically thin regions and when scattering is treated as true absorption are always confined to $<120$\,K and only occur at the optical depths log\,$\tauross\leq-2.0$. For certain low-excitation lines this may lead to modest changes in the abundance corrections that are largest for lines of neutral elements with low ionization energies and for the CO molecule, reaching up to $\sim0.1$\,dex. For lines of other neutral elements, as well as of ionized species, the changes are significantly smaller ($<0.02$\,dex).

We also find that for the weak, lower-excitation lines ($\chi<6$\,eV) there is little dependence of the abundance corrections on the mixing-length parameter, \mlp\, used with the comparison 1D LHD models, both at solar and subsolar metallicities. Differences in the case of high-excitation lines (such as \ion{Fe}{ii} at $\chi=10$\,eV) are, however, non-negligible and may reach to 0.1\,dex. While such sensitivity to \mlp\ may in fact be smaller in case of stronger lines (which form over a wider range of optical depths and reach into the outer atmosphere which is unaffected by the choice of \mlp\ used), these findings nevertheless give an indication that proper care has to be taken to account for the sensitivity on \mlp\ when calculating abundance corrections for high-excitation lines at low metallicities.


\begin{acknowledgements}
{We thank the anonymous referee for very comprehensive comments and suggestions which helped to improve the paper significantly. This work was supported by grant from the Research Council of Lithuania (MIP-101/2011). HGL, and EC acknowledge financial support by the Sonderforschungsbereich SFB\,881 ``The Milky Way System'' (subproject A4) of the German Research Foundation (DFG). AK and HGL acknowledge financial support from the Sonderforschungsbereich SFB\,881 ``The Milky Way System'' (subproject A4) of the German Research Foundation (DFG) that allowed exchange visits between Vilnius and Heidelberg. PB and AK acknowledge support from the Scientific Council of the Observatoire de Paris and the Research Council of Lithuania (MOR-48/2011) that allowed exchange visits between Paris and Vilnius.}
\end{acknowledgements}


\bibliographystyle{aa}

\Online

\begin{appendix}

\section{Excitation-like temperature in the \xtmean{\mathrm{3D}} model?}
\label{sect:AA}

The referee raised the question why an average temperature based on its forth
moment is suitable for constructing the \xtmean{\mathrm{3D}} model for
comparison. Arguably, for representing spectral properties an average
representative of the excitation properties related to the line formation
might be the most suitable choice. In this section we want to emphasize that
the choice of a suitable average is far from obvious, and our actual choice falls into the range of
plausible choices -- besides the argument brought already forward that an
average based on the forth moment conserves the total flux well.

One may construct an excitation-like temperature~$\tex$ in each layer
representing the mean population number of a particular energy level
produced by the temperature distribution in this layer via
\beq
\tex^b e^{-\frac{\chi}{k\tex}} \equiv
\xtmean{T^b e^{-\frac{\chi}{kT}}} ,
\label{eqn:texdefinition}
\eeq
where $\chi$ is the lower level energy, $k$ -- Boltzmann's constant, and $b$ is an arbitrary
exponent. The angular brackets \xtmean{.} denote the spatial
average. Relation~\eref{eqn:texdefinition} is motivated from the
Saha-Boltzmann equations governing the excitation and ionization in LTE. In
this case, $b=3/2$ or $b=5/2$ (depending whether the electron density or
electron pressure stays constant during temperature changes), and $\chi$ might
have to be identified with the difference between ionization energy and actual
level energy in case one deals with a minority species. Factors related to
partition functions are ignored. We want to derive an analytical estimate of
\tex. To this end, we consider the temperature fluctuation around the
mean~\tm\ as small and derive the relation to lowest nontrivial
order. Introducing for abbreviation the parameter $a\equiv\frac{\chi}{k\tm}$
we obtain for small $x$ ($x=\Delta T/\overline{T}$) the expansion:
\begin{eqnarray}
(1+x)^b e^{-\frac{a}{1+x}} & \approx & e^{-a}\,\left[1+\left\{a+b\right\}
  x^{\phantom{1}}\right.\nonumber\\
  &+&\frac{1}{2} \left.\left\{a^2+
  2a (b-1) +b(b-1) \right\} x^2\right] \nonumber\\
  &+& \mathrm{O}[x^3]
\end{eqnarray}
Introducing $\delt\equiv T-\tm$ and $\deltex\equiv \tex-\tm$ we obtain to
leading second order
\begin{eqnarray}
\xtmean{T^b e^{-\frac{\chi}{kT}}} & \approx & \tm^b e^{-\frac{\chi}{k\tm}}
\left[1+\frac{1}{2} \left\{ \left(\frac{\chi}{k\tm}\right)^2 \right.\right.\nonumber\\
  &+& \left.\left.
  2 \frac{\chi}{k\tm} (b-1) +b(b-1) \right\} \,\frac{\var{T}}{\tm^2}\right] .
\label{eqn:texpans}
\end{eqnarray}
$\var{T}=\xtmean{\delt^2}$ is the variance of the temperature distribution. The
linear term in $\delt$ drops out since $\xtmean{\delt}=0$. Similarly, we obtain
to leading first order
\beq
\tex^b e^{-\frac{\chi}{k\tex}} \approx \tm^b e^{-\frac{\chi}{k\tm}} \left[ 1 +
  \left( \frac{\chi}{k\tm} +b \right)\frac{\deltex}{\tm}\right].
\label{eqn:teexpans}
\eeq
Combining the expansions~\eref{eqn:texpans} and~\eref{eqn:teexpans} results in
\beq
\frac{\deltex}{\tm}\approx \frac{1}{2}\left(\frac{\chi}{k\tm}+(b-1)-\frac{1}{1+b/\frac{\chi}{k\tm}}
\right) \,\frac{\var{T}}{\tm^2} .
\label{eqn:dte}
\eeq
As example, for $b=0$ (corresponding to the level population of a majority
species) and $\frac{\chi}{k\tm} > 2$ the excitation temperature is always
greater than the (arithmetic) mean temperature in a given layer. Taken the Sun
as typical example where $k\tm\approx0.5\pun{eV}$ in the line forming region,
this means that for lines with $\chi>1\pun{eV}$ it follows that
$\tex>\tm$. However, the difference between effective excitation temperature
and mean temperature is generally not large since the difference depends on
the square of the temperature fluctuations.

Average temperatures based on higher moments of the temperature also lead to
changes with respect to the mean temperature~\tm. One may ask which moment is
necessary to produce a temperature equal to the effective excitation
temperature. With the auxiliary formula
\beq
(1+x)^n \approx 1 + n x + \frac{1}{2}n(n-1)x^2 +\mathrm{O}[x^3]
\eeq
we obtain
\beq
\xtmean{T^n}= \tm^n \xtmean{\left(1+\frac{\delt}{\tm}\right)^n}
\approx \tm^n \left( 1+ \frac{1}{2}n(n-1)\frac{\var{T}}{\tm^2}\right),
\eeq
and
\beq
\tex^n= \tm^n \left(1+\frac{\deltex}{\tm}\right)^n
\approx \tm^n \left( 1+ n \frac{\deltex}{\tm} \right) .
\eeq
Since we demand $\xtmean{T^n}=\tex^n$ we arrive at
\beq
\frac{\deltex}{\tm}\approx \frac{1}{2} (n-1)\frac{\var{T}}{\tm^2}.
\label{eqn:deltex}
\eeq
Combining the above result with equation~\eref{eqn:dte} we obtain for the
temperature moment~$n$
\beq
n = \frac{\chi}{k\tm}+b-\frac{1}{1+b/\frac{\chi}{k\tm}} \ge  \frac{\chi}{k\tm}+b-1 .
\label{eqn:tmoment}
\eeq
The inequality in relation~\eref{eqn:tmoment} holds for $\chi, b\ge 0$.
Equation~\eref{eqn:tmoment} illustrates that for already moderately high
$\chi$, potentially enhanced by a non-zero $b$, one would need rather high
moments of the temperature to represent the effective excitation temperature
correctly. The consideration does not allow to identify an optimal~$n$ but
nevertheless illustrates that $n=4$ is not an unreasonable choice taking the
excitation-like temperature as criterion.

\section{Base RGB versus tip RGB abundance corrections}
\label{sect:AB}

As pointed out above, the abundance corrections derived in the present work
for the red giants located near the base of the RGB ($\Teff=5000$\,K,
$\log g=2.5$, $\moh=0.0, -1, -2, -3$) show little variation with wavelength,
at least in the range of $400 - 1600$ nm. In contrast,
Paper~II demonstrated for a cooler red giant located near
the tip of the RGB ($\Teff=3600$\,K,  $\log g=1.0$, $\moh=0.0$) that the
theoretical abundance corrections at 1600\,nm were significantly smaller than
those at optical wavelengths. In the following, we shall provide some basic
explanation for the different wavelength dependence of the abundance
corrections in these two type of giants.

As an example, we consider the formation of a weak fictitious high-excitation
\ion{Fe}{ii} line with $\chi = 6$~eV, representative of the ionized atoms
that show significant abundance corrections (at $\moh=0.0$). In case of the
tip RGB giant, the abundance correction derived for this line is
$\Delta_{\rm 3D-\langle3D\rangle}\approx -0.4$~dex at $\lambda=850$~nm, and
$\Delta_{\rm 3D-\langle3D\rangle}\approx -0.02$~dex at $\lambda=1600$~nm.
In case of the less evolved giant of the present work, the corresponding
numbers, for $\moh=0.0$, are
$\Delta_{\rm 3D-\langle3D\rangle} \approx -0.06$~dex ($\lambda=850$~nm),
and $\Delta_{\rm 3D-\langle3D\rangle} \approx +0.02$~dex ($\lambda=1600$~nm).
The metallicity dependence of the corrections is weak at $\lambda=850$~nm
and negligible at $\lambda=1600$~nm (see Figs.\, \ref{fig:dabuIonw850} and
\ref{fig:dabuIonw1600}).
Since the corrections in the infrared are small for both type of giants,
the remaining question is why the corrections in the red at $\lambda=850$~nm
are so much smaller in the `warm' giant than in the more evolved 'cool' giant.

As in Paper~II, the following considerations are restricted to vertical rays
(disk-center intensity) in a single snapshot from the different 3D simulations.
We focus here on the analysis of the `granulation correction',
$\Delta_{\rm 3D-\langle3D\rangle}$.

\begin{table*}[tb]
\caption{Mean height of formation, $\langle\log \tau_{\rm c}\rangle$,
of a fictitious \ion{Fe}{ii} line with $\chi=6$~eV, at
$\lambda=850$ and $1600$~nm, respectively, for the red giant model
discussed in Paper~II, and two of the \COBOLD\ red giant models used
in this work. $\langle\log \tau_{\rm Ross}\rangle$ is the mean height of
formation on the Rosseland optical depth scale; $\langle T\rangle_{\rm line}$
is the average temperature at the mean height of
formation. The last two columns give the relative RMS temperature fluctuation
on the iso-surface $\log \tau_{\rm c}=\langle\log \tau_{\rm c}\rangle$ and
$\log \tau_{\rm Ross}=\langle\log \tau_{\rm Ross}\rangle$, respectively.}
\label{table:height-of-line}
\centering
\begin{tabular}{lcccccccc}
\hline
Model & $\Teff$  & $\log g$ & \moh & $\langle\log \tau_{\rm c}\rangle$ &
$\langle\log \tau_{\rm Ross}\rangle$ & $\langle T \rangle_{\rm line}$     &
\multicolumn{2}{c}{$\delta T_{\rm RMS}$} [\%] \\
name & {[K]}    & {[cgs]}  &      &                                  &
                                   & {[K]}                    &
$\tau_{\rm c}$ & $\tau_{\rm Ross}$                   \\
\hline\noalign{\smallskip}
\multicolumn{4}{c}{} & \multicolumn{5}{c}{$\lambda\,850$~nm} \\
\hline\noalign{\smallskip}
d3t36g10mm00 & 3600 & 1.0 & ~~$0$ &  0.53 &  0.32 & 4480 & 8.15 & 6.33 \\
d3t50g25mm00 & 5000 & 2.5 & ~~$0$ & -0.01 & -0.06 & 5260 & 5.21 & 4.68 \\
d3t50g25mm20 & 5000 & 2.5 &  $-2$ & -0.36 & -0.54 & 4770 & 2.91 & 2.87 \\
\hline\noalign{\smallskip}
\multicolumn{4}{c}{} & \multicolumn{5}{c}{$\lambda\,1600$~nm} \\
\hline\noalign{\smallskip}
d3t36g10mm00 &  3600 & 1.0 & ~~$0$ & -0.03 &  0.68 & 5120 & 6.48 & 7.46 \\
d3t50g25mm00 &  5000 & 2.5 & ~~$0$ & -0.47 &  0.15 & 5600 & 4.31 & 5.23 \\
d3t50g25mm20 &  5000 & 2.5 &  $-2$ & -0.72 & -0.16 & 5120 & 4.00 & 4.49 \\
\hline\noalign{\smallskip}
\end{tabular}
\end{table*}

\subsection{Temperature fluctuations on iso-surfaces of optical depth}

We define the mean height of line formation as the center of gravity of
the \emph{equivalent width contribution function} $\mathcal{B}$ on the
monochromatic continuum optical depth scale $\tau_{\rm c}$:

\begin{equation}
\langle \log \tau_{\rm c}\rangle = \frac{
\int_{-\infty}^{\infty} \log \tau'_{\rm c}\,
\mathcal{B}(\tau'_{\rm c}) \, \mathrm{d}  \log \tau'_{\rm c}}
{\int_{-\infty}^{\infty}
\mathcal{B}(\tau'_{\rm c}) \, \mathrm{d}  \log \tau'_{\rm c}} \, .
\end{equation}

For simplicity, we use the contribution function of the \xtmean{\mbox{3D}}
atmosphere, corresponding to $\mathcal{B}_{\rm 2,2,2}$ in Paper~II
(Eq.\,B.10). The results for the \ion{Fe}{ii} line with $\chi = 6$~eV
are given in Table\,\ref{table:height-of-line} for three different model
atmospheres at $\lambda=850$ and $1600$~nm, respectively.
The mean height of formation on the Rosseland optical depth scale,
$\langle\log \tau_{\rm Ross}\rangle$, is defined such that the average
temperature on the iso-surface
$\log \tau_{\rm Ross}=\langle\log \tau_{\rm Ross}\rangle$ is equal
to the average temperature on the iso-surface
$\log \tau_{\rm c}=\langle\log \tau_{\rm c}\rangle$, which we denote as
$\langle T \rangle_{\rm line}$. For all three models listed in
Tab.\,\ref{table:height-of-line}, the infrared line forms in significantly
deeper atmospheric layers than the red line. This is the expected result since
the continuum opacity is mainly due to H$^-$ and shows a local maximum near
$\lambda=850$~nm, and a minimum close to $\lambda\, 1600$~nm. In the metal-poor
giant, the line contribution function is more asymmetric and extends towards
higher atmospheric layers than in the solar metallicity giant at the same $\Teff$ and
$\logg$. This is related to the fact that the continuum opacity decreases more
steeply with height in the metal-poor atmosphere (due to the reduced electron
pressure). For given stellar parameters, the mean height of line formation is
therefore shifted towards higher atmospheric layers at low metallicity.

The Table also gives the relative RMS temperature fluctuation
$\delta T_{\rm RMS}$, both on the $\tau_{\rm c}$ and the $ \tau_{\rm Ross}$
iso-surface defining the center of the line forming region.
The full depth-dependence of $\delta T_{\rm RMS}$ is shown in
Fig.\,\ref{table:height-of-line}. We find that, at any given mean
temperature, $\delta T_{\rm RMS}$ is always systematically larger
on $\tau_{850}$ iso-surfaces than on $\tau_{1600}$ iso-surfaces.
This is a consequence of the \emph{lower} temperature sensitivity of the
continuum opacity at $\lambda\, 850$~nm compared to $\lambda\, 1600$~nm.
The Rosseland opacity has an intermediate $T$ sensitivity, and hence the
amplitude of the temperature fluctuations lies between the values at
$\lambda\, 850$~nm and  $\lambda\, 1600$~nm.
This behavior is seen for all model atmospheres considered here, but is
particularly pronounced in the cool giant. The general picture emerging
from Table\,\ref{table:height-of-line} and Fig.\,\ref{table:height-of-line}
indicates that the temperature fluctuations in the line forming layers
are largest in the cool giant, and smallest in the warmer metal-poor
giant. The 3D abundance corrections are therefore expected to be potentially
larger for giants near the tip than near the base of the RGB. This is
confirmed by the more detailed analysis below.

\begin{figure}[tb]
\centering
\mbox{\includegraphics[bb=14 56 580 380, width=8.4cm]
      {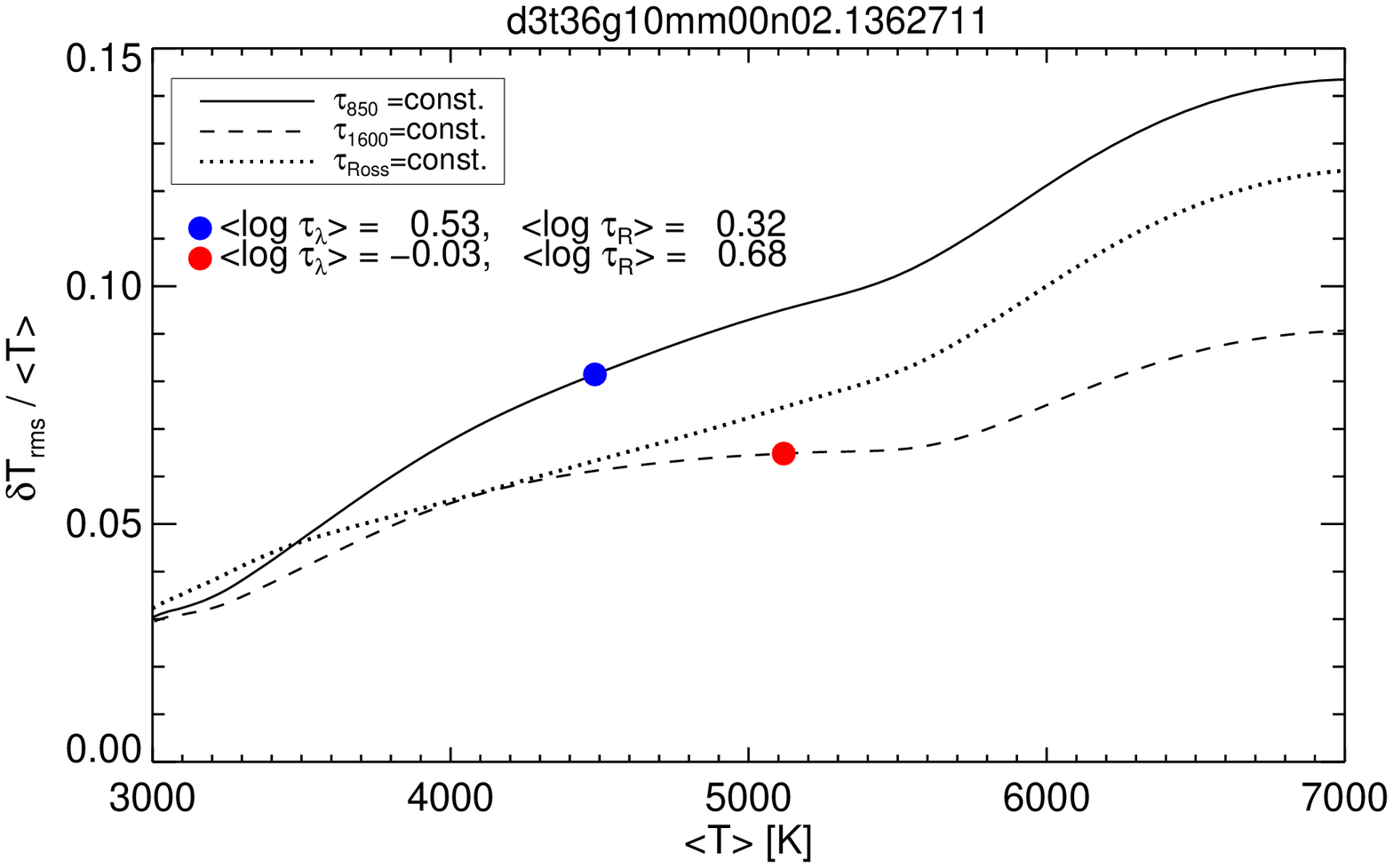}} \\[3mm]
\mbox{\includegraphics[bb=14 56 580 380, width=8.4cm]
      {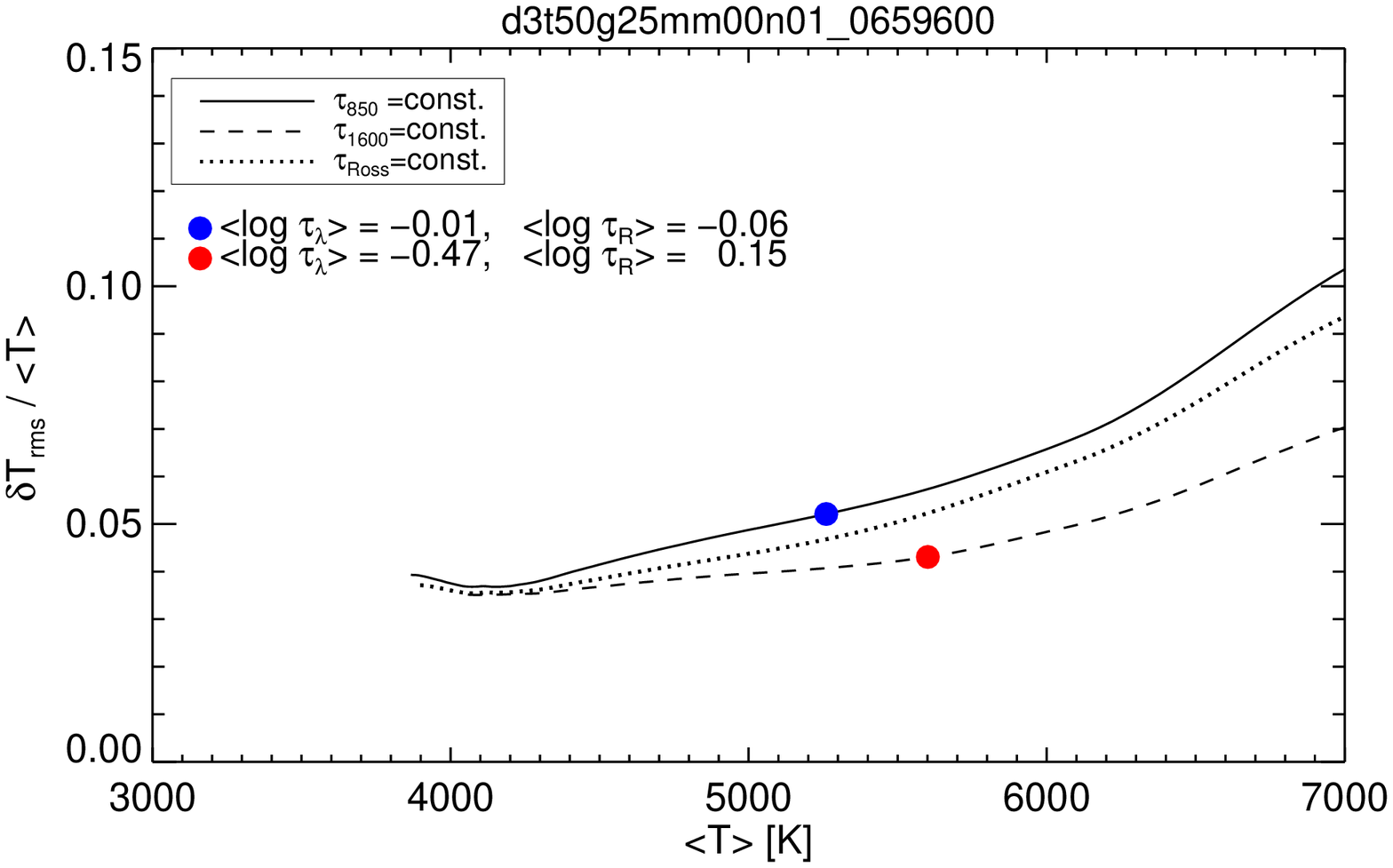}} \\[3mm]
\mbox{\includegraphics[bb=14 56 580 380, width=8.4cm]
      {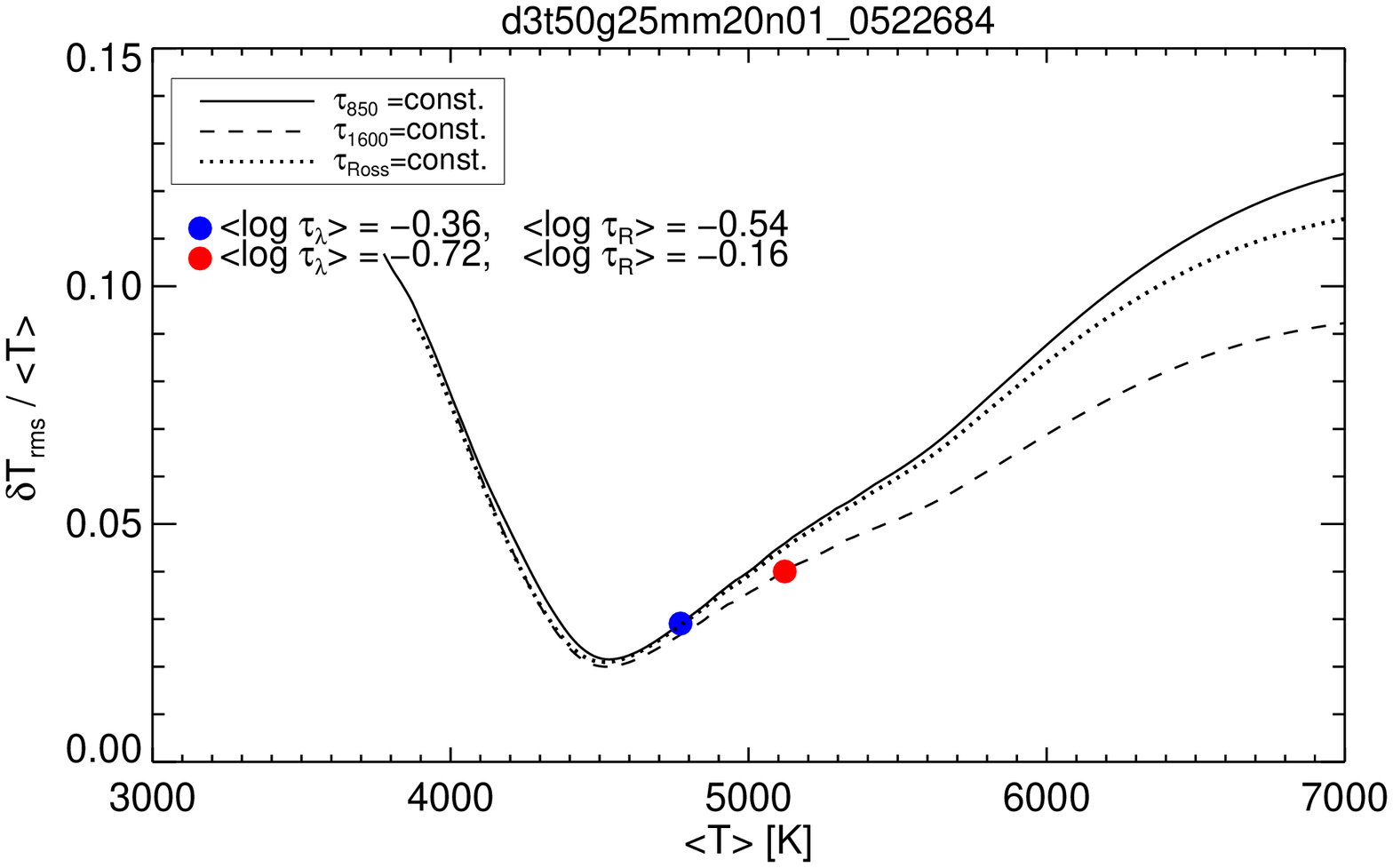}}
   \caption{Amplitude of the relative RMS temperature fluctuations
            $\delta T_{\rm RMS}/\langle T\rangle$ versus the average
            temperature $\langle T\rangle$, evaluated on different iso
            optical depth surfaces: $\tau_{850}$ (solid), $\tau_{1600}$
            (dashed), $\tau_{\rm Ross}$ (dotted). Filled circles indicate
            the values at the mean height of line formation at
            $\lambda\, 850$~nm and $\lambda\, 1600$~nm, respectively,
            as given in Table\,\ref{table:height-of-line}.
           }
      \label{fig:dTrms_tau}
\end{figure}

\subsection{Opacity fluctuations and abundance corrections}
We recall that, in the weak line limit, the abundance correction
$\Delta_{\rm 3D-\langle3D\rangle}$ can be computed as
\begin{equation}
\Delta_{\rm 3D-\langle\mathrm{3D}\rangle} =
-\log\,(W_{\rm 3D}/W_{\rm \langle\mathrm{3D}\rangle})\, ,
\label{eqn:B9}
\end{equation}
where $W_{\rm 3D}$ and $W_{\rm \langle\mathrm{3D}\rangle}$ denote the line
equivalent width obtained from the 3D and the \xtmean{\mbox{3D}}
model, respectively, for the same atomic line parameters and elemental
abundance (see Paper~II, Appendix B). With the help of the
equivalent width contribution function $\mathcal{B}$, we can write
the ratio of equivalent widths as
\begin{equation}
\frac{W_{\rm 3D}}{W_{\rm \langle\mathrm{3D}\rangle}} = \frac{
\int_{-\infty}^{\infty} \mathcal{B}_{3,3,3}(\tau'_{\rm c})
\, \mathrm{d}  \log \tau'_{\rm c}}
{\int_{-\infty}^{\infty} \mathcal{B}_{2,2,2}(\tau'_{\rm c})
\, \mathrm{d}  \log \tau'_{\rm c}} \equiv \frac{W_{3,3,3}}{W_{2,2,2}}
\equiv \mathcal{R}_{3,3,3}\, ,
\end{equation}
with the mixed contribution function
\beq
\mathcal{B}_{i,j,k}\,(\tau_{\rm c}) = \ln 10\,\tau_{\rm c}\,\exp\{-\tau_{\rm c}\}\,
\left\langle\,u_{{\rm c},i}\,(\tau_{\rm c})\,
\frac{\kappa_{\ell, j}\,(\tau_{\rm c})}{\kappa_{{\rm c}, k}\,(\tau_{\rm c})\,}
\right\rangle_{x,y}\,
\label{eqn:B10}
\eeq
as defined in Paper~II, Eq.(B.10). Subscripts $2$ and $3$ refer to the
$\langle$3D$\rangle$ model and the 3D model, respectively.

To isolate the role of the \emph{line opacity fluctuations} for the
abundance corrections, we consider the ratio

\begin{eqnarray}
\mathcal{R}_{2,3,2}\ \equiv \frac{W_{2,3,2}}{W_{2,2,2}} &=& \frac{
\int_{-\infty}^{\infty} \mathcal{B}_{2,3,2}(\tau'_{\rm c})
\, \mathrm{d}  \log \tau'_{\rm c}}
{\int_{-\infty}^{\infty} \mathcal{B}_{2,2,2}(\tau'_{\rm c})
\, \mathrm{d}  \log \tau'_{\rm c}} \nonumber \\
 &=& \frac{
\int_{-\infty}^{\infty} \, \mathcal{B}_{2,2,2}(\tau'_{\rm c})\,
                       \mathcal{A}_{2,3,2}(\tau'_{\rm c})
\, \mathrm{d}  \log \tau'_{\rm c}}
{\int_{-\infty}^{\infty} \mathcal{B}_{2,2,2}(\tau'_{\rm c})
\, \mathrm{d}  \log \tau'_{\rm c}} \, ,
\label{eqn:A5}
\end{eqnarray}
where we have defined the \emph{local amplification factor}
$\mathcal{A}_{2,3,2}$ as

\begin{equation}
\mathcal{A}_{2,3,2}(\tau_{\rm c}) = \frac{ \mathcal{B}_{2,3,2}(\tau_{\rm c})}
                                    { \mathcal{B}_{2,2,2}(\tau_{\rm c})}
                             = \left\langle\frac{\kappa_{\ell,3}(\tau_{\rm c})}
                                                {\kappa_{\ell,2}(\tau_{\rm c})}
                               \right\rangle_{x,y}
                     = \frac{\left\langle\kappa_\ell(T)\right\rangle_{x,y}}
                            {\kappa_\ell\left(\langle T\rangle_{x,y}\right)}\, .
\end{equation}

\begin{figure*}[tb]
\centering
\mbox{\includegraphics[bb=14 56 580 380, width=8.4cm]
      {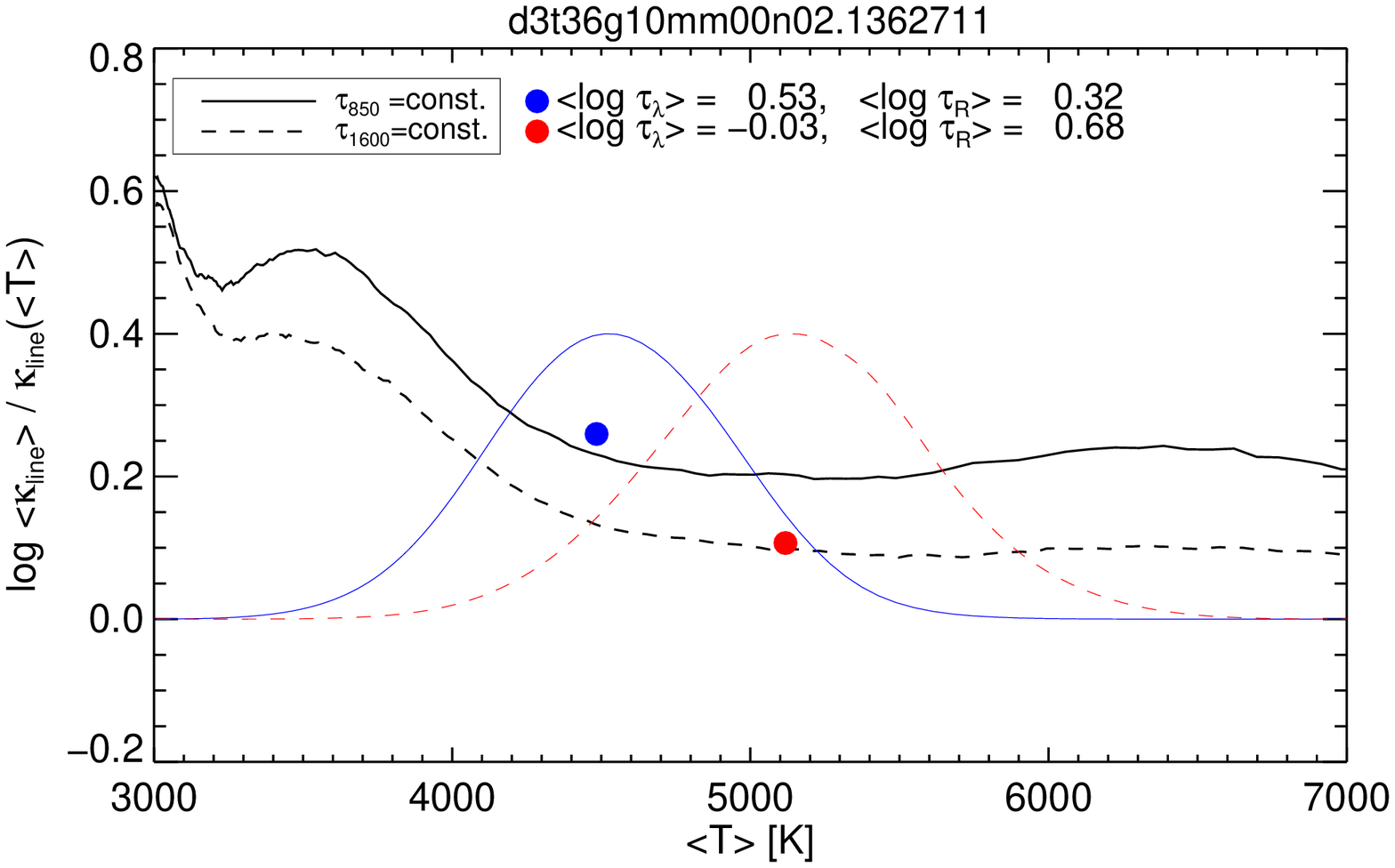}}
\mbox{\includegraphics[bb=14 56 580 380, width=8.4cm]
      {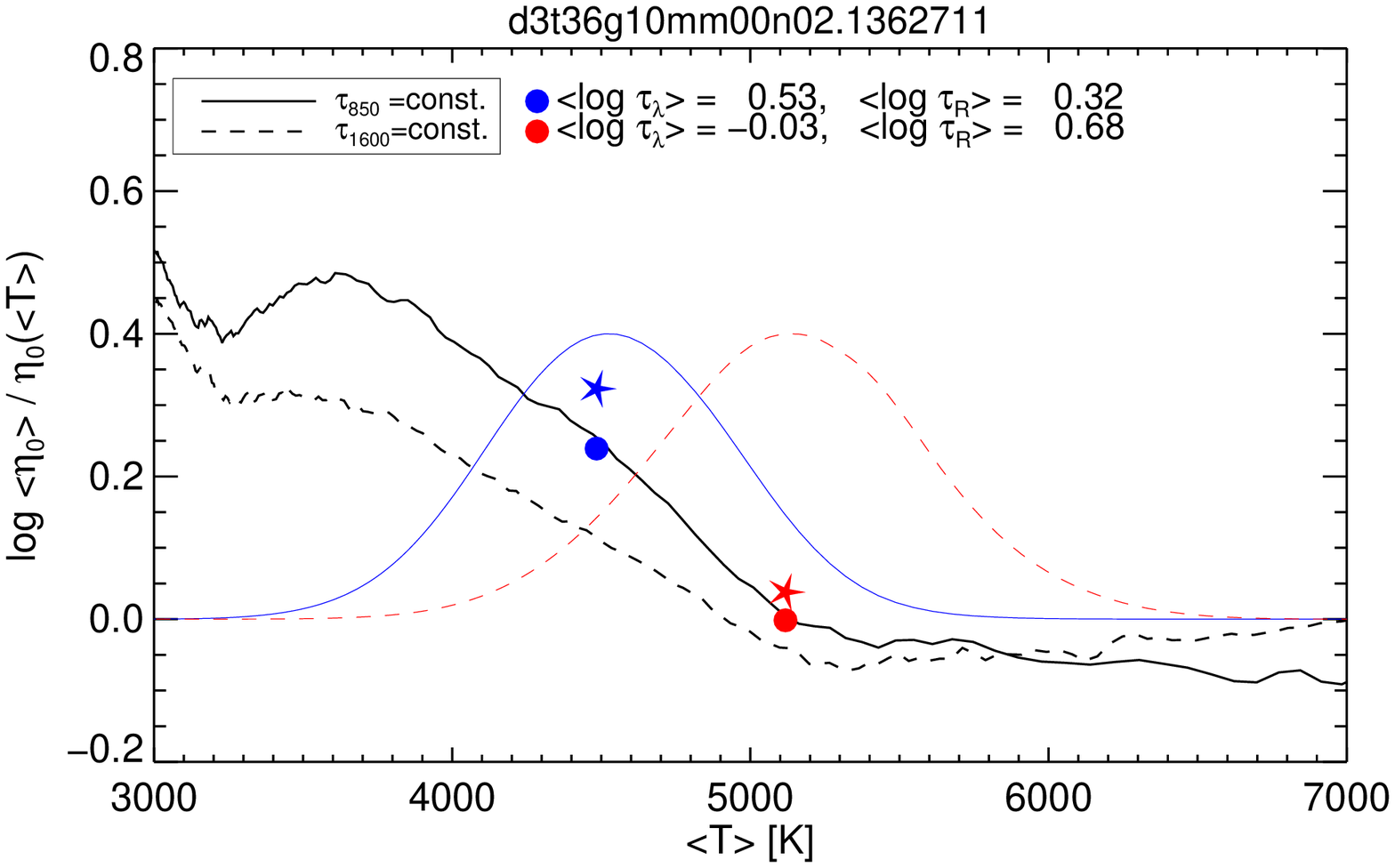}} \\[3mm]
\mbox{\includegraphics[bb=14 56 580 380, width=8.4cm]
      {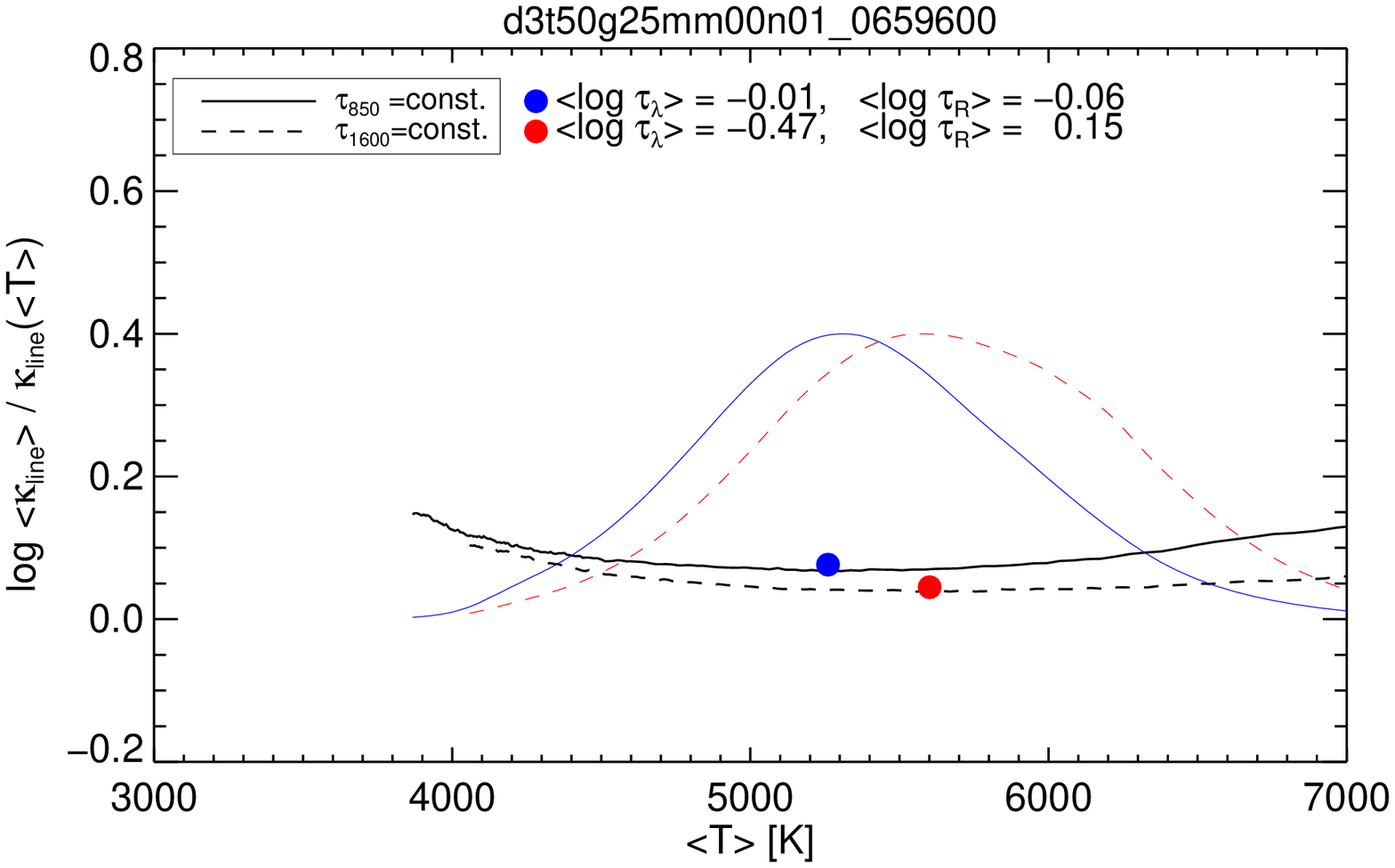}}
\mbox{\includegraphics[bb=14 56 580 380, width=8.4cm]
      {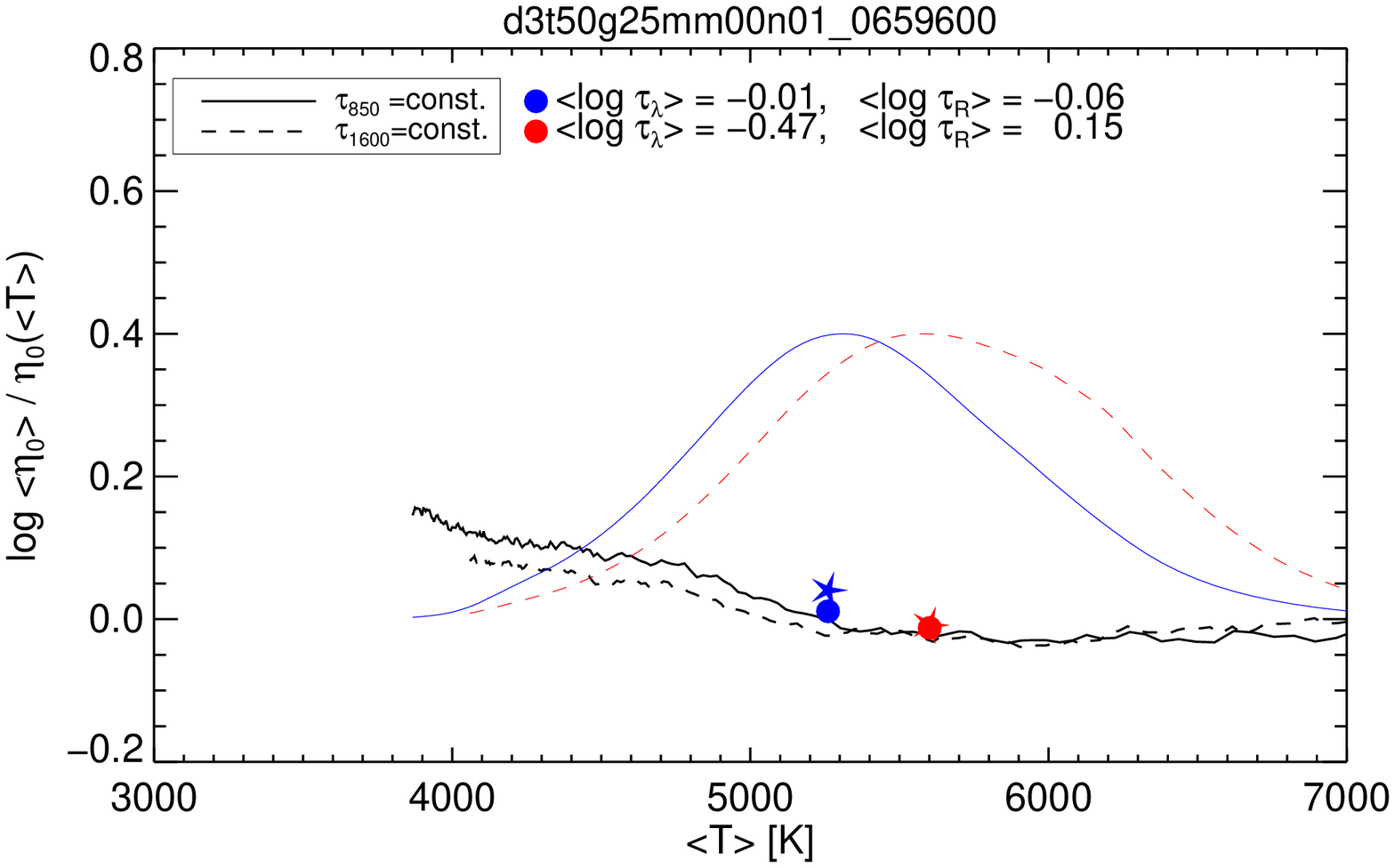}} \\[3mm]
\mbox{\includegraphics[bb=14 56 580 380, width=8.4cm]
      {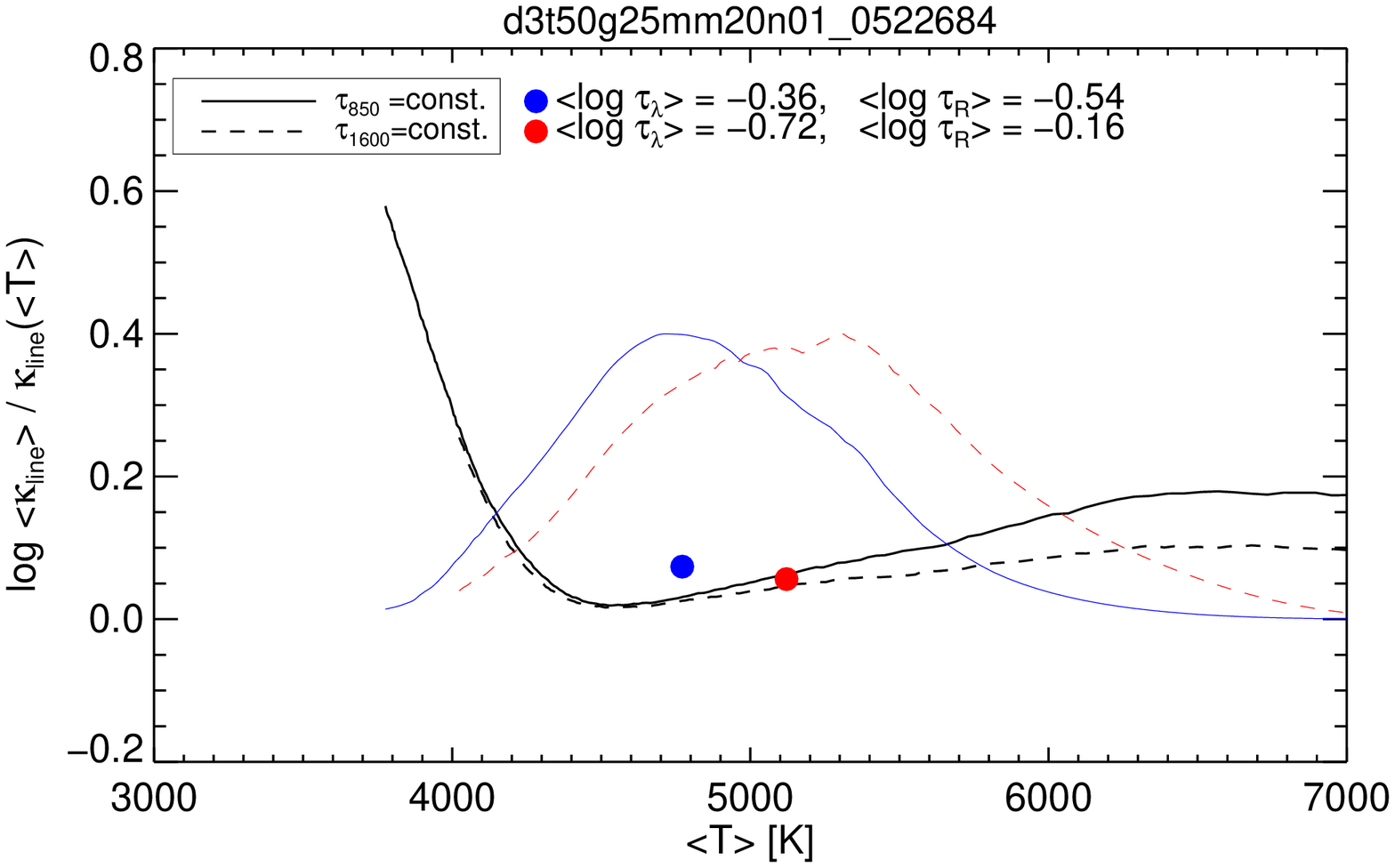}}
\mbox{\includegraphics[bb=14 56 580 380, width=8.4cm]
      {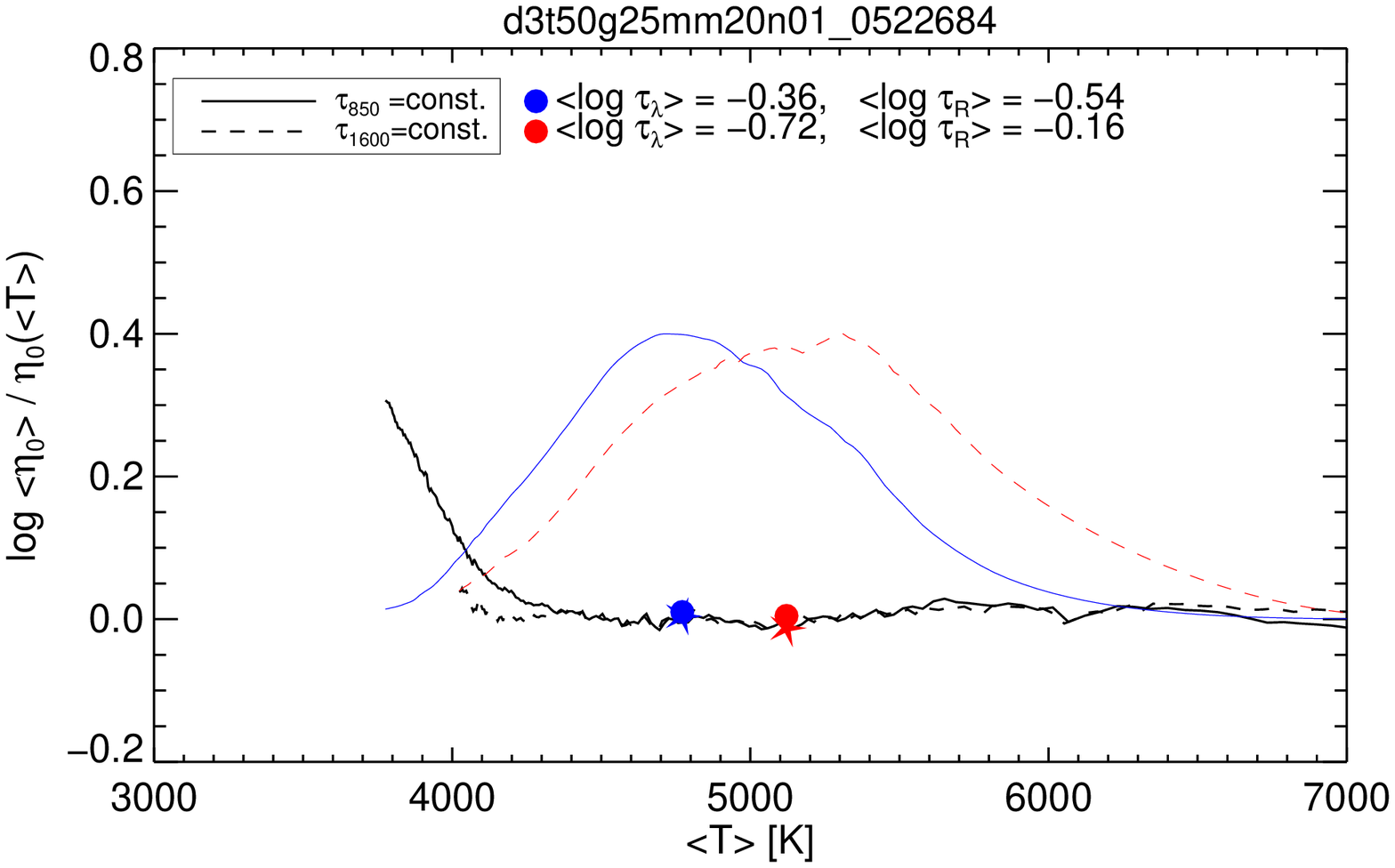}} \\[3mm]
   \caption{Depth dependence of amplification factors $\mathcal{A}_{2,3,2}$
            (left column) and $\mathcal{A}_{2,3,3}$ (right column) of
            \ion{Fe}{ii}, $\chi=6$~eV, at $\lambda\,850$ (solid) and $1600$~nm
            (dashed) for the three red giant models of
            Table\,\ref{table:height-of-line} (top to bottom). The weighting
            function $\mathcal{B}_{2,2,2}$ for both wavelengths is also shown
            (solid blue and dashed red lines). Filled circles are centered at
            $\langle T \rangle_{\rm line}$ and indicate the
            resulting effective amplification factors $\log \mathcal{R}_{2,3,2}$
            (left hand panels) and $\log \mathcal{R}_{2,3,3}$ (right hand
            panels). Star symbols indicate the values of
            $\log \mathcal{R}_{3,3,3}$ (right panels).
                       }
      \label{fig:xl_eta_tau}
\end{figure*}

According to Eq.\,(\ref{eqn:A5}), the equivalent width ratio
$\mathcal{R}_{2,3,2}$ can be expressed as an average of the amplification
factor $\mathcal{A}_{2,3,2}$ over optical depth with weighting function
$\mathcal{B}_{2,2,2}$. Figure\,\ref{fig:xl_eta_tau} (left column) shows the
depth dependence of $\mathcal{A}_{2,3,2}$ at both wavelengths for the three
red giant models under consideration, together with the weighting function
$\mathcal{B}_{2,2,2}$ and the resulting values of $\mathcal{R}_{2,3,2}$. The
largest 3D effects are found in the cool giant. The reason is twofold:
(i) the amplitude of the temperature fluctuations in the line forming layers
of the cool giant is significantly higher than in the warmer giants
(see above); (ii) the \ion{Fe}{ii} ionization fraction is strongly variable in
the atmosphere of the cool giant, especially in the layers where
the $\lambda\, 850$~nm line forms, while \ion{Fe}{ii} is always the
dominating ionization stage in the warmer giants.
Both factors enhance the nonlinearity of the line opacity fluctuations
in the cool giant, and lead to the pronounced 3D abundance corrections.

Next we consider the \emph{combined effect of line and continuum opacity
fluctuations}. For this purpose, we have to evaluate the amplification
factor $\mathcal{A}_{2,3,3}$:

\begin{equation}
\mathcal{A}_{2,3,3}(\tau_{\rm c}) = \frac{ \mathcal{B}_{2,3,3}(\tau_{\rm c})}
                                    { \mathcal{B}_{2,2,2}(\tau_{\rm c})}
                             = \left\langle\frac{\eta_{\ell,3}(\tau_{\rm c})}
                                                {\eta_{\ell,2}(\tau_{\rm c})}
                               \right\rangle_{x,y}
                     = \frac{\left\langle\eta_\ell(T)\right\rangle_{x,y}}
                            {\eta_\ell\left(\langle T\rangle_{x,y}\right)}\, ,
\end{equation}
where $\eta_\ell = \kappa_\ell/\kappa_{\rm c}$ is the ratio of line to
continuum opacity. Again, the effective amplification factor
 $\mathcal{R}_{2,3,3}$ is obtained as the average of the local amplification
factor $\mathcal{A}_{2,3,3}$ over optical depth with weighting function
$\mathcal{B}_{2,2,2}$. The result is shown in Figure\,\ref{fig:xl_eta_tau}
(right column). Obviously, the line and the continuum opacity are
correlated in the sense that both increase with temperature, such that
the fluctuation amplitude of their ratio,
$\eta_\ell$, is always lower than the fluctuation amplitude of the line
opacity, $\kappa_\ell$. Hence, $\mathcal{R}_{2,3,3}$ is smaller than
$\mathcal{R}_{2,3,2}$. For the warmer giants, the reduction of $\mathcal{R}$
is of similar magnitude for both wavelengths,
$\log (\mathcal{R}_{2,3,3}/\mathcal{R}_{2,3,2}) \approx -0.05$~dex.
For the cool giant, however, the reduction factor is much smaller
at $\lambda\,850$~nm ($\approx -0.01$~dex) than at
$\lambda\,1600$~nm ($\approx -0.11$~dex), further increasing the difference
between the 3D corrections derived for the two wavelengths.

The similarity of $\mathcal{R}_{2,3,2}$ and
$\mathcal{R}_{2,3,3}$ for the $\lambda\,850$~nm line in the cool giant
is implied by the similarity of the local amplification factors
 $\mathcal{A}_{2,3,2}$ and $\mathcal{A}_{2,3,3}$ in the line forming
region. The reason for this behavior can be traced back to the low
temperature sensitivity of the continuum opacity in the temperature range
$4000$~K $< T < 5000$~K (cf.\ Fig. B.2. in Paper~II), which coincides
with the formation region of this line (see blue weighting function
in the upper panels of Fig.\,\ref{fig:xl_eta_tau}).
For the warmer giant with solar metallicity, a similar `plateau' in
the $\kappa_{\rm c}$ versus $T$ relation is found in the same temperature
range. However, in this giant the $\lambda\,850$~nm line forms at higher
temperatures (see blue weighting function in the middle panels of
Fig.\,\ref{fig:xl_eta_tau}), and the temperature sensitivity of
$\kappa_{\rm c}$ is thus comparable at both wavelengths. Remarkably,
the `plateau' in the $\kappa_{\rm c}$ versus $T$ relation is not seen in
 the metal-poor giants. Presumably, the `plateau' is related to the
transition from a hydrogen to a metal dominated electron source, the latter
being negligible in a metal-poor atmosphere.

Finally, we have also evaluated the full 3D corrections $\mathcal{R}_{3,3,3}$,
marked by star symbols in  Fig.\,\ref{fig:xl_eta_tau} (right panels).
The additional effect of the fluctuations of the source function
gradient $u_{\rm c}$ is only significant in the cool giant, and
increases the dichotomy between the corrections at $\lambda\,850$ and
$16000$~nm even further. A similar behavior was found for the
\ion{Fe}{ii} line with $\chi=10$~eV investigated in Paper~II.

In summary, the 3D corrections for high-excitation lines of
ions at wavelengths near $\lambda\,850$~nm are significantly larger
in the cool tip RGB giant discussed in Paper~II than in the warmer giants
near the base of the RGB studied in the present work because, in the cool
giant, the lines form in a region where
(i) the temperature fluctuations are larger, (ii) the temperature sensitivity
of the continuum opacity is particularly low, and (iii) the ionization
fraction of typical ions like \ion{Fe}{ii} varies considerably, thus enhancing
the temperature sensitivity of the line opacity.

\section{Formation of molecular lines}
\label{sect:AC}

\begin{figure}[tb]
\centering
\mbox{\includegraphics[width=8.5cm,clip=true]{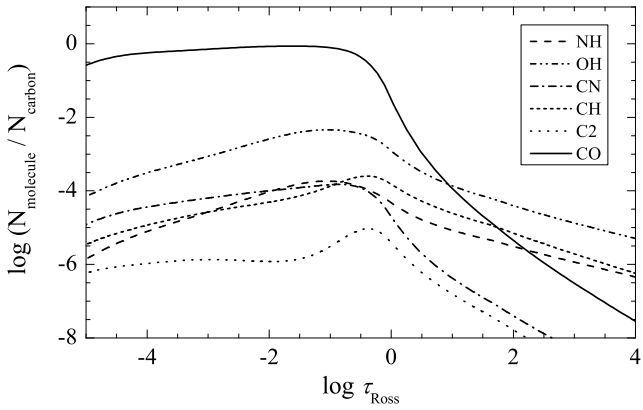}}
\caption{Number density of different molecules, normalized to the total
number density of carbon, plotted as a function of the Rosseland optical depth in the 1D LHD model at $\moh=-3.0$.
\label{fig:frac-molec}}
\end{figure}

\begin{figure*}[tb]
\centering
\mbox{\includegraphics[width=8.5cm,clip=true]{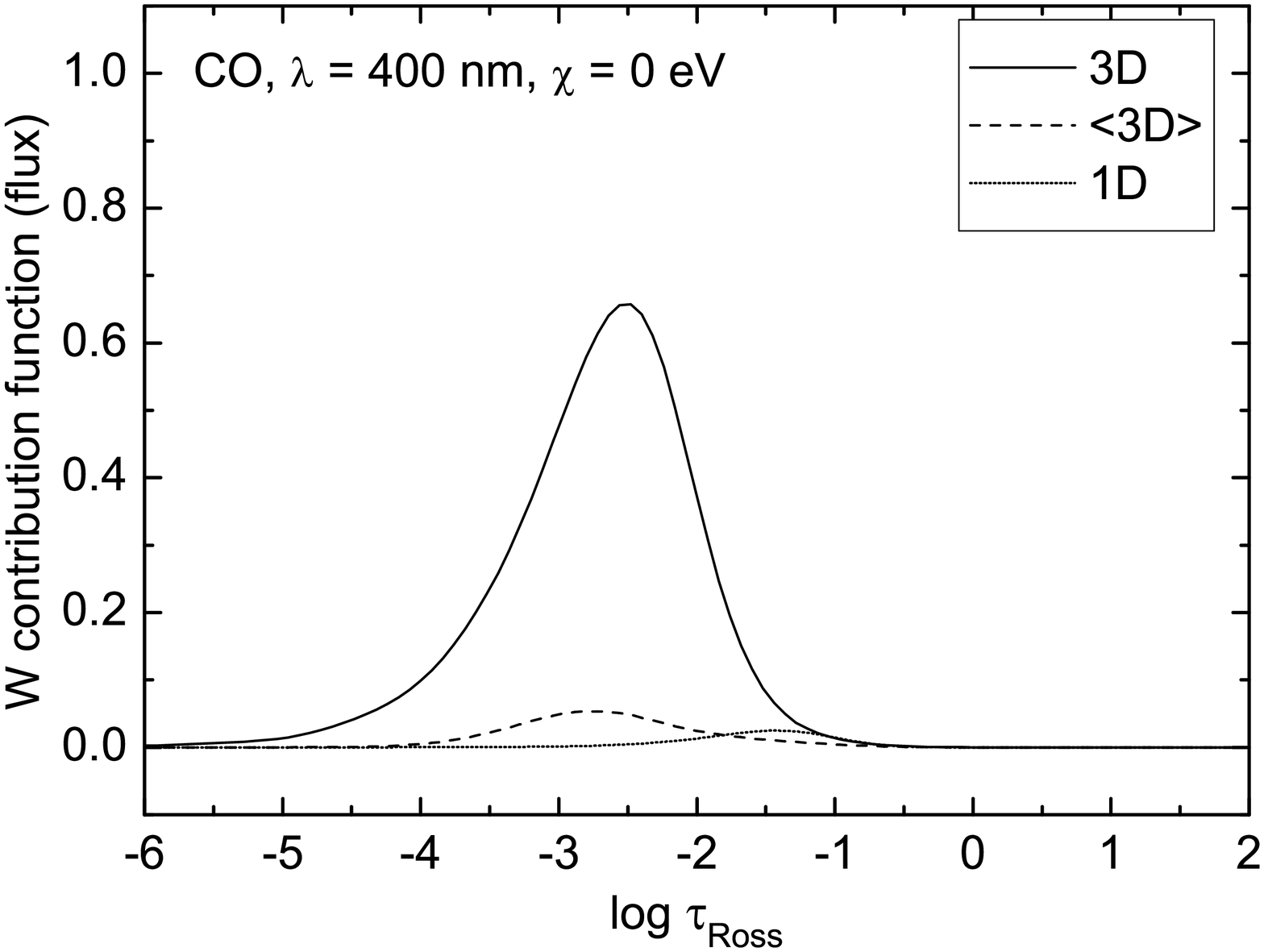}}
\mbox{\includegraphics[width=8.5cm,clip=true]{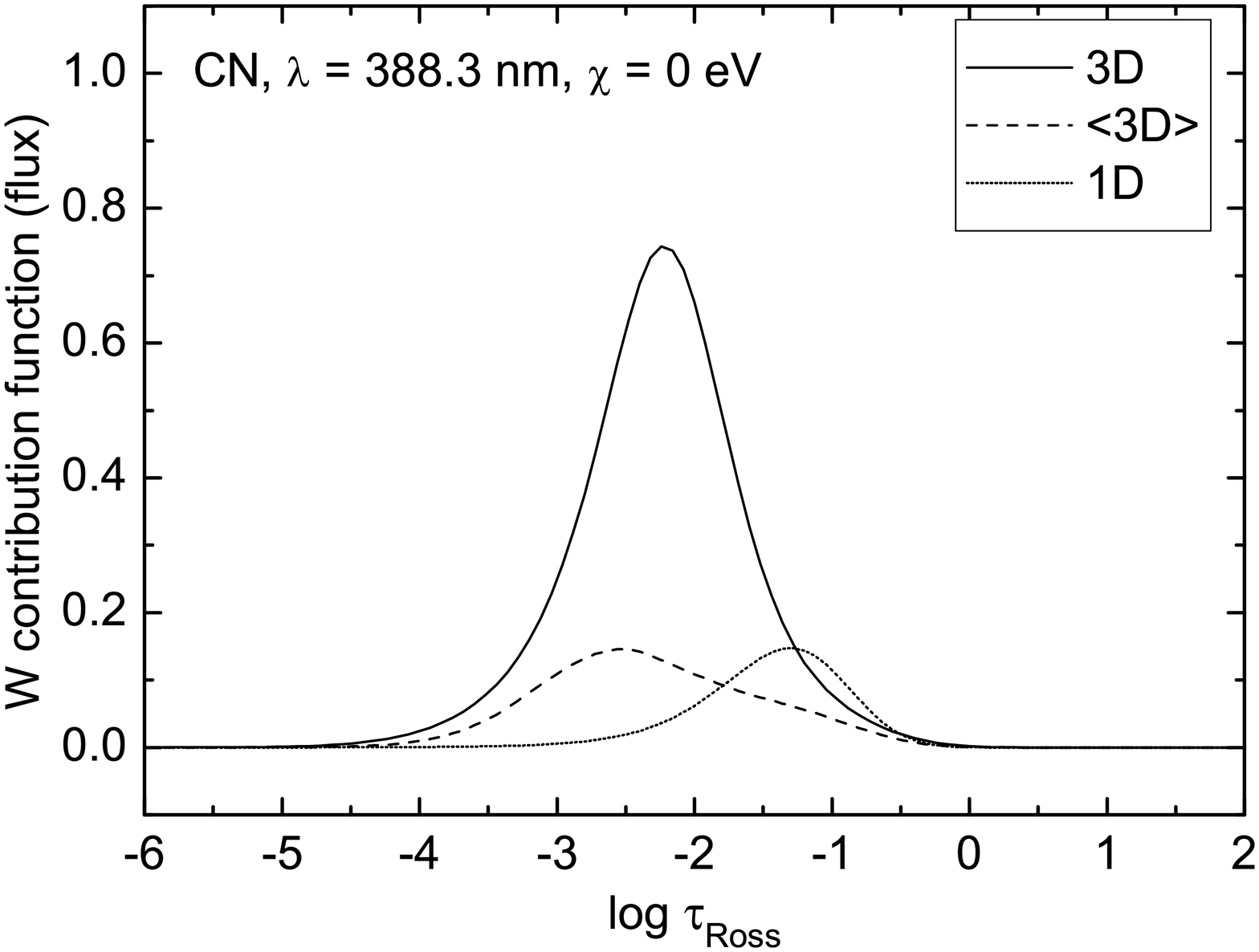}}
\mbox{\includegraphics[width=8.5cm,clip=true]{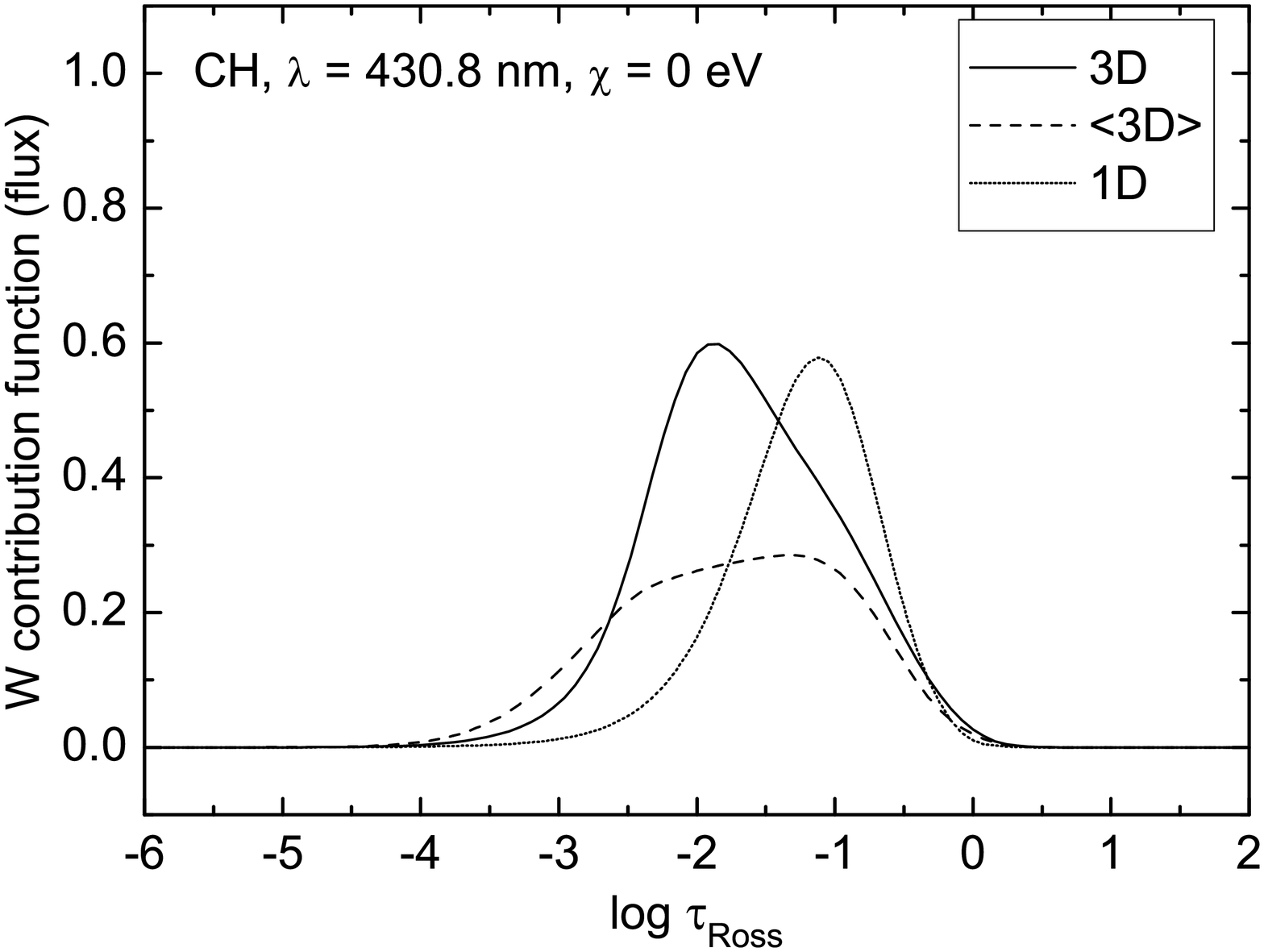}}
\mbox{\includegraphics[width=8.5cm,clip=true]{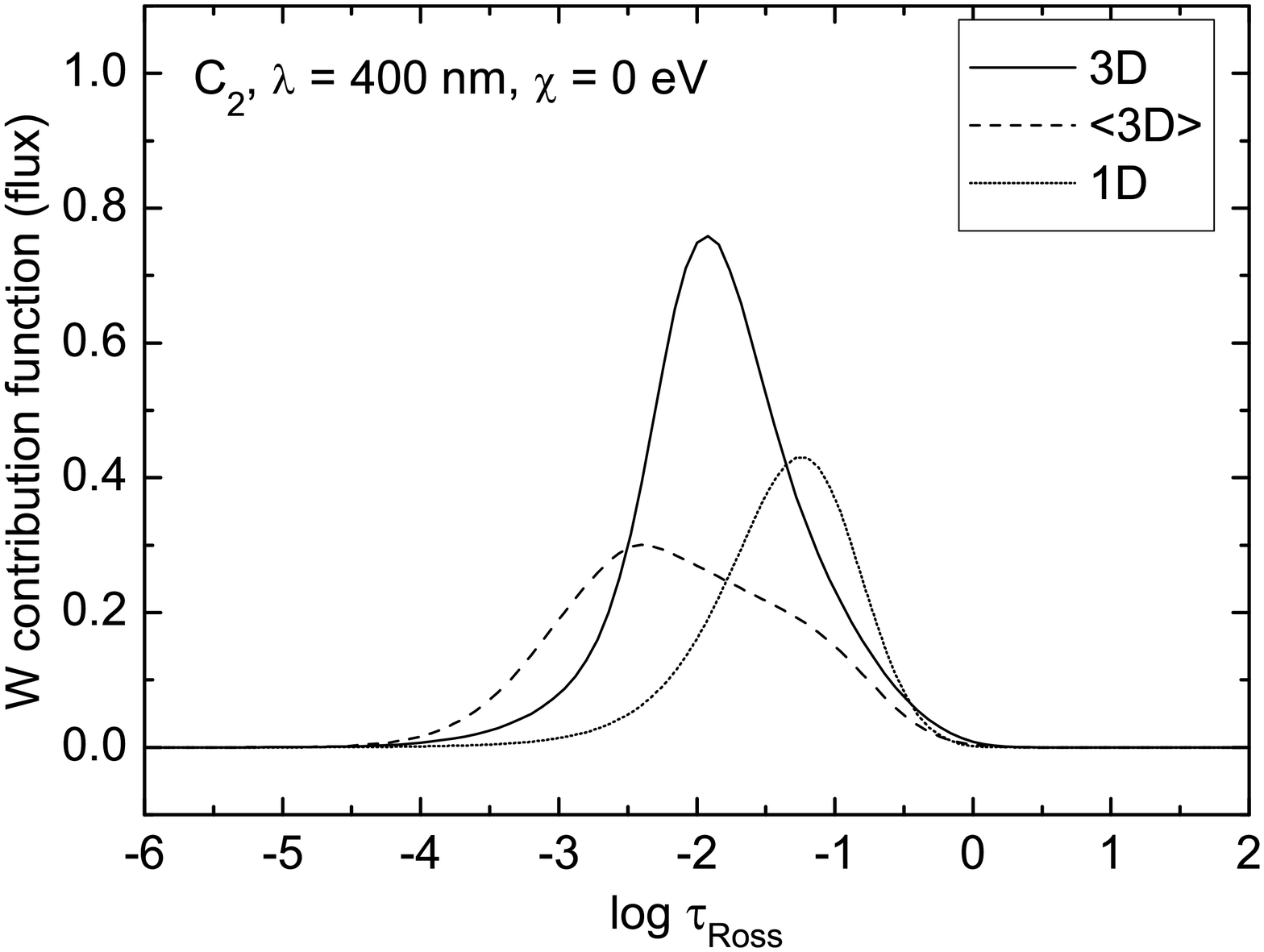}}
\mbox{\includegraphics[width=8.5cm,clip=true]{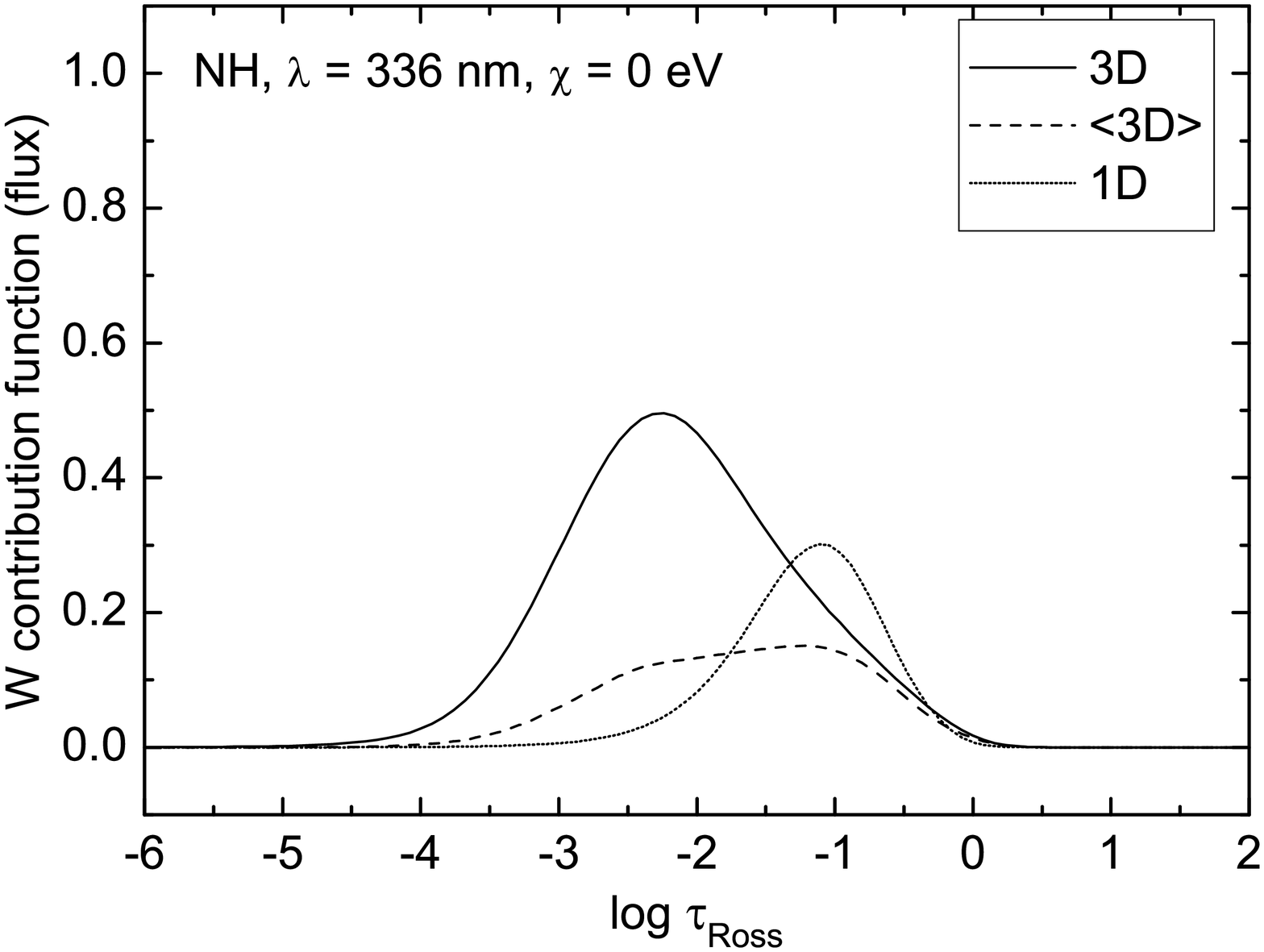}}
\mbox{\includegraphics[width=8.5cm,clip=true]{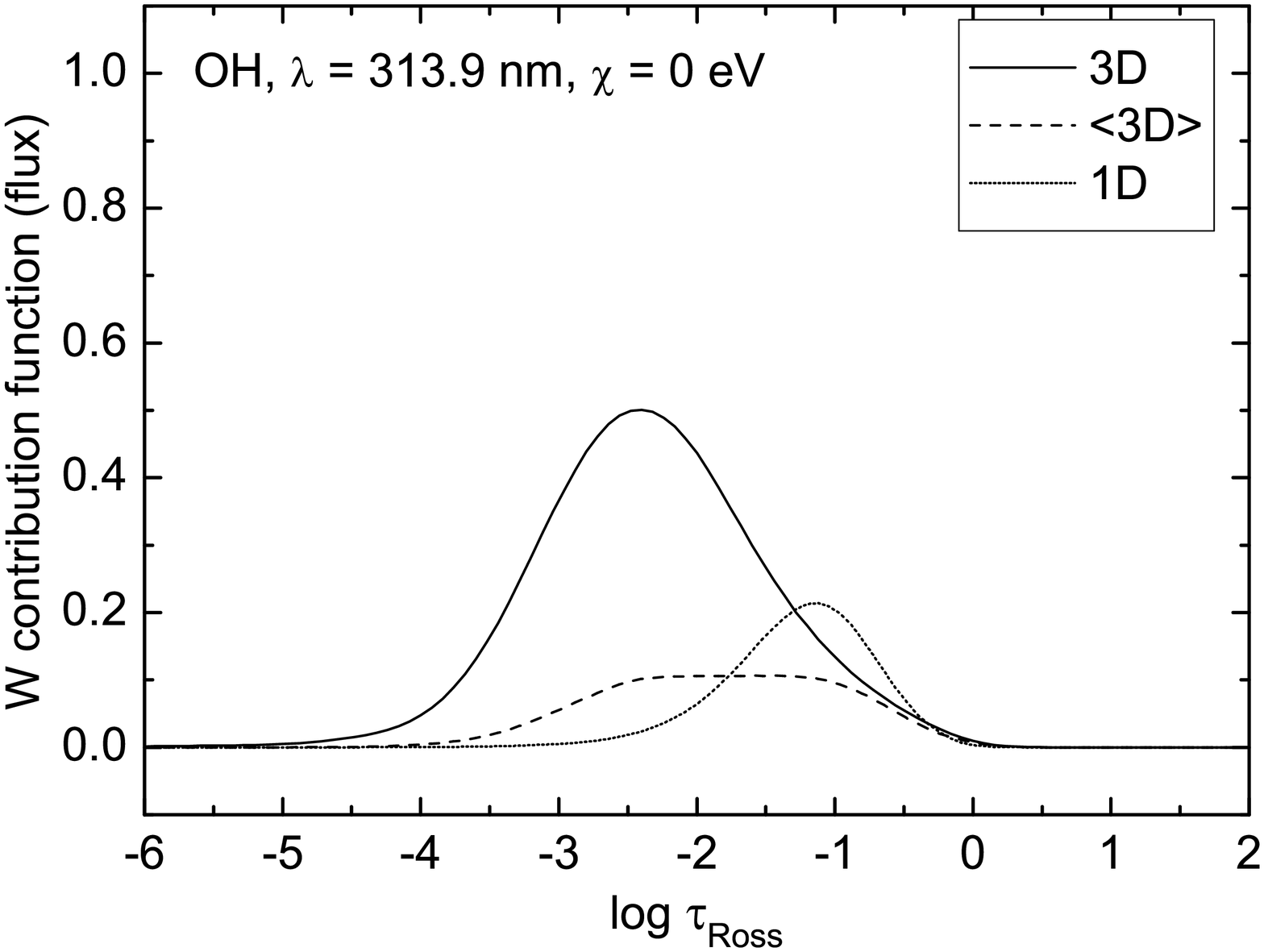}}
\caption{Equivalent width contribution function (flux) of weak ($W \leq 0.5$\,pm) artificial
molecular lines of CO, CN, CH, C$_2$, NH, OH (top left to bottom right) with
lower level excitation potential $\chi=0$\,eV in the full 3D, average
$\xtmean{\mbox{3D}}$, and 1D model atmospheres with metallicity $\moh=-3.0$.
 Contribution functions were normalized to the equivalent width of the line computed using 3D models.
\label{fig:cf-molec}}
\end{figure*}

Horizontal temperature fluctuations play an essential role in defining
abundance corrections for the molecules. Firstly, they are larger in
the 3D hydrodynamical models at lower metallicity
(cf. Fig.~\ref{fig:Ttaumm00}--\ref{fig:Ttaumm30}), which leads to the
highest molecular number density and line opacity fluctuations in the
lowest metallicity models. Secondly, in a model of a given metallicity
different molecules form at different optical depths which are
characterized by different amplitude of the horizontal temperature
fluctuations. For example, CO formation occupies the outermost range
in terms of Rosseland optical depth of all molecules studied here
(Fig.~\ref{fig:cf-molec}). Consequently, its lines form in the
atmospheric layers characterized by the largest horizontal temperature
fluctuations (cf. Fig.~\ref{fig:Ttaumm30}) in comparison to
those in the layers where other molecules form. The fluctuations of
the line opacity is therefore largest in the case of CO in the 3D model
at $\moh=-3.0$, which is part of the explanation of the large abundance
corrections for CO. However, further factors are important.

According to Paper~II, the number density of molecules per
unit mass (calculated as $X_i = N_i/\rho$, here $\rho$ is the mass
density) can be computed using the following Saha-like equation

\beq
X_{AB} = \rho\,\,Q_{AB}(T)\,\frac{X_A}{U_A(T)}\,\frac{X_B}{U_B(T)}\,
\left(\frac{h^2}{2\pi\, m_{AB}\, k T}\right)^{3/2}\,\mathrm{e}^{~D_0/kT}\, .
\label{eqn:C1}
\eeq

\noindent where $Q_{AB}$ is partition function of the molecule $AB$,
$X_A$ and $X_B$ are number densities per unit mass of the constituent
species $A$ and $B$, with their partition functions $U_A$ and $U_B$,
respectively, mass of the molecule $m_{AB}$, its dissociation energy
$D_0$, and temperature, $T$. The number densities of the molecules
studied here are shown in Fig.~\ref{fig:frac-molec} plotted versus
Rosseland optical depth in the 1D \LHD\ model at $\moh=-3.0$ (number
densities are normalized to the total number density of carbon nuclei,
$X_{AB}/\sum X_{\rm C}$). Qualitatively, the situation is very similar
to that in the cool red giant studied by Paper~II: because
the abundance of oxygen is higher than that of carbon, most of the
available C is locked into CO beyond $\log \tau_{\rm
  Ross}\sim-0.5$. In the deeper atmosphere, the molecules are quickly
destroyed due to dissociation.

\begin{figure*}[tb]
\centering
\mbox{\includegraphics[width=8.5cm,clip=true]{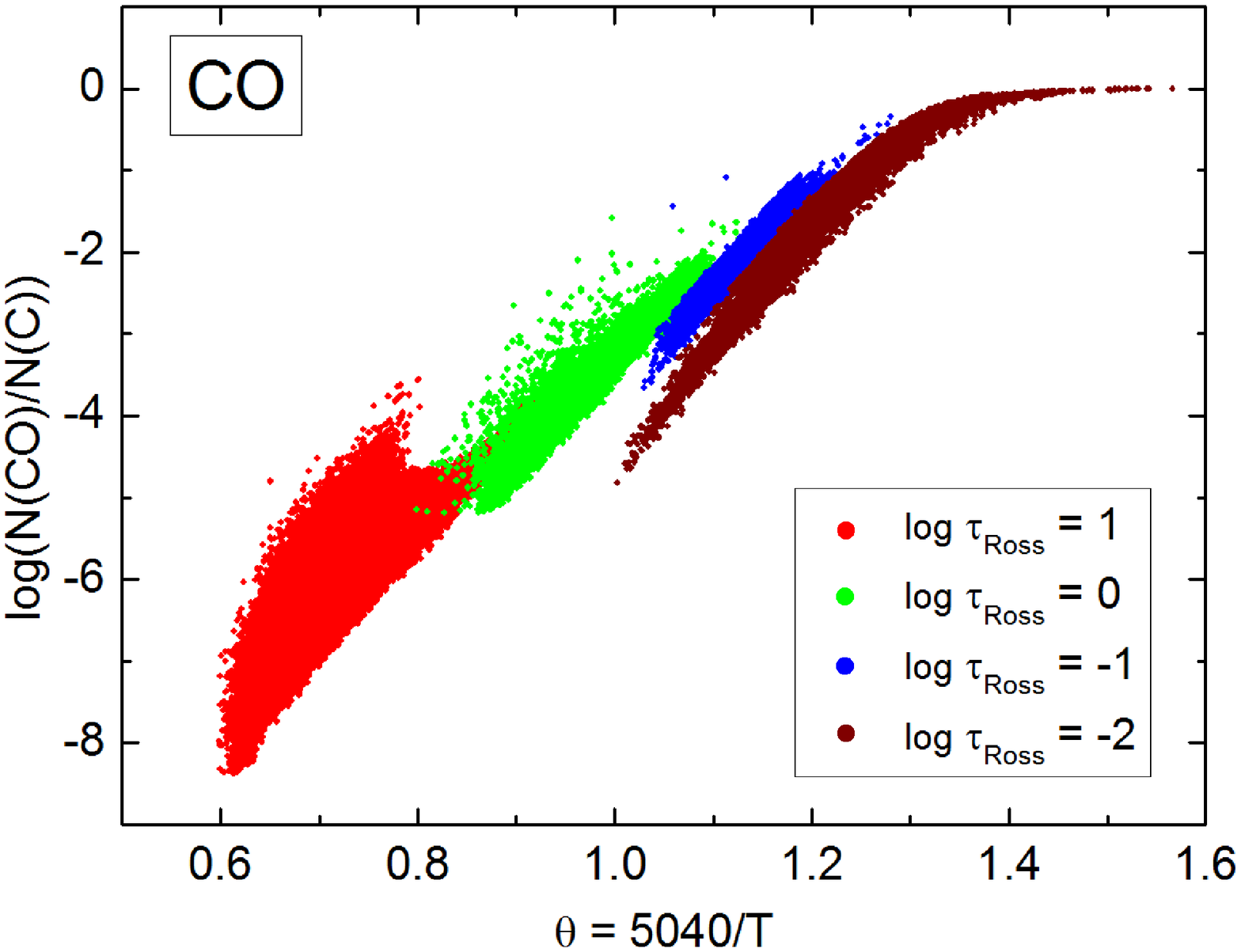}}
\mbox{\includegraphics[width=8.5cm,clip=true]{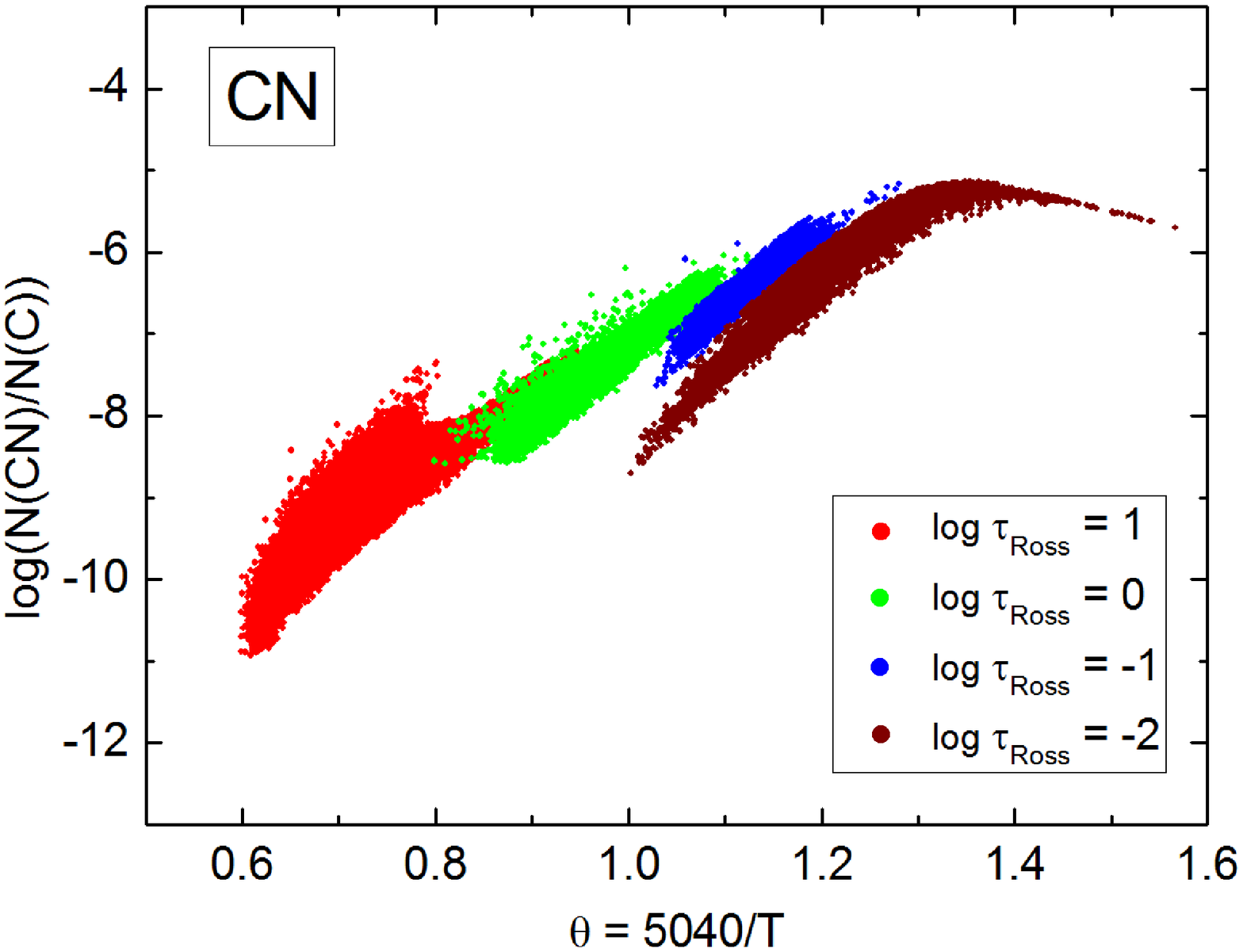}}
\mbox{\includegraphics[width=8.5cm,clip=true]{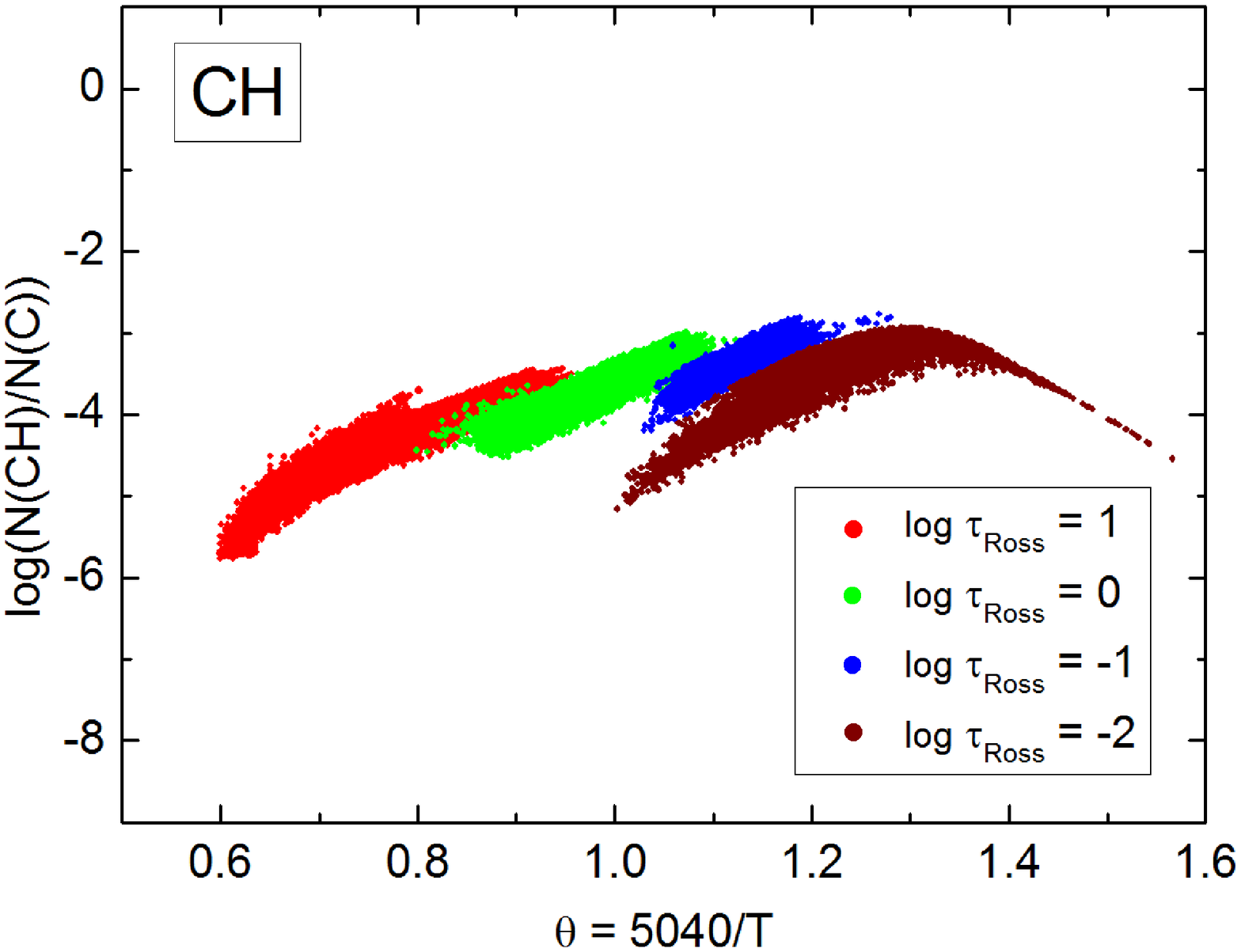}}
\mbox{\includegraphics[width=8.5cm,clip=true]{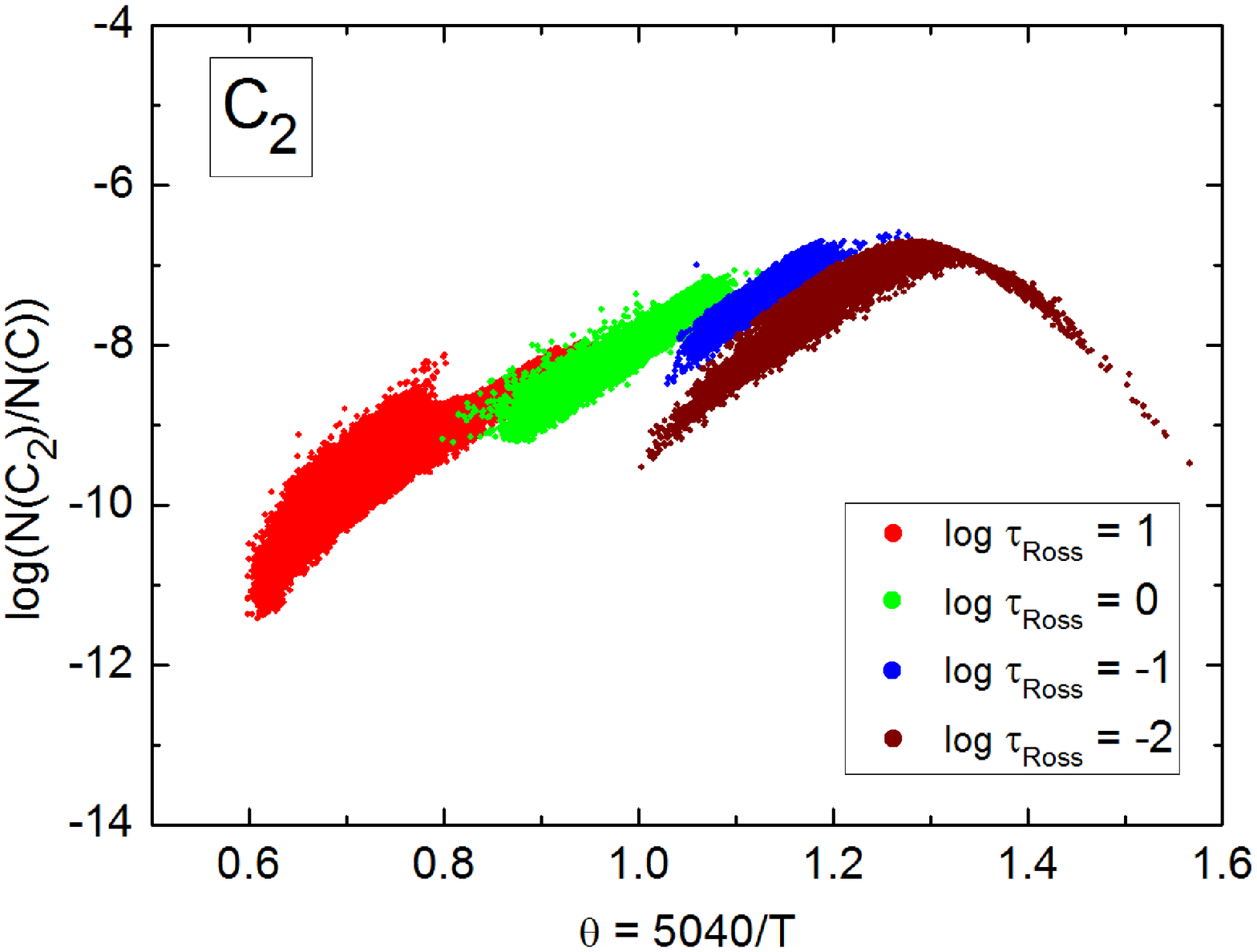}}
\mbox{\includegraphics[width=8.5cm,clip=true]{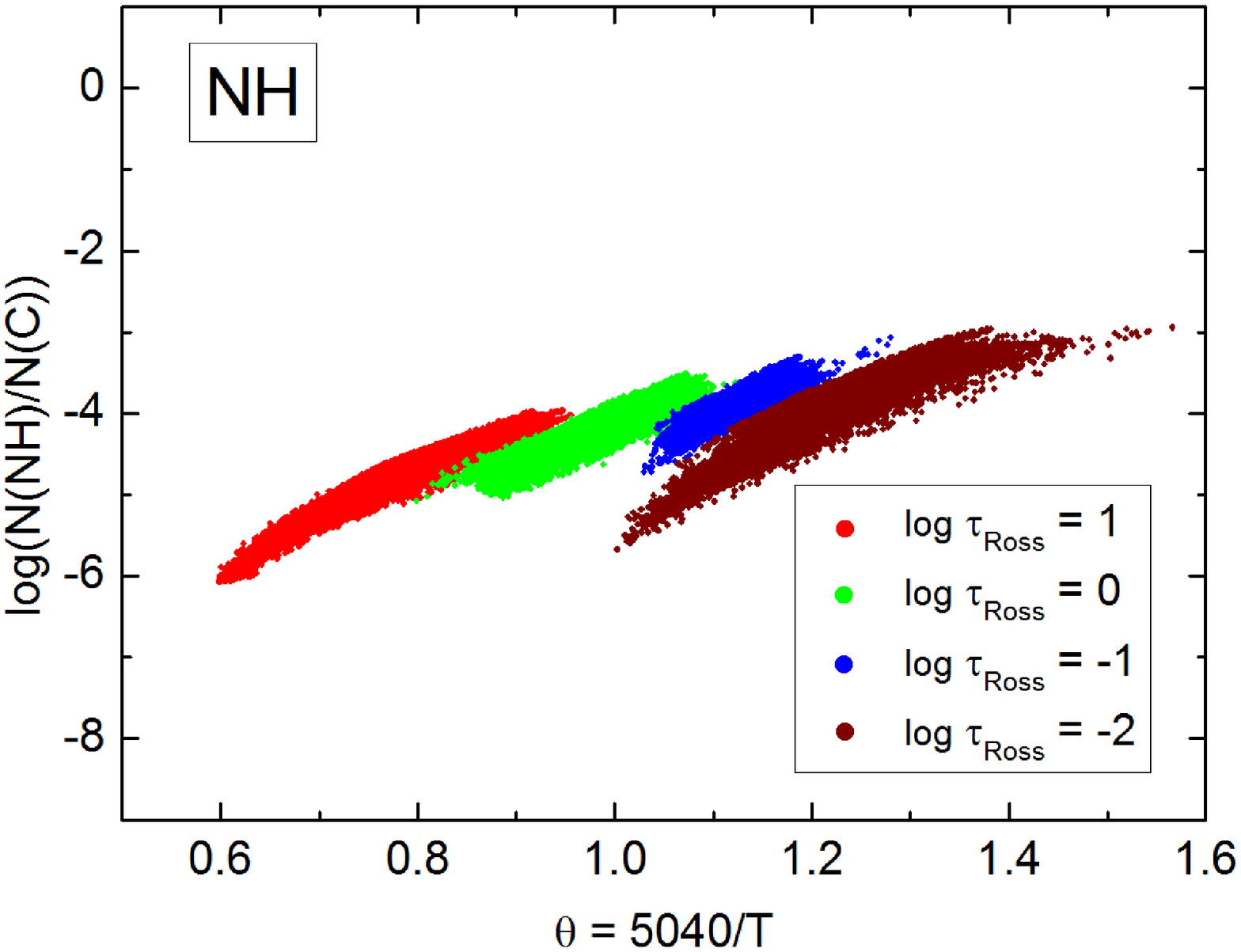}}
\mbox{\includegraphics[width=8.5cm,clip=true]{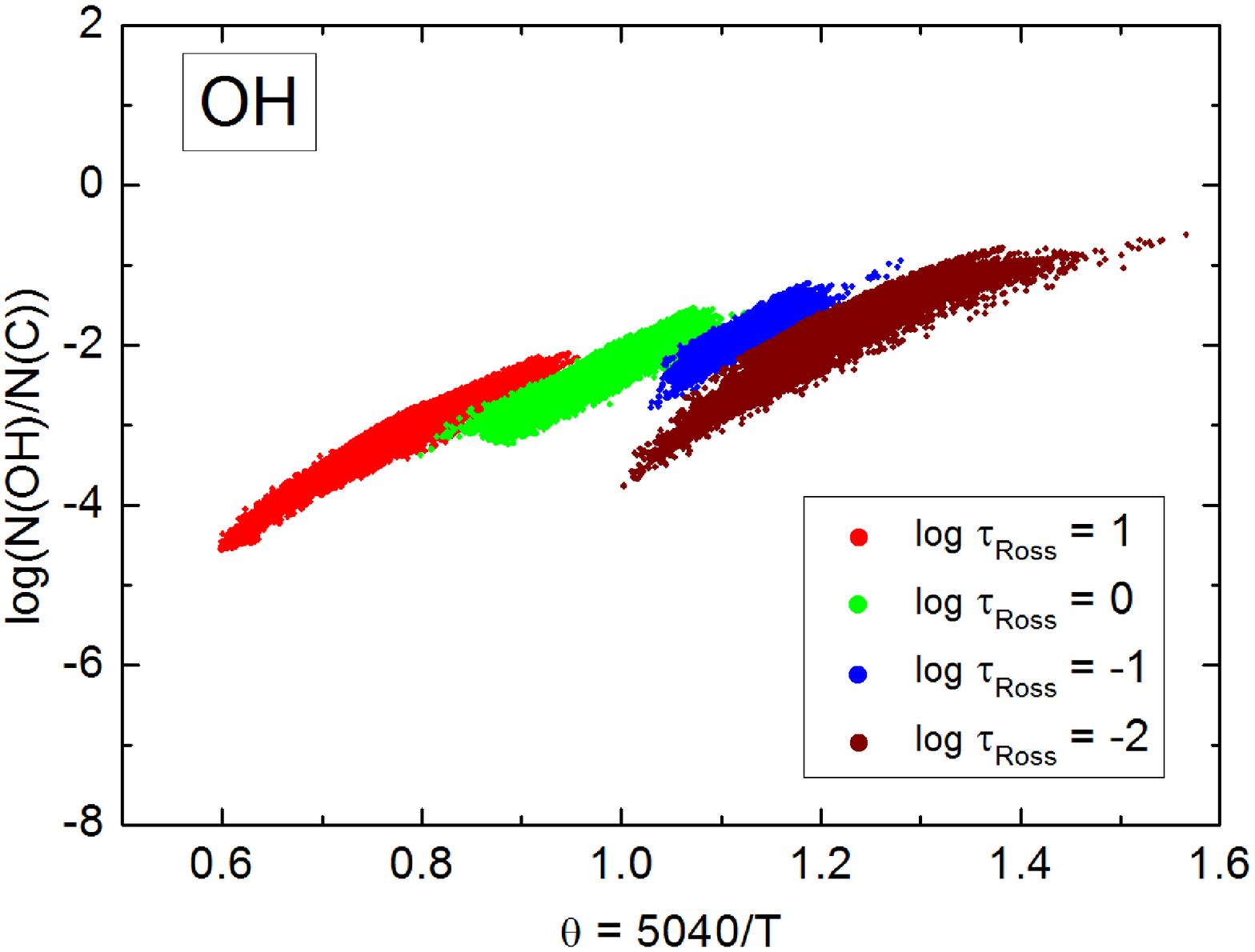}}
\caption{Normalized number densities of molecules CO, CN, CH, C$_2$, NH, OH
(top left to bottom right) in the 3D model atmosphere ($\moh=-3.0$)
plotted at different levels of constant Rosseland optical depth
(identified by color) as a function of inverse temperature $\theta = 5040/T$.
\label{fig:nd-molec}}
\end{figure*}

From Eq.~(\ref{eqn:C1}) one can expect rather different temperature
sensitivity of the number densities of different molecules. Indeed,
this is clearly seen in Fig.~\ref{fig:nd-molec} where we show number
(normalized) densities of CO, CN, CH, C$_2$, NH and OH in the 3D model
($\moh=-3.0$) at different levels of constant Rosseland optical depths,
plotted versus inverse temperature $\theta=5040/T$. It is obvious
that CO is significantly more sensitive to changes in $T$ than the
other molecules, essentially due to the high dissociation energy of
CO, $D_0=11.1$\,eV.
In fact, the temperature sensitivity of the molecular number densities
scales with $D_0$ in the hotter parts of the atmosphere where $T \ga 4000$~K.
In the cooler regions, however, where almost all carbon is locked up in CO
and atomic carbon is a minority species, the temperature dependence of
the number density of the other carbon-bearing molecules is reversed:
here a temperature increase leads to the destruction of
CO, and in turn, to a larger concentration of atomic carbon that enables
an enhanced formation of carbon-bearing molecules. Carbon-free molecules
do not show this anomalous behavior at low temperature (see lower panels
of Fig.~\ref{fig:nd-molec}).

Since the line opacity is directly proportional to $X_{AB}$,
larger fluctuations of the number densities should lead to larger
fluctuations and larger nonlinearities in the line opacities, and thus
to larger abundance corrections. This is basically what one sees in case of
CO: the broadest range of fluctuations in the case of CO
leads to the largest abundance corrections amongst all molecules
investigated here.

Interestingly, in the case of cool red giant studied in
Paper~II, the abundance corrections were largest for C$_2$.
In contrast, the abundance corrections for C$_2$ obtained in this
work are in fact smallest amongst all molecules. A closer look at the
dependence of C$_2$ number density on temperature reveals that
$X_{\rm C2}$ increases steeply up to $T \sim4000$\,K and then decreases
towards the higher temperatures (Fig.~\ref{fig:nd-molec}, middle right).
This is because the formation of C$_2$ is tightly coupled to that of CO,
as explained above, and makes the number density of C$_2$ extremely sensitive
to changes in temperature at low $T$. In the cool red giant analyzed by
Paper~II the atmosphere is significantly cooler and the C$_2$
line forms entirely the lower temperature regime where $X($C$_2)$ is highly
sensitive to the temperature fluctuations, leading to the largest abundance
corrections for C$_2$. In the warmer red giant studied here, however,
the C$_2$ line forms in the transition region between both temperature
regimes, such that the amplitude of the fluctuations is limited and their
net effect on the abundance corrections is much smaller.

NH ($D_0=3.5$~eV) and OH ($D_0=4.4$~eV)
have similar molecular properties and line
formation regions. The 3D abundance correction is slightly more negative
for OH because this molecule has a somewhat higher dissociation energy.
The correction for CN ($D_0=7.8$~eV) falls in the same range, although this
molecule has a much higher dissociation energy.
Again, the explanation for this unexpected behavior is related to the
change of sign in the temperature sensitivity of the number density of CN
in the line forming region ($\log \tau_{\rm Ross} = -2$), due to the coupling
with CO at low temperatures. The coupling with CO is missing in the case of
carbon-free molecules NH and OH, where the slope
$\partial{\log N}/\partial{\theta}$ is therefore essentially
constant over the whole temperature range (see Fig.~\ref{fig:nd-molec},
lower panels).


\end{appendix}

\end{document}